\documentclass[aps,twocolumn,superscriptaddress]{revtex4-1}
\usepackage{amsmath}
\usepackage{amssymb}
\usepackage{graphicx}
\usepackage{dcolumn}
\usepackage{bm}
\usepackage{textcomp}
\usepackage{booktabs}
\usepackage{float}
\usepackage{hhline}
\usepackage{braket}
\usepackage[dvipsnames]{xcolor}
\usepackage{filecontents}

\graphicspath{{figures/}}

\usepackage{xr}

\begin{document}

\title[]{Influence of non-adiabatic effects on linear absorption spectra in the condensed phase: Methylene blue}

\author{Angus J. Dunnett}
\affiliation{Sorbonne Universit\'{e}, CNRS, Institut des NanoSciences de Paris, 4 place Jussieu, 75005 Paris, France}

\author{Duncan Gowland}
\affiliation{Department of Physics, King's College London, London WC2R 2LS, United Kingdom}

\author{Christine M. Isborn}
\affiliation{Chemistry and Chemical Biology, University of California Merced, Merced, CA 95343, USA}

\author{Alex W. Chin}
\affiliation{Sorbonne Universit\'{e}, CNRS, Institut des NanoSciences de Paris, 4 place Jussieu, 75005 Paris, France}

\author{Tim J. Zuehlsdorff}
\email{zuehlsdt@oregonstate.edu}
\affiliation{Department of Chemistry, Oregon State University, Corvallis, Oregon 97331, USA}

\begin{abstract}
  Modeling linear absorption spectra of solvated chromophores is highly challenging as contributions are present both from coupling of the electronic states to nuclear vibrations and solute-solvent interactions. In systems where excited states intersect in the Condon region, significant \emph{non-adiabatic} contributions to absorption lineshapes can also be observed. Here, we introduce a robust approach to model linear absorption spectra accounting for both environmental and non-adiabatic effects from first principles. This model parameterizes a linear vibronic coupling (LVC) Hamiltonian directly from energy gap fluctuations calculated along molecular dynamics (MD) trajectories of the chromophore in solution, accounting for both anharmonicity in the potential and direct solute-solvent interactions. The resulting system dynamics described by the LVC Hamiltonian are solved exactly using the thermalized time-evolving density operator with orthogonal polynomials  algorithm (T-TEDOPA). The approach is applied to the linear absorption spectrum of methylene blue (MB) in water. We show that the strong shoulder in the experimental spectrum is caused by vibrationally driven population transfer between the bright S$_1$ and the dark S$_2$ state. The treatment of the solvent environment is one of many factors which strongly influences the population transfer and lineshape; accurate modeling can only be achieved through the use of explicit quantum mechanical solvation. The efficiency of T-TEDOPA, combined with LVC Hamiltonian parameterizations from MD, leads to an attractive method for describing a large variety of systems in complex environments from first principles.
\end{abstract}

\date{\today}

\maketitle

\section{Introduction}

\noindent
Spectroscopy is a key step in the screening of materials and molecules for technological applications such as photovoltaics, in understanding photochemical reactions, and in the investigation of biological processes. \cite{Zucca2017,Rivera-Torrente2020} In condensed phases, the environment can have a significant influence on the spectral peak position and intensity, causing a change in the electronic and nuclear quantum states of the molecule.\cite{Marini2010, Provorse2016, Zuehlsdorff2020}  Modeling the effect of the environment on a molecule in an accurate and computationally affordable way is a persistent challenge. \cite{Puzzarini2019,Zuehlsdorff2021}  Additionally, the accurate and affordable quantum treatment of nuclear dynamics and non-adiabaticity is in constant development. \cite{Zener1932, Meyer1979, Baer2002,  Curchod2018} The exact calculation of experimental spectra in the condensed phase requires a unification of accurate treatment of the non-adiabatic quantum dynamics with proper inclusion of the effects of the environment. \cite{Tavernelli2011}

In many cases, transitions involve bright, energetically well-separated states, justifying the use of the Condon approximation. Similarly, if the chromophore is relatively rigid, the harmonic approximation can be made for the shape of the initial and final state adiabatic Born-Oppenheimer potential energy surfaces (PESs).\cite{Charaf-Eddin2013, Stendardo2016} This harmonic Franck-Condon approach has been employed using time-dependent (effective for large multi-mode systems) and time-independent (sum over states, effective for smaller systems) techniques, and can be easily extended to include linear effects of the structure on transition dipole moment (Herzberg-Teller effects).\cite{Doktorov1975,Santoro_2008,Baiardi2013,DeSouza2018, Berger1998} When the environment interacts weakly with the ground and excited states, combining Franck-Condon calculations with polarizable continuum models for the solvent environment can yield highly effective environment corrected vibronic lineshapes,\cite{Ferrer2011, Avila_Ferrer2014, Cerezo2015b} and for systems with stronger coupling to the environment, some of the authors have presented combined ensemble - Franck-Condon approaches to improve lineshapes calculation. \cite{Zuehlsdorff2018b, Zuehlsdorff2018, Shedge2021, Zuehlsdorff2021}

However, there are many instances where such a Franck-Condon treatment of the absorption spectrum is insufficient. For example, if there is a region of the PES in which excited states intersect, causing a breakdown of the Born-Oppenheimer approximation, electronic states may mix, leading to non-adiabatic effects such as intensity-borrowing between states. \cite{Domcke2012, Orlandi1973} For polyatomic systems these crossings are ubiquitous, but the Condon approximation relies on such crossings being far from the initial state equilibrium, representing rare events. If such crossings are close to the region of the potential sampled by the ground state, and thus contribute to the absorption spectrum via non-adiabatic effects such as vibrationally driven population transfer, it may become effective to utilize the linear vibronic coupling (LVC) Hamiltonian to describe the system dynamics. \cite{Koppel1984} 

Historically the importance and efficacy of the LVC model in describing spectral lineshapes is well reported. \cite{Domcke1981} One seminal example is the calculation of the second singlet excitation of pyrazine,\cite{Worth1996} and a recent egregious example is the UV spectra of cyclopropane. \cite{Neville2018}  In the latter case, the authors show extreme intensity borrowing from the optically dark $A’_2$ and $A’_1$ states from the $E’$ state, completely transforming the absorption lineshape. Generally, these studies have focused on small high symmetry molecules in the gas phase, where expensive numerical approaches are both feasible and effective. \cite{Yarkony2012} The previously highlighted examples of pyrazine and cyclopropane were successfully modeled using the multiconfigurational time-dependent Hartree (MCTDH) method, which is often regarded as the gold standard in many-body approaches to non-adiabatic problems.\cite{Meyer1990} Unfortunately, these methods suffer from the ‘curse of dimensionality’, meaning inclusion of only a few nuclear and electronic degrees of freedom is feasible, and they generally utilize precalculated potential energy surfaces with high level electronic structure methods.\cite{Meyer2009} Recently, Santoro and co-workers have parameterized the LVC Hamiltonian for a dense manifold of states for larger molecules using more affordable time-dependent density-functional theory. However, these calculations still limit the number of modes in the system by constructing LVC Hamiltonians in vacuum or relying on polarizable continuum models to represent the environment. \cite{Yaghoubi_Jouybari2020, Aranda2021, Green2021}

There are a number of approaches able to circumvent the exponential scaling of many body states through the use of efficient and compact representations of wave functions, such as multilayer (ML-) MCTDH, matrix product states (MPS), and tensor network states. \cite{Wang2003,Schroter2015, Schroder2019} A tensor network-based, many body approach that has been developed to handle complex open quantum systems (OQS) problems is the time-evolving density operator with orthogonal polynomials (TEDOPA) algorithm, based on a mapping that transforms one or more environments into a 1D chain of effective oscillator modes with nearest-neighbour coupling (\textit{vide infra}) \cite{prior2010efficient,Chin2010,chin2013role}. This geometry unlocks the full power of the MPS/Tensor networks ansatz that perform optimally for 1D systems with locally short-range couplings \cite{schollwock2011density}. Indeed, a recent comparison of ML-MCTDH and MPS methods for 1D dipolar chains found that the MPS approach was both faster and required less computational memory to obtain accurate ground states for this class of models. \cite{gatti21}

TEDOPA is in principle numerically exact, can be combined with machine learning and entanglement renormalization methods, and also gives full access to observable bath dynamics, as recently verified through time-resolved excited-state vibrational spectroscopy in bipentacenes \cite{Schroder2019,schnedermann2019molecular}. However, until recently, accessing finite temperatures required time-consuming sampling over bath configurations. This limitation was overcome, first with the introduction of the thermofield approach, in which a bath of negative energy modes that act like a source of thermal energy are introduced, and later with the mapping of Tamascelli \textit{et al.} to treat a thermal environment by defining a temperature dependent spectral density, leading to so called T-TEDOPA. \cite{tamascelli2019efficient,dunnett2021matrix} These advances now allow finite temperature OQS dynamics to be extracted from many-body pure state wave function simulations at 0K - offering substantial advantages over other methods in terms of efficiency \cite{10.3389/fchem.2020.600731}. In this article we demonstrate how this new capability allows us to predict optical spectra of real-world molecules in realistic environments that can be directly compared to experiments in solution. In achieving this, we will also show how the Dirac-Frenkel variational principle applied to the TEDOPA state allows us to include long-range couplings that prove to be essential for including the energy gap~\emph{correlations} that play a determining role in the excited state dynamics and spectra. 

We apply this method to the curious case of the linear absorption spectra of the cationic methylene blue (MB) chromophore in aqueous solution. Both the molecular structure and excitation have been shown to be environmentally sensitive, and there are many open questions about the nature of solvent and aggregation of this molecule and how it influences the spectral lineshape. \cite{Ovchinnikov2016, Heger2005} Focusing on the monomeric solution, it shows a singular peak ($\lambda_\textrm{max}=$ 664 nm, 1.86 eV) with a broad, almost square, higher energy shoulder (610 nm, 2.03 eV) at half the absorption maximum intensity.\cite{Shahinyan2019} Despite its simple lineshape, previous models for this system appear incomplete.  The broad shoulder intensity has been previously both significantly under- and over- estimated depending on the electronic structure, solvent model, and vibronic coupling method of choice. \cite{Dean2015} For example, a recent study by de Queiroz \textit{et al.} indicates the S$_2$ state gains some intensity when examining vertical excitations in explicit solvent, suggesting non-Condon effects. Yet in the same study their computation of the S$_0 \rightarrow$ S$_1$ Franck-Condon spectra \textit{in vacuo} overestimates the shoulder intensity. \cite{deQueiroz2021} 

Here we expand on previous studies of MB by including explicitly quantum mechanical treatment of the solvent polarization on excitation and nuclear dynamics for vibronic lineshapes. Linear absorption spectra are computed using second and third order truncation of a cumulant expansion of the linear response function constructed from energy gap fluctuations along a molecular dynamics trajectory. We also extend the lineshape calculations to include the strong non-adiabatic effects of higher excited states by solving the LVC problem for three electronic states, finding that the S$_2$ state of MB plays a large role in determining the excited state dynamics and absorption lineshape. In contrast to previous studies parameterizing LVC Hamiltonians from electronic structure theory we show that LVC model parameters can be obtained directly from correlation functions\cite{Mavros2016} calculated along generally anharmonic molecular dynamics simulations in explicit solvent environments, capturing solute-solvent configuration and polarization effects.\cite{Zuehlsdorff2019} 

This paper is structured as such. First, we present background theory defining the linear vibronic coupling problem in the context of our solvated chromophore, methylene blue. Then we present background on the T-TEDOPA method and the novel derivation of the response function for the T-TEDOPA mapping. Next, we take special care to address the nature of the bath correlations for our solvated chromophore, and how they are addressed in the T-TEDOPA framework and tensor network calculations. The computational method for a) ab initio calculation of various linear vibronic coupling parameters of the chromophore, and b) the tensor network calculation of the T-TEDOPA dynamics are then described. In the results section we provide analysis of our chosen molecule's excitations and dynamics, establishing both the importance of considering the dark S$_{2}$ state and the efficacy of our choice of Hamiltonians. We then turn to the question of the importance of including correlation between the excited state dynamics and their influence on the spectral lineshape. From these results we identify the role of the S$_{1}$-S$_{2}$ energy gap, and so explore the range of this value obtained from electronic structure calculations. 

\section{Theory}

\noindent
In this work, we consider a three-level electronic system consisting of an electronic ground state S$_0$ and two electronic excited states, S$_1$ and S$_2$, coupled to nuclear degrees of freedom. The nuclear degrees of freedom are described through the spin-boson model (SBM) or Brownian oscillator model (BOM), which makes the assumption that ground- and excited state PESs are harmonic surfaces with the same curvature that only differ by a displacement of their respective minima.\cite{Mukamel-book, Cho08} The BOM Hamiltonian can then be written as:
\begin{equation}
  \hat{H}_\textrm{BOM}=   \begin{pmatrix}
    {H}_0 & V_{01} & V_{02} \\
    V_{10} & H_1 & 0 \\
    V_{20} & 0 & H_2 
  \end{pmatrix},
\end{equation}
where $V_{0n}=V^\dagger_{n0}$ denotes the transition dipole operator between the electronic ground and $n^{th}$ excited state. We also assume that the Condon approximation is valid, such that the dependence of the transition dipole operator on nuclear degrees of freedom can be ignored  and $V_{0n}(\boldsymbol{\hat{q}})\approx V_{0n}$. The nuclear Hamiltonian of the electronic ground state is denoted $H_0(\boldsymbol{\hat{q}})$, and in the BOM the nuclear Hamiltonians for a system with $N$ vibrational modes can be written as:
\begin{eqnarray}
  H_0(\boldsymbol{\hat{p}},\boldsymbol{\hat{q}})&=&\sum_j^N  \left(\frac{\hat{p}_j^2}{2}+\frac{1}{2}\omega^2_j \hat{q}^2_j \right) \\
  H_1(\boldsymbol{\hat{p}},\boldsymbol{\hat{q}})&=&\sum_j^N  \left(\frac{\hat{p}_j^2}{2}+\frac{1}{2}\omega^2_j \left(\hat{q}_j-K_j^{\{1\}}\right)^2 \right)+\Delta_{01} \\
  H_2(\boldsymbol{\hat{p}},\boldsymbol{\hat{q}})&=&\sum_j^N  \left(\frac{\hat{p}_j^2}{2}+\frac{1}{2}\omega^2_j \left(\hat{q}_j-K_j^{\{2\}}\right)^2 \right) +\Delta_{02}
\end{eqnarray}
where $\Delta_{01}$ and $\Delta_{02}$ are the adiabatic energy gaps between the ground and first electronic excited state, and ground and second excited state, respectively, and atomic units are used throughout.  $\textbf{K}^{\{1\}}$ denotes the displacement vector of the minimum of the first electronic excited state relative to the electronic ground state, and $\omega_j$ is the angular frequency of mode $j$. For the simple BOM Hamiltonian of a three-level system outlined above, the system-dynamics is completely specified by the two spectral densities of system-bath coupling for the first and second electronic excited state: 
\begin{eqnarray}
  \mathcal{J}_{01}(\omega) &=&\frac{\pi}{2}\sum^N_j \omega^3_j \left(K^{\{1\}}_j\right)^2\delta (\omega -\omega_j),  \\ 
  \mathcal{J}_{02}(\omega) &=&\frac{\pi}{2}\sum^N_j \omega^3_j \left(K^{\{2\}}_j\right)^2\delta (\omega -\omega_j).
\end{eqnarray}
Given that the first and second excited state are completely decoupled, the Condon approximation is assumed, and the nuclear degrees of freedom are described by the BOM Hamiltonian, the linear absorption spectrum $\sigma(\omega)$ can then be evaluated exactly in the cumulant formalism\cite{Mukamel-book,Zuehlsdorff2019} 
\begin{equation}
  \sigma(\omega)\propto \omega \int_{-\infty}^\infty \textrm{d}t\, e^{\textrm{i}\omega t} \left(\chi_{01}(t)+\chi_{02}(t)\right)
\end{equation}
with 
\begin{equation}
  \label{eq:linear_response_BOM}
  \chi_{01}(t)=|V_{01}|^2e^{\textrm{i}\omega_{01}^\textrm{av}t}\textrm{exp}\left(-g\left[\mathcal{J}_{01}\right](t)\right).
\end{equation}
Here, $\omega_{01}^\textrm{av}=\left\langle U_{01}\right\rangle$ is the ground state thermal average of the energy gap operator $U_{01}=H_{1}-H_0$ between the electronic ground- and first excited- states, and $g\left[\mathcal{J}_{01}\right](t)$ is the second-order cumulant lineshape function.\cite{Mukamel-book} Truncation of the cumulant expansion at second order is exact for a system with Gaussian fluctuations of the energy gap, as we have for this BOM Hamiltonian. For the purely linear coupling to nuclear degrees of freedom considered in the BOM, $g(t)$ is completely determined by the spectral density and can be written as\cite{Mukamel-book}
\begin{equation}\label{eq:2nd-order-cumulant-lineshape-function}
  \begin{split}
    g\left[\mathcal{J}\right](t) &= \frac{1}{\pi}\int_{0}^{\infty}d\omega\ \frac{\mathcal{J}(\omega)}{\omega^2}\Big[\coth\Big(\frac{\beta \omega}{2}\Big) [1 - \cos(\omega t)] \\
    &\qquad \qquad \qquad \qquad \qquad - \textrm{i} [\sin(\omega t) - \omega t] \Big].
  \end{split}
\end{equation}

For the simple BOM Hamiltonian,  Eqns.~\ref{eq:linear_response_BOM} and \ref{eq:2nd-order-cumulant-lineshape-function} yield the exact linear spectrum. However, $\hat{H}_\text{BOM}$ neglects a number of important effects on an optical absorption spectrum that are present in realistic systems. Even when retaining a harmonic approximation to the ground and excited state potential energy surfaces, allowing the ground- and excited- state vibrational frequencies to differ introduces non-linear energy gap fluctuations for which the second-order cumulant lineshape is no longer exact.\cite{Zuehlsdorff2019} The most general Hamiltonian based on harmonic approximations to the PESs is the Generalized Brownian Oscillator Model (GBOM)\cite{Zuehlsdorff2019}, which allows for differences between the ground and excited state curvature and mode-mixing effects described by the Duschinsky rotation.\cite{Duschinsky1937} The GBOM Hamiltonian can be constructed from ground- and excited- state normal mode analysis and can be solved exactly in the Franck-Condon approach.\cite{Santoro_2008,Baiardi2013,DeSouza2018} Such Franck-Condon calculations are implemented in a range of standard electronic structure packages. Furthermore, the Condon approximation can be relaxed in this formalism through the inclusion of Herzberg-Teller effects.\cite{Santoro_2008,Baiardi2013} It was recently shown by some of the authors that the nonlinear coupling effects introduced in the GBOM Hamiltonian can also be approximately recovered in the cumulant formalism by including a third-order cumulant correction term.\cite{Zuehlsdorff2019}

Both the cumulant approach and the Franck-Condon approach based on the GBOM Hamiltonian can be used to construct the linear absorption spectrum in systems where the individual excited states are separated far enough in energy that they can be treated as fully decoupled. However, if electronic excited states intersect in the Condon region, this intersection leads to strong coupling between the two excited states that is dependent on nuclear coordinates, resulting in a breakdown of the Condon approximation and the intensity-borrowing between electronic excited states that is commonly observed in linear absorption spectra of small organic chromophores\cite{Worth1996, Neville2018,Yaghoubi_Jouybari2020,Aranda2021}.

A simple model Hamiltonian that can describe these dynamics is the linear vibronic coupling (LVC) Hamiltonian. For the purpose of the present work, we only consider linear coupling between the two electronic excited states, such that the LVC Hamiltonian can be written as
\begin{equation}
  \hat{H}_\textrm{LVC}=\hat{H}_\textrm{BOM}+\sum_j^N
  \begin{pmatrix}
    0 & 0 & 0 \\
    0 & 0 & \Lambda_j \hat{q}_j  \\
    0 & \Lambda_j \hat{q}_j  & 0  \\
  \end{pmatrix},
\end{equation}
where the couplings $\Lambda_j$ to the bath in the off-diagonal elements of the Hamiltonian that mixes the first and the second excited state is then specified by the coupling spectral density
\begin{equation}
  \mathcal{J}_{12}(\omega)=\frac{\pi}{2} \sum_j^N \frac{\Lambda^2_j}{\omega} \delta(\omega-\omega_j).
\end{equation}
In contrast to the BOM Hamiltonian, in the LVC model the electronic and nuclear Hamiltonians no longer commute, rendering an analytical description of time-evolution unobtainable. Therefore, more sophisticated methods are required, in this case T-TEDOPA. The details of this description of our system dynamics and the linear response function $\chi(t)$ needed for absorption spectra are contained in Sec.~\ref{ssec:ttedopa}). 

\subsection{The LVC Hamiltonian in complex condensed-phase environment}
\label{ssec:lvc_ham}

\noindent
For a molecule in the gas phase, the LVC Hamiltonian can be described in terms of 3$N$-6 well-defined vibrational modes coupling to the electronic excitations. In the condensed phase,  such as for a  chromophore in solution or embedded in a protein environment, a large number of bath modes couple to the system, including collective chromophore-environment motion. In this case it becomes convenient to consider the spectral densities introduced in the previous section as continuous functions that can be constructed from equilibrium quantum correlation functions of fluctuation operators\cite{Mavros2016}. The spectral density $\mathcal{J}_{01}(\omega)$ describing the coupling to the first excited state can then be written as 
\begin{equation}\label{eq:spectral-density}
  \mathcal{J}_{01}(\omega) = \textrm{i}\theta(\omega) \int \textrm{d}t\ e^{\textrm{i} \omega t} \ \mathrm{Im}\ C_{01}(t),
\end{equation}
where $\theta(\omega)$ is the Heaviside step function and the quantum autocorrelation function of the energy gap fluctuation operator is given by
\begin{equation}\label{eq:energy-gap-fluctuation-autocorrelation-function}
  C_{01}(t) = \langle \delta U_{01}(\hat{\mathbf{q}}, t) \delta U_{01}(\hat{\mathbf{q}}, 0) \rangle,
\end{equation}
and $\delta U_{01}=\left(H_1-H_0\right)-\omega_{01}^\textrm{av}=U_{01}-\omega_{01}^\textrm{av}$.

The exact quantum correlation function $C_{01}(t)$ is, in general, impossible to compute for anything but the most simple model systems. A practical approach for complex condensed phase systems is to approximately reconstruct $C_{01}(t)$ from its classical counterpart using quantum correction factors\cite{Bader1994,Egorov1999,Kim2002b}. In this work, we use the harmonic quantum correction factor\cite{Egorov1999,Valleau2012} that can be derived by equating the classical correlation function with the Kubo-transformed quantum correlation function possessing the same symmetries as its counterpart\cite{Craig2004,Ramirez2004}. This choice yields
\begin{equation}\label{eq:spectral-density-from-classical-md}
  \mathcal{J}_{01}(\omega) \approx \theta(\omega) \frac{\beta \omega }{2} \int \textrm{d}t\ e^{\textrm{i} \omega t} \ C^{\mathrm{cl}}_{01}(t),
\end{equation}
where $\beta=1/k_\textrm{B}T$ and $\theta(\omega)$ is the Heaviside step function.

Eqn.~\ref{eq:spectral-density-from-classical-md} enables the computation of spectral densities in complex condensed phase systems directly by evaluating classical correlation functions of electronic excitation energy fluctuations computed along a molecular-dynamics (MD) trajectory on the ground state potential energy surface. However, computing excitation energies from electronic structure methods such as time-dependent density functional theory (TDDFT) yields \emph{adiabatic} electronic states, i.e. states that are electronically decoupled but can change their electronic character along the MD trajectory. Instead, to parameterize the LVC Hamiltonian outlined in the previous section, it is necessary to construct coupled \emph{diabatic} states that do not change character along the trajectory, and thus have constant ground- to excited state transition dipole moments. 

As with the choice of quantum correction factor\cite{Egorov1999,Valleau2012}, the choice of diabatic states is not unique and several approaches exist to construct quasi-diabatic states from adiabatic states\cite{Pracher1988,Subotnik2008,VanVoorhis2010,Hoyer2014}. For the present example, we are interested in exploring the effect of the coupling between one bright and one dark state close in energy on the optical absorption spectrum. In this case, an efficient diabatization strategy has been outlined previously by Subotnik and co-workers. \cite{Subotnik2017} Following this approach, we define the transition dipole matrix $\textbf{D}$ as
\begin{equation}
  \textbf{D}=
  \begin{pmatrix}
    V_{01}\cdot V_{01} & V_{01}\cdot V_{02} \\
    V_{01}\cdot V_{02} & V_{02}\cdot V_{02}
  \end{pmatrix} 
\end{equation}
Diagonalizing $\textbf{D}$ results in two states with maximally different transition dipole moments and oscillator strengths. The eigenvectors of $\textbf{D}$ can then be used to rotate the diagonal matrix of adiabatic excitation energies into a matrix with two diabatic excitation energies on the diagonal and their electronic coupling as the off-diagonal elements. Performing the diabatization procedure for every snapshot along an MD trajectory, it is then straightforward to construct the classical autocorrelation functions $C^\textrm{cl}_{01}(t)$, $C^\textrm{cl}_{02}(t)$ and $C^\textrm{cl}_{12}(t)$, which, after substitution into Eqn.~\ref{eq:spectral-density-from-classical-md}, yields the spectral densities defining the LVC Hamiltonian. 

The final parameters needed to define the LVC Hamiltonian are the electronic energy gaps $\Delta_{01}$ and $\Delta_{02}$. Assuming a linear coupling to vibrational degrees of freedom, these can be constructed from the classical thermal averages of the energy gap fluctuations, such that $\Delta_{01}=\omega_{01}^\textrm{av}-\lambda^\text{R}_{01}$. Here, $\lambda^\text{R}_{01}$ is the nuclear reorganization energy of electronic state S$_{1}$ that can be computed from the spectral density $\mathcal{J}_{01}(\omega)$ as:
\begin{equation}
  \lambda^\text{R}_{01}=\frac{1}{\pi}\int_0^\infty \frac{\mathcal{J}_{01}(\omega)}{\omega}\textrm{d}\omega =\frac{1}{2}\sum_{j}\omega_{j}^{2}\left ( K_{j}^{\{1\}}\right ) ^2 
  \label{eq:solvent_reorg}
\end{equation}
Using the above parameters, we can define two distinct measures of the energy gap between S$_1$ and S$_2$, which is expected to be a key parameter in this system determining the linear absorption spectrum. First, $\Delta_{12}=\Delta_{02}-\Delta_{01}$, which is the difference between the minima of the diabatic S$_1$ and S$_2$ potential energy surfaces, and second, $\omega^\text{av}_{12}=\omega^\text{av}_{02}-\omega^\text{av}_{01}$, the thermal average of the diabatic S$_1$-S$_2$ energy gap difference, which represents a measure of the separation of the surfaces in the Condon region. 

\subsection{T-TEDOPA \& the Response Function}
\label{ssec:ttedopa}

\noindent
In this section we will briefly describe the T-TEDOPA method and show how it may be employed to
calculate the response function.
We begin by splitting $\hat{H}_\textrm{LVC}$ into three parts
\begin{equation}
  \label{eq:H2ndquant}
  \hat{H}_\textrm{LVC}=H_\text{S}+H_\text{I}+H_\text{B}.
\end{equation}
$H_\text{S}$ is a system Hamiltonian governing the electronic states
\begin{equation}
  \label{eq:Hs}
  H_\text{S}=\sum_{\alpha=1}^{2}(\lambda^\text{R}_{0\alpha}+\Delta_{0\alpha})\ket{S_{\alpha}}\bra{S_{\alpha}} + \delta(\ket{S_{1}}\bra{S_{2}}+\text{h.c.}),
\end{equation}
where h.c. denotes the Hermitian conjugate of the preceding terms and $\lambda^\text{R}_{0\alpha}$ denotes the reorganization energies of the  S$_{1}$ and S$_{2}$ baths. We have also included for completeness a coupling $\delta$ between S$_{1}$ and S$_{2}$, although in the case of MB this coupling is extremely small. The Hamiltonian $H_\text{I}$ is the interaction Hamiltonian which describes the coupling to the nuclear degrees of freedom (DOF). We further decompose this interaction Hamiltonian as follows 
\begin{equation}
  H_\text{I}=H_\text{I}^\text{EL}+H_\text{I}^{\delta},  
\end{equation}
where the energy level interaction Hamiltonian $H_\text{I}^\text{EL}$ describes the coupling which causes fluctuations in the S$_{1}$ and S$_{2}$ energies, and where $H_\text{I}^{\delta}$ is responsible for the fluctuations in the S$_{1}$-S$_{2}$ coupling matrix element. To begin with, we make the assumption that the three fluctuation motions in the system are uncorrelated, and thus we treat the spectral densities $\mathcal{J}_{01}$, $\mathcal{J}_{02}$ and $\mathcal{J}_{12}$ as pertaining to three independent bosonic baths which we label with $\alpha=1$, 2 and 3
respectively. We will go on to refine this assumption in section \ref{sec:corr}. Making the transformation to second quantized creation and annihilation operators, we obtain the following forms for the interaction Hamiltonians 
\begin{equation}
  \label{eq:HI_EL_uncorrelated}
  H_\text{I}^\text{EL}=-\frac{1}{\sqrt{2}}\sum_{j,\alpha=1}^{2}K_{j}^{\{\alpha\}}\omega_{j}^{\frac{3}{2}}(a_{j\alpha}^{\dagger}+a_{j\alpha})\ket{S_{\alpha}}\bra{S_{\alpha}},
\end{equation}
\begin{equation}
  \label{eq:HId}
  H_\text{I}^{\delta}=\sum_{j}\frac{\Lambda_{j}}{\sqrt{2\omega_{j}}}(a_{j3}^{\dagger}+a_{j3})(\ket{S_{1}}\bra{S_{2}}+\text{h.c.}).
\end{equation}

Finally, the bath Hamiltonian describing the free motion of the nuclei is given by
\begin{equation}
  \label{eq:Hb}
  H_\text{B}=\sum_{j,\alpha=1}^{3}\omega_{j}a_{j\alpha}^{\dagger}a_{j\alpha}.
\end{equation}

The initial condition is the of product density matrices
\begin{equation}
  \label{eq:rho0}
  \rho(0)=\rho_\text{S}\otimes\rho^{\{1\}}\otimes\rho^{\{2\}}\otimes\rho^{\{3\}},
\end{equation}
where $\rho^{\{\alpha\}}$ is a thermal equilibrium density matrix for bath $\alpha$ with inverse temperature  $\beta_{\alpha}$ and $\rho_\text{S}=\ket{S_{0}}\bra{S_{0}}$ is the electronic ground state. These thermal density matrices may appear a major problem since they represent statistical mixtures - implying an averaging over a large  number of initial states. However, by applying Tamascelli \textit{et al.}'s T-TEDOPA mapping to each bath, we transform  these density matrices into vacuum states $\rho^{\{\alpha\}}\rightarrow\ket{0}_{\alpha}$ - single pure state wave  functions - while the bath spectral density picks up a temperature dependence $J(\omega)\rightarrow J_{\beta}(\omega)$.  \cite{tamascelli2019efficient,dunnett2021matrix} This new \emph{thermal} spectral density, which encodes the detailed balance of thermal emission and absorption in the mode coupling strengths, is defined on the domain $\omega\in[-\infty,\infty]$ and thus introduces effective \emph{negative} energy modes into the bath. The zero-temperature equilibrium correlation function of this spectrally extended environment is identical to that of the original finite temperature environment, and as the reduced system dynamics defined by $\text{Tr}_\text{E}\{\rho(t)\}$ can be shown to be uniquely determined by the bath correlation function, the proxy, extended zero temperature environment can be used to obtain finite temperature results \cite{tamascelli2019efficient}. Moreover, the mapping can also be inverted on the bath modes to provide thermal expectations for the nuclei in the original basis \cite{dunnett2021matrix,10.3389/fchem.2020.600731}.

In general, the optical dipole operator is
\begin{equation}
  \label{eq:V}
  \hat{V}=V_{01}\ket{S_{0}}\bra{S_{1}}+V_{02}\ket{S_{0}}\bra{S_{2}}+\text{h.c.},
\end{equation}
and the response function is given by
\begin{equation}
  \label{eq:rf}
  \chi(t)=\langle\hat{V}(t)\hat{V}(0)\rangle_{\rho(0)}.
\end{equation}
Although the T-TEDOPA mapping preserves the dynamics of the system's reduced density matrix, it is not clear that the same can be said for arbitrary multi-time correlation functions,although some progress has recently been made for the case of third-order response within the thermofield approach \cite{gelin2021simulation}. However, in the present case the response function can be shown to depend on a  reduced density matrix for the system which has been time-evolved from a particular initial state; therefore, we can correctly employ the mapping. Firstly, we expand the thermal expectation of Eq.~\ref{eq:rf} and rearrange the contents using the cyclic property of the trace
\begin{equation}
  \label{eq:rfexp}
  \begin{split}
    \langle\hat{V}(t)\hat{V}(0)\rangle_{\rho(0)} &= \text{Tr}\{e^{\text{i}\hat{H}t}\hat{V}e^{-\text{i}\hat{H}t}\hat{V}\rho(0)\}\\
    &= \text{Tr}\{\hat{V}e^{-\text{i}\hat{H}t}\hat{V}\rho(0)e^{\text{i}\hat{H}t}\},
  \end{split}
\end{equation}
where Tr denotes the trace over all electronic and nuclear coordinates and we have dropped the subscript from  $\hat{H}_{\text{LVC}}$. Making use of the Condon approximation, we perform the trace over the nuclear DOF first giving 
\begin{equation}
  \label{eq:rfexp2}
  \begin{split}
    \langle\hat{V}(t)\hat{V}(0)\rangle_{\rho(0)}&=\text{Tr}_\text{S}\{\hat{V}\text{Tr}_\text{E}\{e^{-\text{i}\hat{H}t}\hat{V}\rho(0)e^{\text{i}\hat{H}t}\}\}\\
    &=\text{Tr}_\text{S}\{\hat{V}\rho_\text{R}'(t)\}\\
    &=\langle\hat{V}\rangle_{\rho_{R}'(t)}.
  \end{split}
\end{equation}
\noindent
Thus, the response function can be expressed as the expectation value of $\hat{V}$ with respect to the reduced system
density matrix $\rho_\text{R}'(t)$, which has been evolved from an initial state $\hat{V}\rho(0)$. Unfortunately, $\hat{V}\rho(0)$ does not represent a valid initial state, since it contains only off-diagonal components: $\ket{S_{1}}\bra{S_{0}}$ and $\ket{S_{2}}\bra{S_{0}}$; therefore, it is not possible to construct a simulation with $\hat{V}\rho(0)$ as an initial condition. Indeed, because $\hat{V}$ is a Hermitian operator, its expectation cannot equate to that of a multi-time correlation function since the latter may be complex valued. However, using the initial state $\rho'_\text{S}(0)=\ket{\psi}\bra{\psi}$, where
$\ket{\psi}=c(\ket{S_{0}}+V_{01}\ket{S_{1}}+V_{02}\ket{S_{2}})$ and measuring the \emph{non-Hermitian} operator $\hat{V}'=V_{01}\ket{S_{0}}\bra{S_{1}}+V_{02}\ket{S_{0}}\bra{S_{2}}$, we find exactly the same expectation value as we would have found had we measured $\hat{V}$ with the invalid initial condition $\hat{V}\rho(0)$. This is because the system's Hamiltonian does not mix S$_{1}$ or S$_{2}$ with S$_{0}$ and thus the additional terms in $\rho_\text{S}'$, such as $\ket{S_{1}}\bra{S_{2}}$, will be projected out by the measurement of $\hat{V}'$. As $\rho_\text{S}'$ is a valid initial state, and as nothing prevents us from measuring a non-Hermitian operator, it is now straightforward to construct a simulation for $\chi(t)$.

\subsection{Bath Correlations}
\label{sec:corr}

\noindent
In the previous section we made the assumption that the three fluctuation motions, corresponding to the two energy levels and the coupling between them, were completely uncorrelated and could thus be treated as arising from three independent baths. While it is normally justified to assume that there is no correlation between the fluctuations of the coupling and those of the energy levels, the bath motions required for these two kinds of fluctuations being of very different natures, the same is not in general true for the fluctuations of the energy levels between themselves. Indeed, by measuring the cross-correlator of the S$_{1}$ and S$_{2}$ energy fluctuation operators along the MD trajectory
\begin{equation}
  C_{\text{cross}}(t)=\langle\delta U_{01}(\hat{\mathbf{q}},t)\delta U_{02}(\hat{\mathbf{q}},0)\rangle,
\end{equation}
we find a strong, principally positive, correlation between these two motions for MB. This leads, via Eq.~\ref{eq:spectral-density}, to the cross-correlation spectral density $\mathcal{J}_{\text{cross}}$. 

As a means of assessing quantitatively the strength and parity of these correlations, we define the normalized cross-correlation spectral density as $\tilde{\mathcal{J}}_{\text{cross}}=\mathcal{J}_{\text{cross}}/ \sqrt{\mathcal{J}_{01}\mathcal{J}_{02}}$. This function takes values in the range $[1,-1]$ and has the following interpretation: if $\tilde{\mathcal{J}}_{\text{cross}}=1$, we have fully positively correlated modes; the bath induced fluctuations of the S$_{1}$ and S$_{2}$ energies are perfectly in phase - a raising of the S$_{1}$ energy being associated with a simultaneous raising of the S$_{2}$ energy (although the amplitudes, which are determined by $\mathcal{J}_{01}$ and $\mathcal{J}_{02}$, need not be the same). Similarly, if $\tilde{\mathcal{J}}_{\text{cross}}=-1$, the fluctuations will be perfectly anti-correlated - a raising of the S$_{1}$ energy being associated with a lowering of the S$_{2}$ energy (again the amplitudes need not be identical). Finally, if $\tilde{\mathcal{J}}_{\text{cross}}=0$, the fluctuations are uncorrelated; i.e., the energy fluctuations behave as two independent sources of Gaussian noise. 

The normalized cross-correlation spectral density for MB from our energy gap sampling is plotted in SI Fig.~\ref{fig:normcross}. We find that, with a few exceptions, the environmental modes are between 40\% and 100\% positively correlated (further analysis is presented in SI Sec.\ref{SI:mode_assign}). In order to include these correlations in the dynamical simulations we must generalize the energy level interaction Hamiltonian $H_\text{I}^\text{EL}$ to take the following form: 
\begin{equation}
  \label{eq:HI_EL}
  H_\text{I}^\text{EL}=\sum_{k}\sum_{\alpha\beta}g_{k}^{\alpha\beta}\ket{S_{\alpha}}\bra{S_{\alpha}}(a_{\beta k}^{\dagger}+a_{\beta k}),
\end{equation}
where now, each harmonic bath couples to both the S$_{1}$ and S$_{2}$ energies. This interaction Hamiltonian is capable of describing arbitrary correlations between the S$_{1}$ and S$_{2}$ energy fluctuations. The parameters $g_{k}^{\alpha\beta}$ are determined so as to reproduce the calculated spectral densities $\mathcal{J}_{01}$, $\mathcal{J}_{02}$ and $\mathcal{J}_{\text{cross}}$. The details of the procedure for obtaining $g_{k}^{\alpha\beta}$ are contained in SI Sec.~\ref{SI:bath_parameters}. LVC calculations performed using the exact S$_{1}$-S$_{2}$ cross-correlation are labeled MDCC (molecular dynamics cross-correlated). 

There are two important limiting cases of this Hamiltonian. The first occurs when the off-diagonal coupling coefficients vanish ($g_{k}^{12}=g_{k}^{21}=0$), which corresponds to the uncorrelated limit introduced in section \ref{ssec:ttedopa}. In this case the two excited states are each coupled to their own independent bosonic bath with no communication between them. The second limiting case, which we refer to as the fully positively correlated (FPC) limit, occurs when the two columns of the matrix $g_{k}^{\alpha\beta}$ are identical; i.e., when $g_{k}^{12}=g_{k}^{11}$ and $g_{k}^{21}=g_{k}^{22}$. In this case, the coupling matrix $g_{k}^{\alpha \beta}$ possesses a zero eigenvalue and thus $H_\text{I}^\text{EL}$ reduces to a coupling of the collective motion of S$_{1}$ and S$_{2}$ to a \emph{single, shared} bath (whose creation and annihilation operators are the linear combinations  $b_{k}^{(\dagger)}=\frac{1}{\sqrt{2}}\left(a_{1k}^{(\dagger)}+a_{2k}^{(\dagger)}\right)$). The coupling Hamiltonian for this case takes the form
\begin{equation}
  H_\text{I}^\text{EL}=\sum_{k}\left(g_{k}^{11}\ket{S_{1}}\bra{S_{1}}+g_{k}^{22}\ket{S_{2}}\bra{S_{2}}\right)\left(b_{k}^{\dagger}+b_{k}\right). 
  \label{eqn:H_FPC}
\end{equation}

In the FPC limit the energy fluctuations of the two excited states are induced by the same set of modes and thus have no independent character. We interest ourselves in this FPC limit for two reasons: first, it is interesting to consider, from a theoretical point of view, the effect of correlated energy fluctuations on the absorption spectra and excited state dynamics; second, given that for MB most modes are strongly positively correlated, the FPC limit represents a reasonable approximation which carries with it a significant reduction in computational overhead, because of the need to simulate only one bath. One could also consider the fully negatively correlated limit where the coupling would be to the system operator $g_{k}^{11}\ket{S_{1}}\bra{S_{1}}-g_{k}^{22}\ket{S_{2}}\bra{S_{2}}$, however this is nonphysical for MB.

\section{Computational details}
\subsection{Molecular Dynamics and Electronic Structure calculations}

\noindent
To sample the energy gap fluctuations needed to generate the necessary correlation functions and spectral densities of MB in water, four independent trajectories of 8~ps length were generated. The same trajectories as generated for a previous study by some of the authors were used,\cite{Zuehlsdorff2020b} and the full computational details can be found therein. Here, we summarize the main computational details. 

To obtain independent starting points for the four trajectories, force field based molecular dynamics simulations were performed in OpenMM\cite{openMM}, where water was represented by the TIP3P\cite{TIP3P} water model and the MB force field parameters were generated using the QUBEKit package.\cite{QUBEKit} The system was equilibrated as described in Ref.~\onlinecite{Zuehlsdorff2020b} and a 4~ns production run in the NVT ensemble was performed, where atomic positions and velocities were extracted every 1~ns to yield independent starting points for mixed quantum mechanical/molecular mechanical (QM/MM) simulations.

Using the independent starting points, four 10~ps QM/MM trajectories were generated using the inbuilt QM/MM functionality of the TeraChem package\cite{Ufimtsev2009}. For dynamics, the chromophore and its counter-ion were treated quantum mechanically with the CAM-B3LYP exchange-correlation functional \cite{Yanai2004} and 6-31+G* basis set, and all water molecules were described by the TIP3P force field. Calculations were performed in the NVT ensemble using a Langevin thermostat with a collision frequency of 1~ps$^{-1}$ and a time-step of 0.5~fs was used throughout. The first 2~ps of each trajectory were discarded to allow for the system to equilibrate after switching the chromophore Hamiltonian from the force field Hamiltonian to the DFT Hamiltonian in the QM/MM simulation, resulting in 8~ps of usable trajectory for each independent trajectory. From these trajectories, snapshots were extracted every 2~fs for calculating vertical excitation energies, yielding a total of 16,000 snapshots from which the classical correlation functions were constructed.

Adiabatic excitation energies on each snapshot were computed using time-dependent density-functional theory (TDDFT) as implemented in the TeraChem code.\cite{Isborn2011} To evaluate the influence of different choices of TDDFT functional on the S$_1$/S$_2$ coupling, vertical excitation energies were either computed at the CAM-B3LYP/6-31+G* level of theory in the Tamm-Dancoff approximation or at the B3LYP/6-31+G* level of theory using full TDDFT.\cite{Yanai2004, b3lyp} This choice is motivated by the fact that the relative S$_1$/S$_2$ energy is very sensitive to the treatment of long range Hartree-Fock exchange in the density functional, with the CAM-B3LYP functional predicting a larger energy difference between S$_1$ and S$_2$ (See SI Sec.~\ref{SI:vertical_func} for a discussion of the influence of different density functionals on the calculated excited state energies for MB). We note that the use of the B3LYP functional for computing vertical excitation energies does introduce a mismatch between the Hamiltonian generating the ground state dynamics and the Hamiltonian generating the energy gap fluctuations. Such a mismatch can create artifacts in the computed correlation functions and spectral densities of system-bath coupling, as is commonly observed for spectral densities computed with TDDFT using force-field-based MD trajectories.\cite{Zwier2007,Rosnik2015,Chandraskaran2015,Kim2015,Lee2016,Andreussi2017} However, the ground state properties of MB predicted by CAM-B3LYP and B3LYP  are expected to be similar enough that this mismatch has a relatively minor influence on the computed results. We further validate this choice in SI Sec.~\ref{SI:dynamic_mismatch}, where we compare single trajectory spectral densities of B3LYP ground state dynamics in MM water to the mixed CAM-B3LYP/B3LYP and CAM-B3LYP/CAM-B3LYP spectral densities. The spectral densities are fairly consistent, with a small red shift of frequencies for B3LYP ground state dynamics compared to the CAM-B3LYP dynamics. 

To fully capture the influence of dynamic polarization of the environment on the energy gap fluctuations, excitation energies are computed by treating every solvent molecule with a center of mass within 6~\AA \, from any chromophore atom fully quantum mechanically in the TDDFT calculation, with the remaining solvent atoms represented by classical point charges. This treatment leads to QM region sizes of the order of $\approx400$~atoms for the TDDFT calculations.  Previous research by some of the authors has shown that for some systems, computed couplings of nuclear vibrations to electronic excited state can be very sensitive to the treatment of polarization effects in the environment, thus making large QM regions necessary.\cite{Zuehlsdorff2020} To assess whether the coupling between the S$_1$ and S$_2$ excited states in MB shows a similar sensitivity to environmental polarization, we recompute all excitation energies using both the B3LYP and the CAM-B3LYP functional, where only the chromophore is treated quantum mechanically and the full solvent environment is represented by classical point charges (see Fig.~\ref{fig:schematic_qm_regions} for an example of the two QM regions considered in this work). 

For all computed data sets, the quasi-diabatic states are computed from the adiabatic energies and transition dipole moments following the approach outlined in Ref.~\onlinecite{Subotnik2017}. This approach yields diabatic S$_1$ and S$_2$ energies, as well as their coupling, for every snapshot. Computing classical autocorrelation functions for the diabatic states and the coupling, we can then parameterize an LVC Hamiltonian that contains the full coupling between the chromophore and its complex environment (See Sec.~\ref{ssec:lvc_ham}). To avoid numerical issues in the Fourier transforms necessary to compute the relevant spectral densities, a decaying exponential of the form $\exp(-|t|/\tau)$ is applied to all classical correlation functions $C^{\textrm{cl}}(t)$, where $\tau=500~\textrm{fs}$. Further details of the formalism and implementation of cumulant lineshape calculations based upon $C^{\textrm{cl}}(t)$ can be found in recent publications by some of the authors, and are available in the MolSpeckPy package. \cite{pythonspeccode}

\subsection{Tensor Network dynamics}

\noindent
In this section we provide the computational details of the simulations carried out to determine the response function $\chi(t)$. Time-evolution of the density matrix under the LVC Hamiltonian was  carried out using the one-site Time-Dependent-Variational-Principle method (1TDVP) on tree and chain Matrix-Product-States (MPS). \cite{tdvp,kloss2018time,Schroder2019} 

An MPS, or tensor train as they are known within the mathematics community, is a data structure that can be used as an efficient representation of many-body quantum states satisfying the one-dimensional form of the area law. Although an MPS can in principle represent a generic wave-function in any number of dimensions, they are most successfully employed when the system in question possesses a chain-like topology with open boundary conditions. One may also extend the MPS concept to consider systems with a quasi-  one-dimensional topology, i.e., a tree structure, using so called tree-MPS, provided that there are no loops.\cite{Schroder2019} The accuracy of the MPS approximation is controlled by a parameter known as the bond-dimension, with a larger bond-dimension providing a more accurate but more expensive representation. For the simulations used to produce the absorption spectra we present here, a maximum bond-dimension of 20 was found to be sufficient.
\begin{figure}
  \centering
  \includegraphics[width=0.9\columnwidth]{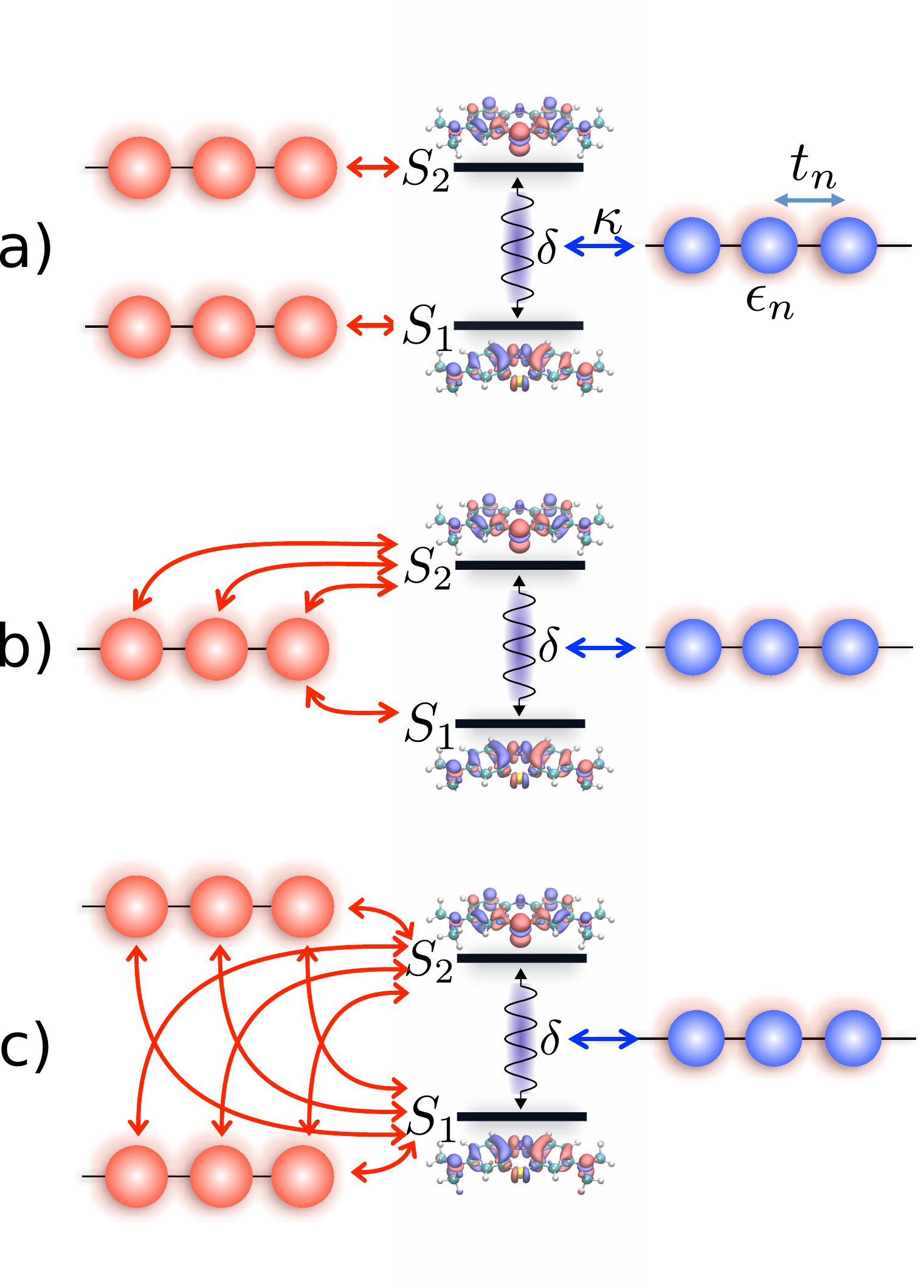}
  \caption{Tensor network structures used for the various bath configurations considered. (a) Uncorrelated baths, tree-MPS with local interactions only. (b) FPC baths, chain-MPS with some long-range interactions. (c) MDCC baths, tree-MPS with long-range couplings. Bath modes involved in $H_\text{I}^{\delta}$ (coupling) are shown in blue, while bath modes involved in $H_\text{I}^\text{EL}$ are shown in red (tuning). While interactions between the system and the tuning modes may become long-ranged, all intra-chain couplings are nearest-neighbour ($t_n$) for oscillators of frequency $\epsilon_n$.}
  \label{fig:tnstruct}
\end{figure}

By employing the chain mapping,\cite{Chin2010} it is possible to transform the Hamiltonians described in sections \ref{ssec:ttedopa} and \ref{sec:corr} into Hamiltonians with the desired topology. We refer to SI Sec.~\ref{SI:chain_mapping} for details on this procedure.  The tensor network structures resulting from the chain mapping are shown in Fig.~\ref{fig:tnstruct} for the uncorrelated and FPC limits introduced in section \ref{sec:corr}, and also for the general MD cross-correlated (MDCC) case. In the case of uncorrelated S$_{1}$-S$_{2}$ energy fluctuations, the three harmonic baths are transformed to three chains of harmonic oscillators with nearest-neighbour couplings, each coupled to the central system site, resulting in a loop-free tree topology. On the other hand, in the FPC limit, there are only two baths and so one obtains a chain topology. However, since one of the baths couples to both S$_{1}$ and S$_{2}$ with different spectral densities, one cannot avoid long-range couplings. For example, if one performs a chain mapping with respect to $\mathcal{J}_{01}$, while S$_{1}$ will be coupled to the first site only, S$_{2}$ will be coupled to every site along the chain. In practice, these long-range couplings introduce only a modest increase in computational complexity within 1TDVP \cite{tdvp}. Specifically, the bond-dimension of the Matrix-Product-Operator (MPO) representation of the Hamiltonian only increases by one. Finally, in the general MDCC case, one again obtains a tree, but now long-range couplings are present on two of the chains. 

Two controllable approximations are necessary to make the simulations of these chain mapped Hamiltonians possible. These are: 1) the truncation of the local Fock space of each chain mode to a finite set of states $d$ and 2) the truncation of the semi-infinite chains to a finite number of chain sites $N$. Both of these approximations introduce errors that are confined by rigorously derived bounds. \cite{woods_simulating_2015} The linear absorption spectra presented here were found to converge with chain modes truncated to $d=20$ Fock states and $N=150$ chain modes for each chain. The observable $\hat{V}'(t)$, used as a proxy for the response function $\chi(t)$, was calculated at 1000 time steps from $t=0$ up to $t=240$~fs. The response function was found to decay to a steady state on the time scale of $\sim 50$~fs, which is physically reasonable for a molecular optical coherence at room temperature.  Simulations were performed on nodes consisting of two 12-core Intel Xeon Haswell (E5-2670v3) processors. Approximate simulation times for the uncorrelated, FPC and MDCC limits were, respectively, 9, 8 and 13 hours.

\section{Results}

\subsection{Excited States and Spectra Computed Within the Condon Approximation}
\begin{figure*}
  \centering
  \includegraphics[width=1.85\columnwidth]{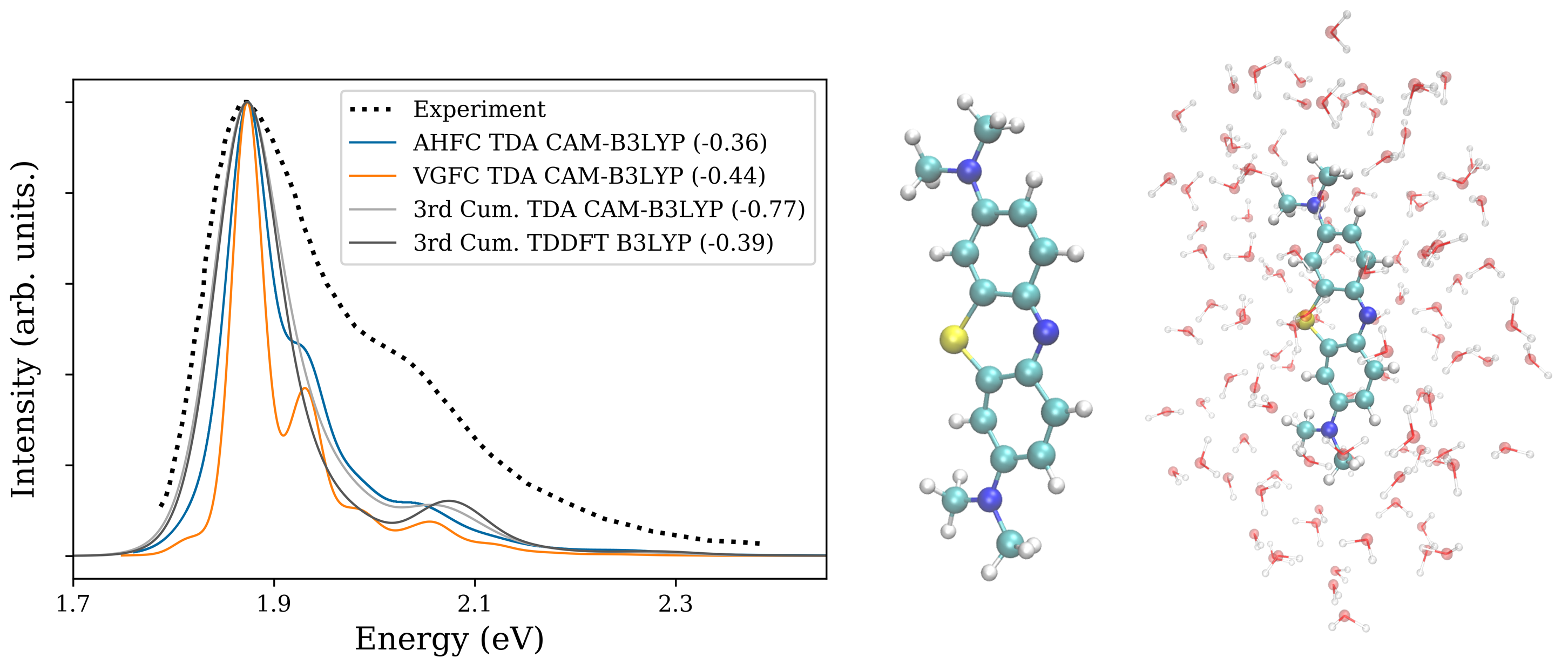}
  \caption{Methylene blue experimental absorption lineshape and calculated Condon type spectra for the bright S$_{1}$ state. Spectra measured in water at low (monomeric) concentration, taken from Ref.~\onlinecite{FERNANDEZPEREZ2019448}.  Structure of methylene blue (middle), and methylene blue in 6\AA \, water (right) represent the two quantum mechanical environments calculated in this study. }
  \label{fig:structure_lineshape}
  \label{fig:schematic_qm_regions}
\end{figure*}

\noindent
Before presenting results computed with the LVC Hamiltonian, we first discuss the results of TDDFT excited state calculations and spectra computed within the Condon approximation for the S$_{0}$ $\rightarrow$ S$_{1}$ transition. These results are jointly presented in the calculated absorption lineshapes in Fig.~\ref{fig:structure_lineshape} and in SI Sec.~\ref{SI:vertical_func}.

For all functionals examined, when excitations are calculated at the S$_0$, S$_{1}$ and S$_{2}$ minima, the S$_{0}$ $\rightarrow$ S$_{1}$ transition of MB is bright and the S$_{0}$ $\rightarrow$ S$_{2}$ oscillator strength is consistently low. S$_{0}$ $\rightarrow$ S$_{2}$ shows $\sim$0.5\% the intensity of the S$_{0}$ $\rightarrow$ S$_{1}$, indicating that any Franck-Condon approach for modeling S$_{0}$ $\rightarrow$ S$_{2}$ will lead to negligible contributions to the total lineshape. Despite the consistency of the oscillator strengths, the S$_{1}$-S$_{2}$ vertical energy gap is highly sensitive to functional choice, ranging from 0.06 to 0.67 eV. 

Fig.~\ref{fig:structure_lineshape} shows the significant difference between the experimental absorption lineshape of MB in water and various Franck-Condon and cumulant lineshape calculations. In all four calculated lineshapes, the shoulder blue shifted 0.15 eV from the 0-0 maxima is underestimated, and the general lineshape significantly under-broadened. For the adiabatic Hessian Franck-Condon (AHFC)\cite{Ferrer2012} calculation, the S$_{1}$ excited state geometry and normal modes have been evaluated (at the TDA/CAM-B3LYP/6-31+G* level) and mode mixing effects are accounted for through a Duschinsky rotation\cite{Duschinsky1937}. This is equivalent to representing the nuclear degrees of freedom through the generalized Brownian oscillator model (GBOM)\cite{Zuehlsdorff2019}. The resultant lineshape exhibits two shoulders; one local to the 0-0 peak, and one separated by 0.15~eV and in line with the broad experimental shoulder. As presented in the SI Sec.~\ref{SI:AHFC_func}, this AHFC lineshape is reproduced by a range of functionals, for both full TDDFT and TDA, for various solvent models, and when including Herzberg-Teller effects. The main differences between functionals are absolute spectral position (which is pinned to the zero-point corrected energy difference between the ground and S$_{1}$ minima), and the presence of the local shoulder. The local shoulder's intensity is correlated with increasing the fraction of long-range exact exchange in the functional. The main shoulder 0.15~eV from the 0-0 transition that is in line with the broad experimental shoulder however is consistently underestimated by all DFT functionals employed here, except by B3LYP with full TDDFT in vacuum, as we discuss further below. 

The other linear absorption lineshapes contained in Fig.~\ref{fig:structure_lineshape} are computed in the vertical gradient Franck-Condon (VGFC)\cite{Ferrer2012} and third order cumulant approaches.\cite{Zuehlsdorff2019} In the vertical gradient method the underlying vibronic Hamiltonian is the BOM parameterized using the ground state vibrational modes and a single excited state gradient calculation at the ground state minimum. The VG approach thus provides a connection between the AHFC GBOM-type calculations and the effective mapping to the BOM in the cumulant method. The key spectral features are maintained between the AHFC and VGFC methods, indicating that there is only moderate improvement from including Duschinsky rotations or differences between ground and excited state PES curvatures for high frequency modes. However, the VGFC is significantly less broadened, suggesting that changes in frequency and Duschinsky rotation are important for treating low frequency modes. The cumulant lineshapes presented here are calculated in the third order approximation from the diabatized S$_{1}$ spectral densities for the system with 6~\AA \, of QM solvent. As will be discussed, the diabatic S$_{2}$ dipole intensity along these trajectories are practically zero, leading to no contribution to the total cumulant lineshape. These lineshapes do not display the local shoulder, with the 0-0 peak instead appearing as a broad Lorentzian profile, but do display a secondary vibronic peak in line with the broad experimental shoulder. The thermalization and broadness of the peak appear slightly better modeled by the cumulant than the Franck-Condon approaches, which can be attributed to the extensive sampling which capture low frequency dynamics (including solvent effects) and some contributions from sampling anharmonic regions of the PES.

Although Fig.~\ref{fig:structure_lineshape} displays 3rd order cumulant lineshapes, which include corrections due to non-linear couplings to nuclear degrees of freedom that go beyond the simple BOM Hamiltonian, these lineshapes are very similar to the second order cumulant lineshapes (see SI Sec.~\ref{SI:BOM_2nd_3rd}).  This similarity shows that non-linear couplings to nuclear degrees of freedom are small, such that the BOM is a good approximation to the underlying dynamics of the system. The appropriateness of the model is echoed by the small skewness values (see SI Sec.~\ref{SI:BOM_2nd_3rd}) for the energy gap fluctuations of both the diabatic S$_{1}$ and S$_{2}$ surfaces. As the higher order moments of the energy gap fluctuations go to zero, their statistics become exactly Gaussian, then become wholly described by the 2nd order cumulant approximation, and are thus equivalent to a perfect mapping to a set of displaced harmonic oscillators.

In a recent study, de Queiroz \emph{et. al.} performed AHFC calculations of MB in vacuum at the TDDFT/B3LYP/def2-SVP level, which produces a large shoulder in line with experimental results, stemming purely from vibronic contributions to the S$_1$ state.  \cite{deQueiroz2021} Our calculations over a range of functionals and lineshape approaches suggest that this large shoulder is anomalous, occurring only for this specific combination of full (non-TDA) TDDFT and B3LYP in vacuum, and could be due to the excited state geometry optimization yielding a state that is of mixed S$_1$-S$_2$ character. Many results in their work indicate state mixing. For example, they report difficulties in optimizing excited states, as well as intensity-borrowing effects of the S$_2$ in snapshot calculations that include explicit solvent. In our studies, this large vibronic shoulder in the lineshape is not reproduced by TDA/B3LYP in vacuum or any methodology in PCM solvent. Similarly, if we perform a vertical gradient Franck-Condon calculation at the TDDFT/B3LYP level in vacuum, the shoulder is removed. We present results for this analysis in SI Sec~\ref{SI:dequeiroz}.   

In summary, we find that despite considering different approaches to computing the lineshape, a range of density functionals, and the inclusion of environmental effects through the cumulant approach, there is a consistently poor reproduction of the broad experimental spectral shoulder when only considering the transition from the ground state to the bright S$_{1}$ state. Furthermore, in static FC calculations, the transition from the ground state to the S$_2$ state is consistently dark and does not contribute significantly to the lineshape.  We also find that the fluctuation statistics of the excited states are closely Gaussian and the BOM is a robust model for the problem at hand. Lastly, TDDFT calculations presented in the SI show that the S$_1$-S$_2$ gap is very sensitive to the choice of electronic structure method, and for several functionals the gap is so small that  decoupling in the calculation of linear absorption spectra is likely inaccurate. In addition, previous results by de Queiroz \emph{et. al.} show that significant intensity-borrowing between S$_1$ and S$_2$ can occur for selected snapshots of MB in explicit solvent\cite{deQueiroz2021}, suggesting that both non-adiabatic and solvent effects might play an important role in the experimental lineshape.

We now turn to analysis of the QM/MM ground state ab initio molecular dynamics and the energy gap fluctuations that act as input for constructing the spectral densities that we use to evaluate cumulant spectra and parameterize the LVC dynamics. 
\begin{figure*}
  \centering
  \includegraphics[width=0.9\textwidth]{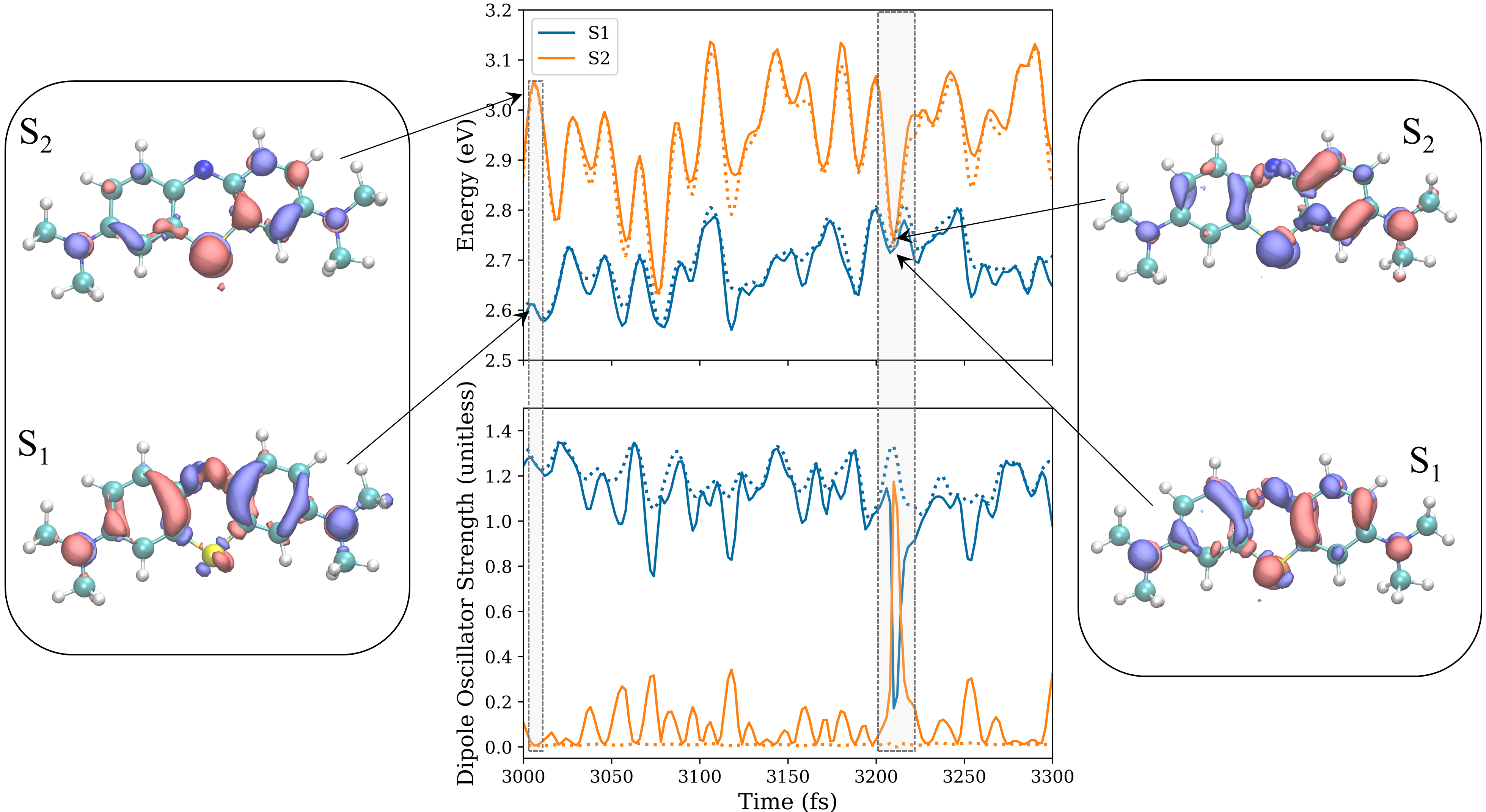}
  \caption{Excitation energies and oscillator strengths for a segment of trajectory 1 as computed with CAM-B3LYP and a 6~\AA\,QM region. Adiabatic energies and oscillator strengths are shown as solid lines, corresponding diabatic values are shown as dotted lines. The transition densities of the adiabatic S$_1$ and S$_2$ states for two specific snapshots along the MD trajectory are also shown.}
  \label{fig:schematic_diabatization}
\end{figure*}
In Fig.~\ref{fig:schematic_diabatization} we show a 300~fs window of a single trajectory's energy gaps and corresponding oscillator strengths. On the left of Fig. \ref{fig:schematic_diabatization} we show the transition densities when the adiabatic states are well-separated, with the S$_{1}$ excitation having a lower energy and high oscillator strength, whilst the S$_{2}$ is higher in energy and has low oscillator strength.  There are dips in the oscillator strength of the S$_{1}$ state at 3076 and 3120 fs which coincide with peaks in the S$_{2}$ intensity.  As can be seen in the oscillator strengths at around 3200 fs, there are also crossings between the S$_{1}$ and S$_{2}$ states; the bright and dark states energetically reorder. On the right of Fig. \ref{fig:schematic_diabatization}, we show transition densities of adiabatic electronic excited states at a point of degeneracy. These represent the S$_{1}$ and S$_{2}$ states becoming 'left' and 'right' mirrored degenerate excitations. The increase in S$_2$ dipole intensity for certain regions of the potential energy surface is consistent with the results of de Queiroz \emph{et. al.}
seen for individual MD snapshot calculations in explicit solvent. \cite{deQueiroz2021} Strong state mixing (as measured by adiabatic S$_{2}$:S$_{1}$ oscillator strength ratio of 1:3 or greater) occurs for approximately 3.3 \% of snapshots at the CAM-B3LYP/TDA/6\AA\,QM solvent level. Thus, non-adiabatic mixing of excited states  occurs semi-frequently around the ground state equilibrium. However, this value is significantly enhanced to 16.8\% for the B3LYP/TDDFT/6\AA \, QM data set, and would likely further increase when accounting for nuclear quantum effects within the MD sampling of the ground state PES\cite{Zuehlsdorff2018}. 

\begin{figure}
  \centering
  \includegraphics[width=\columnwidth]{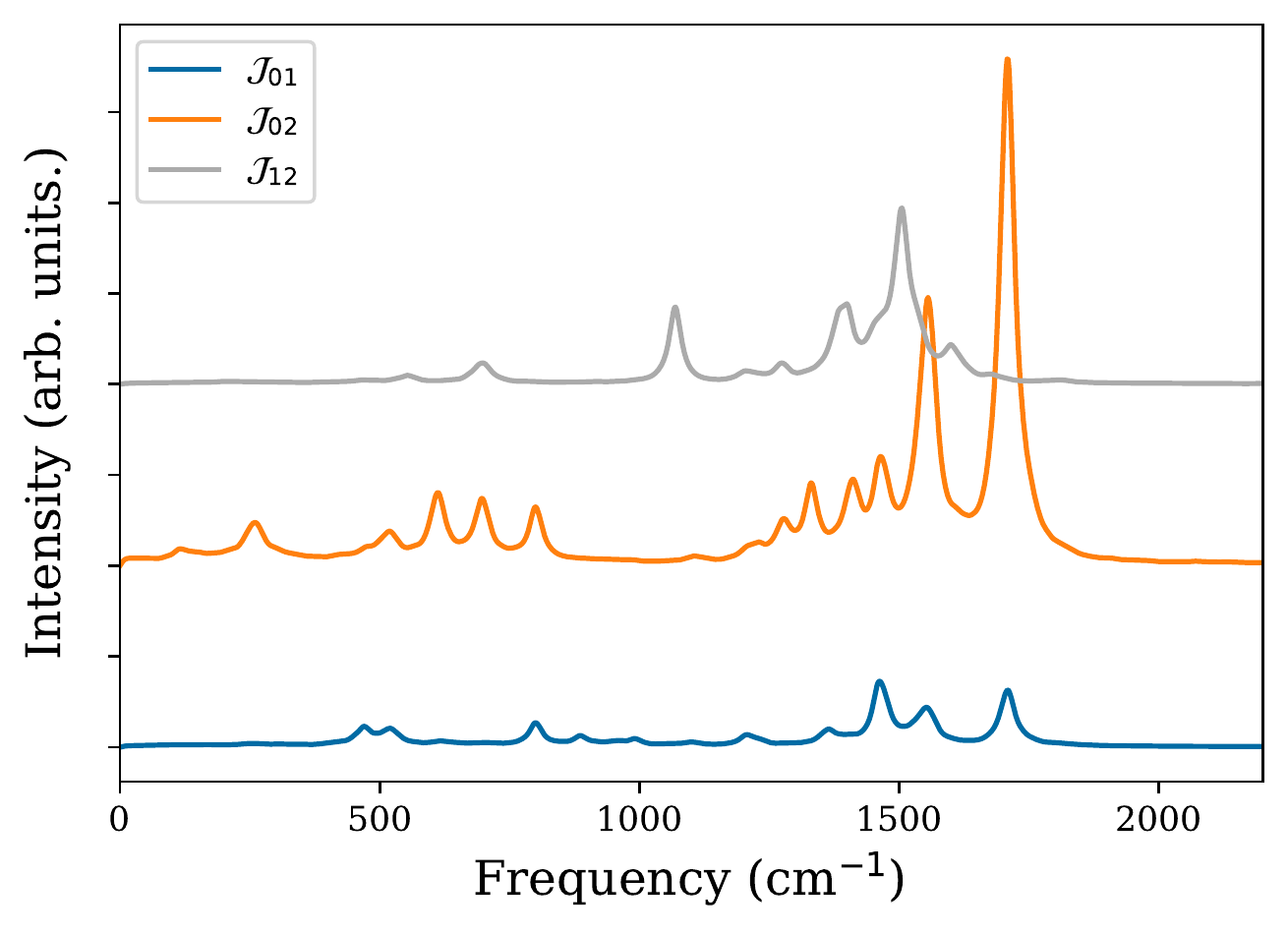}
  \caption{Spectral densities of the diabatic S$_1$ and S$_2$ states, as well as the S$_{1}$/S$_{2}$ coupling spectral density as computed for the CAM-B3LYP data set in 6 \AA \, explicit QM solvent. }
  \label{fig:spectral_dens}
\end{figure}

Application of the diabatization scheme outlined in Ref.~\onlinecite{Subotnik2017} to adiabatic snapshot data is very successful in separating the adiabatic S$_{1}$ and S$_{2}$ into a bright diabatic S$_{1}$ and a dark diabatic S$_{2}$ state. These diabatic states are indicated by the dashed lines in  Fig. \ref{fig:schematic_diabatization}. The consistency of these states restores the Condon approximation, at the price of introducing an explicit coupling between the diabatic S$_1$ and S$_2$ states that has to be accounted for by solving the LVC Hamiltonian. Applying the quantum correction factor and Fourier transform of the classical correlation functions of diabatic energy gap fluctuations leads to the spectral densities for the ground to S$_{1}$, ground to S$_{2}$, and coupling spectral densities shown in Fig. \ref{fig:spectral_dens}. $\mathcal{J}_{01}(\omega)$ and $\mathcal{J}_{02}(\omega)$ detail the ground state vibrational frequencies that couple to the electronic states, whilst $\mathcal{J}_{12}(\omega)$ describes the ground state vibrational frequencies that couple the excited states. These peaks can be assigned approximately to vibrational normal modes calculated for the isolated molecule, which we detail in the SI Sec.~\ref{SI:mode_assign}. To summarize, the diabatic S$_1$ and S$_2$ are shown to couple to A$_1$ symmetric modes, whilst the diabatic coupling is driven by asymmetric B$_2$ modes. Significant contributions to the spectral density range from 110 to 1710 cm$^{-1}$, with the most intense coupling occurring between the S$_2$ and the mode at 1710  cm$^{-1}$, which is a C-C symmetric ring stretching mode. 

Diabatic S$_1$ and S$_2$ spectral densities and their diabatic coupling spectral density are presented in SI Sec.~\ref{SI:all_parameters}, along with the exhaustive table of parameters for the LVC Hamiltonian. The key result of this data is that the average transition dipole moments are quite consistent between electronic structure methods, with the S$_2$ remaining consistently dark for both CAM-B3LYP and B3LYP in both QM and MM solvent. Similarly, the solvent reorganization energy obtained by integrating the spectral densities (eq. \ref{eq:solvent_reorg})  is consistent, and there is minimal change in spectral densities when different functionals are used for the ground state ab initio molecular dynamics. However, the energy gap between the minima of the diabatic potential energy surfaces $\Delta_{12}$ is quite sensitive to the functional and environment. This is seen in $\Delta_{12}$ having a value of 0.087 eV for CAM-B3LYP in QM solvent, and -0.026 eV in B3LYP MM solvent. This energy gap strongly influences the amount of mixing that occurs between the states in the LVC dynamics.

\subsection{Bath correlation}
\begin{figure*}
  \centering
  \includegraphics[width=\columnwidth]{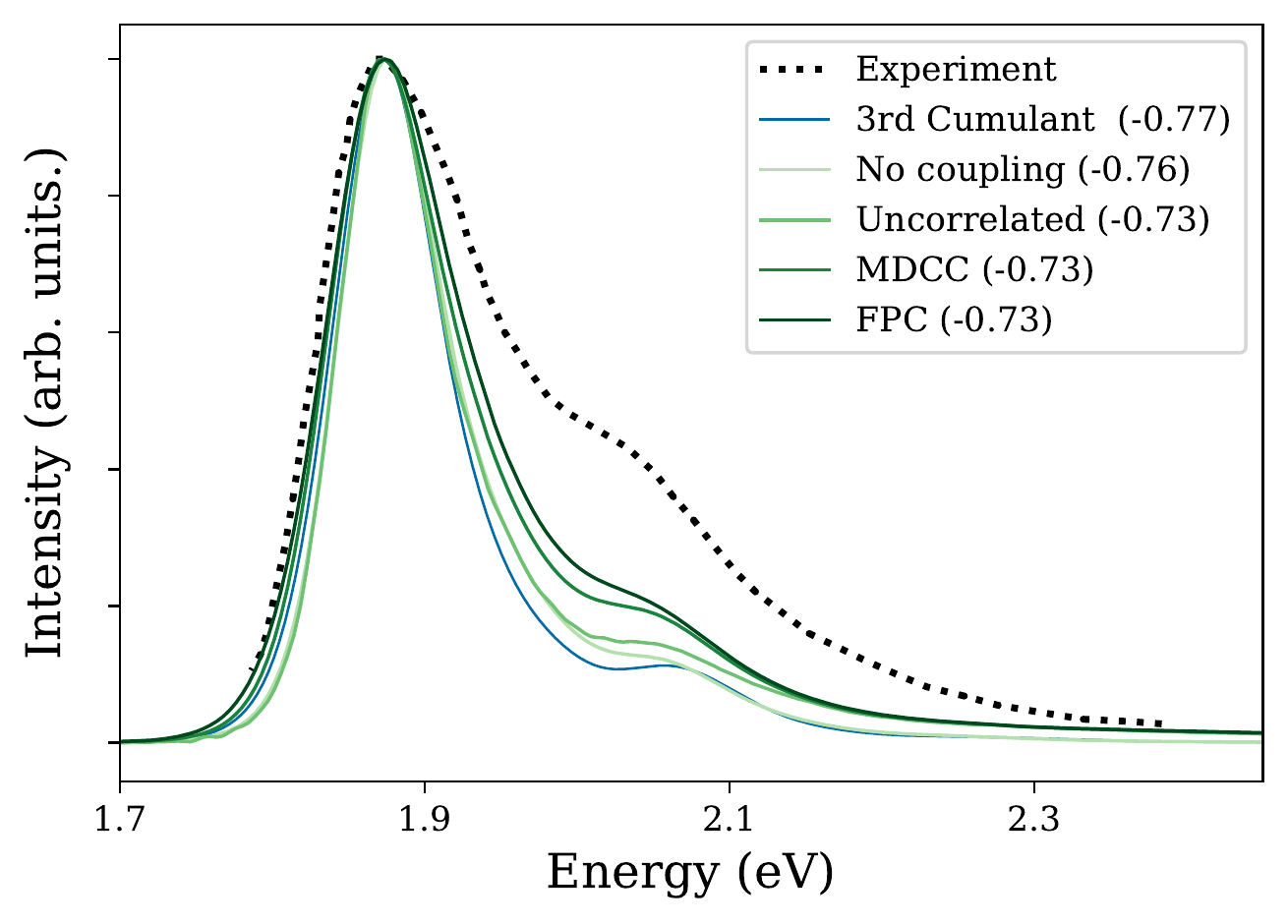}
  \includegraphics[width=\columnwidth]{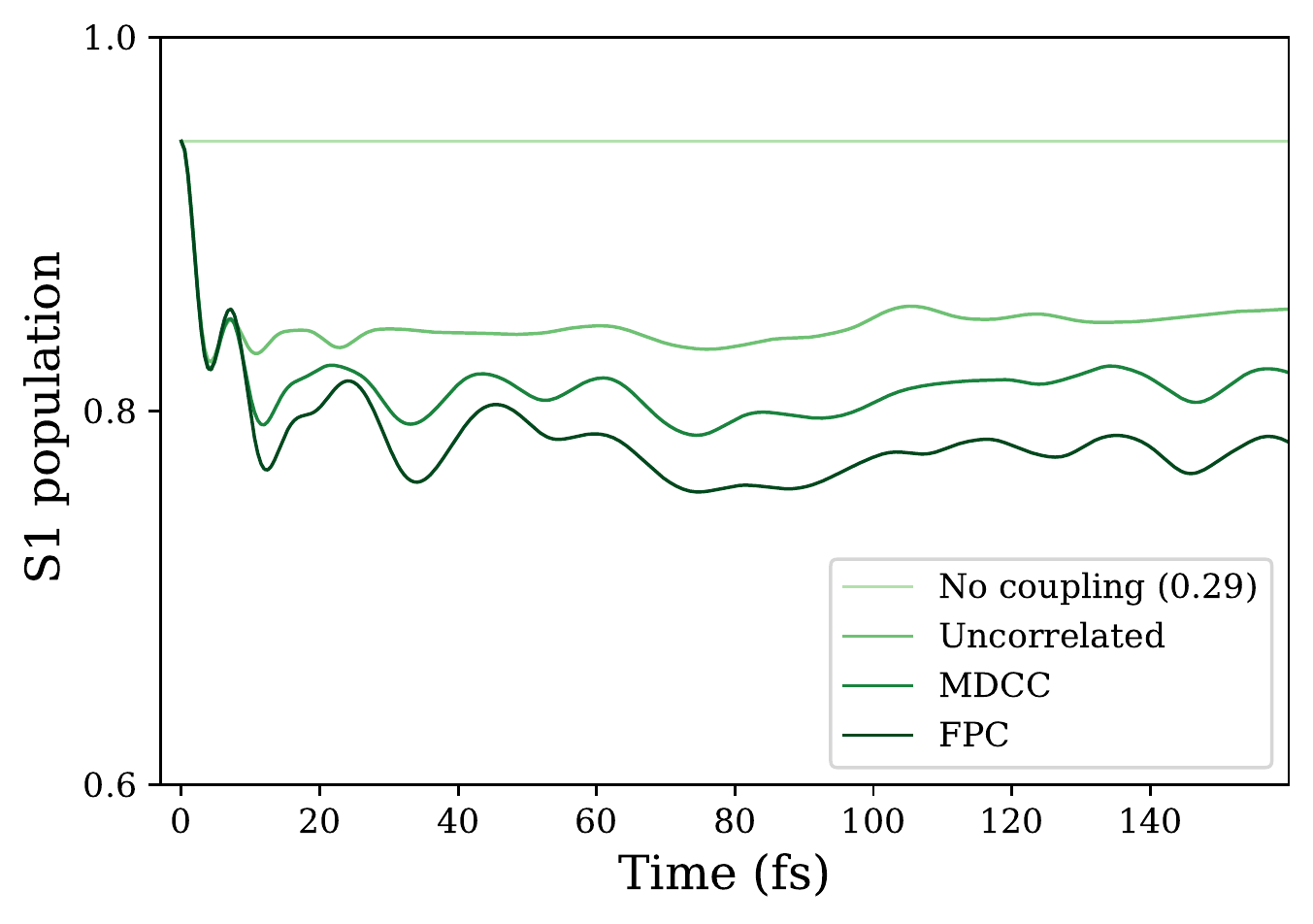}
  \caption{Effect of tensor network bath configurations on lineshape (left) and S$_1$ population (right) using the 6\AA \, QM water TDA/CAM-B3LYP/6-31+G* data. Correlated vibrations (up to fully positively correlated - FPC) between S$_{1}$ and S$_{2}$ drive fast non-adiabatic transitions to the dark S$_{2}$ state and increase the broad shoulder intensity. Lineshapes have been shifted by the left legend value to match experiment. Legend value on the right is the average S$_{1}$-S$_{2}$ gap $\omega^\text{av}_{12}$ in eV. }
  \label{fig:correlation}
\end{figure*}

\noindent
In Fig.~\ref{fig:correlation} we present calculated lineshapes (left) and excited state population dynamics (right) from T-TEDOPA applied to the LVC Hamiltonian parameterized using the spectral densities and parameters from the CAM-B3LYP data set in 6~\AA \, explicit QM solvent, examining different approaches to treating the bath correlation. To first test the T-TEDOPA method without coupling, we compare the T-TEDOPA approach applied to the LVC Hamiltonian with deactivated S$_1$-S$_2$ coupling to the third-order cumulant lineshape. Both these methods have the same pure diabatic spectral densities; however, for the uncoupled LVC approach the total lineshape is given by the sum of the S$_1$ and the low intensity S$_2$ contribution, which slightly increases and smooths the high energy shoulder, whereas the third order cumulant approach includes some effects due to non-linear energy gap fluctuations. As expected, the peak maxima are effectively identical (seen in the legend shift values), the onsets are very similar, and they both significantly underestimate the experimental shoulder. 

As discussed in the methodology and graphically illustrated in Fig.~\ref{fig:tnstruct}, we utilize three different regimes for describing the nature of the couplings between S$_1$ and S$_2$ in the LVC Hamiltonian. The computationally cheapest is the uncorrelated (labeled UC) bath, where the tree-MPS has only local interactions and the cross-correlation between the electronic state fluctuations is assumed to be zero. In the other extreme, the fully positively correlated (FPC) limit, the cross-correlation between fluctuations is unity, such that fluctuations in S$_1$ drives an equivalent change in S$_2$ and \textit{vice versa}. There is a bath shared by the states, leading to a chain-MPS with some long-range interactions, and an increase in computational expense over the UC system. Lastly, the more expensive and physical approach is to include the exact structure of the normalized cross-correlation between electronic states. Examining the short-time (first 50 fs) regime of the population dynamics, which strongly influences the linear absorption spectrum, we note significantly different amounts of fast population transfer to S$_2$ depending on the coupling model. If the states are uncoupled, the S$_1$ population stays pure. However, allowing any kind of S$_1$-S$_2$ coupling leads to a fast ($<$20 fs) and significant ($\sim$15\%)  population transfer from S$_1$ to S$_2$. This population transfer decreases the main peak intensity in the linear absorption spectrum and increases the intensity of the shoulder. The first 8 fs are very similar between models, indicating a fast transition of a portion of the wavepacket from the S$_1$ to the S$_2$ PES via the CI close to the Condon region. The populations quickly differ after this point; for the UC system the population dynamics evolves into a steady state and further mixing with S$_2$ is limited, whilst the correlations in the FPC bath persist and drive further population transfer to and from the S$_2$ state. This population of S$_2$ is reflected in the absorption lineshapes, with the FPC model displaying a stronger shoulder feature, moving closer to experiment, whereas the change in shoulder intensity for the UC bath is moderate. Lowering the main peak intensity through population transfer also decreases the sharpness of the absorption onset, and FPC most closely matches the experimental onset. 

The dynamics and lineshape using the MDCC bath closely mirror FPC, which is rationalized by the fact that the average MD sampled cross-correlation value is 0.8. The key differences in the dynamics of these two models are that the secondary dip is not driven as low as in the FPC dynamics, and the following peak structure is slightly dampened. This dampening could be due to the small cross-correlation value of specific modes or from general lowering of cross-correlation across the entire frequency range. The lower population transfer for the MDCC manifests in the section of the lineshape between the main S${_1}$ and S${_2}$ features and the intensity of the high energy shoulder feature is preserved by the MDCC method. 

Therefore, we find that if cross-correlation between states has an average value close to one, the FPC model is a good approximation to the MDCC dynamics and can lead to appropriate lineshapes for the LVC Hamiltonian with reduced computational cost. However, in this case the inclusion of the exact structure of the fluctuations (in MDCC) between electronic states comes with only a modest increase in computational cost due to the efficiency of MPS/T-TEDOPA. Overall, including non-adiabatic transitions appears essential, as significant fast population transfer occurs even when ignoring correlation between the electronic states. 

\subsection{Influence of the S$_{1}$-S$_{2}$ gap}
\subsubsection{Functionals and environment}

\noindent
Having found the spectral shoulder intensity to be strongly sensitive to the population transfer between the electronic excited states, we next consider the parameters in our LVC Hamiltonian. Noting the consistency of the spectral densities, diabatic dipole moments, and reorganization energies between different solvent models and electronic structure methods, the adiabatic energy gap $\Delta_{12}$ between the two excited state surfaces presents itself as the most important parameter for this problem. 

\begin{figure*}
  \centering
  \includegraphics[width=\columnwidth]{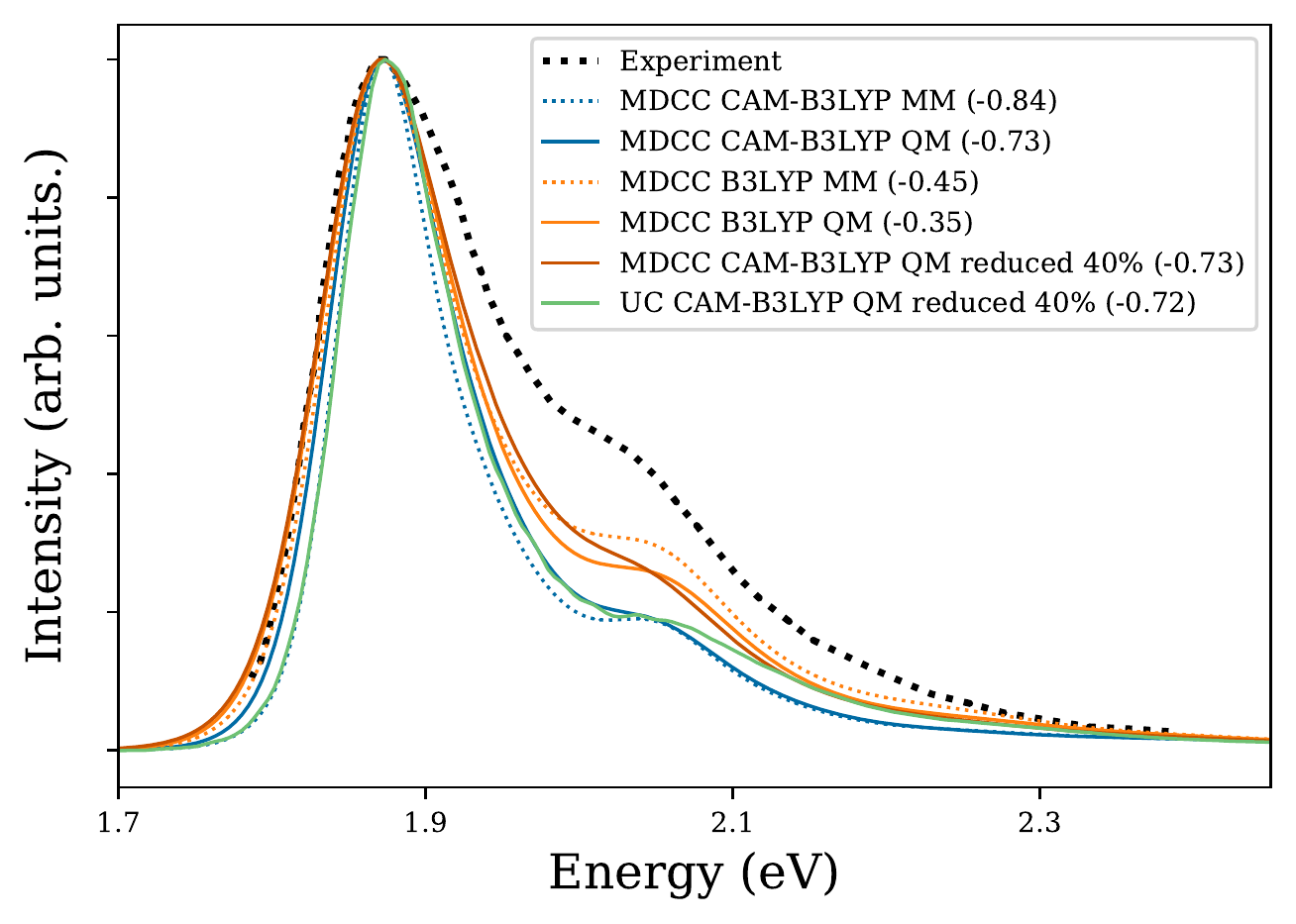}
  \includegraphics[width=0.925\columnwidth]{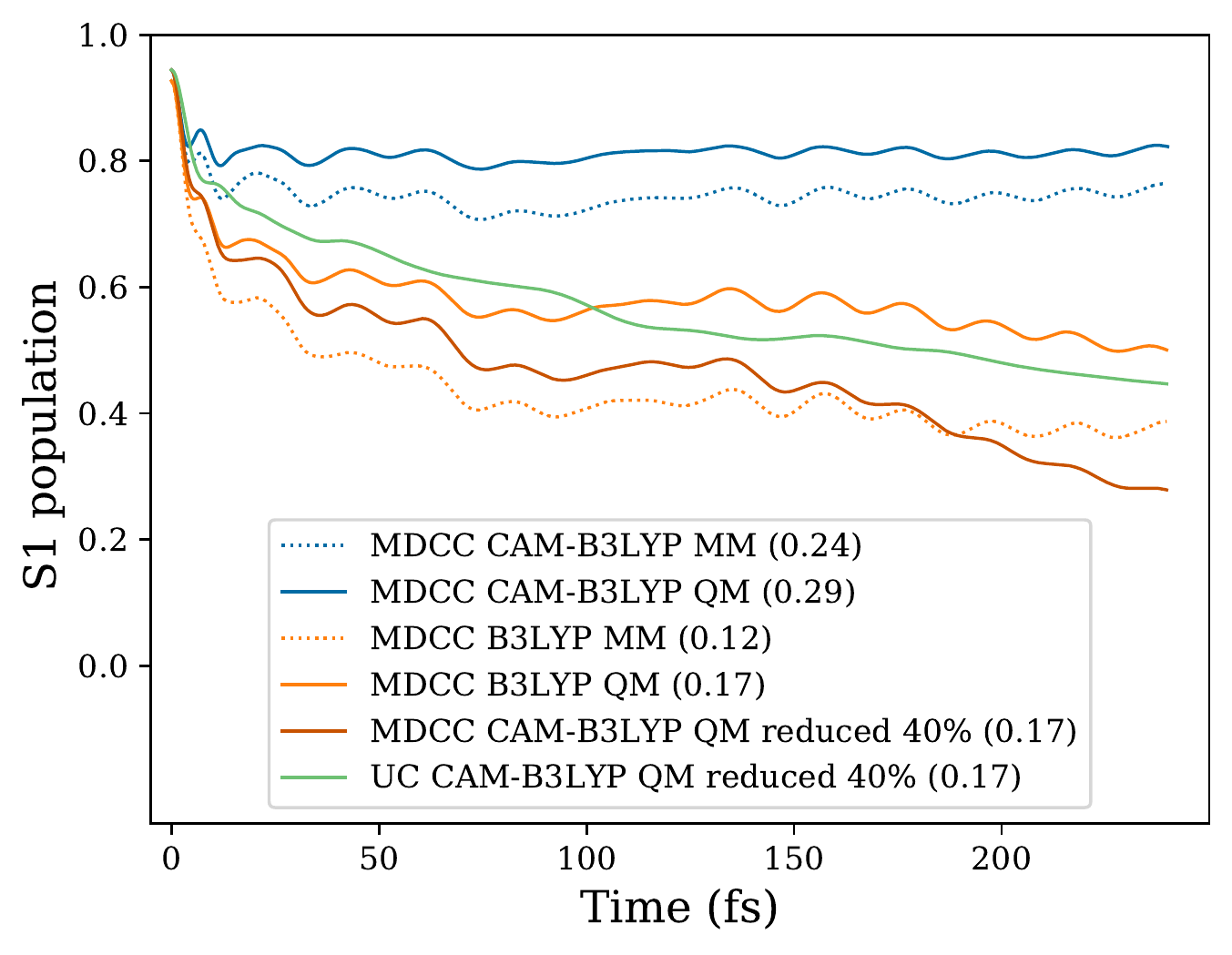}
  \caption{Using molecular dynamics cross-correlation (MDCC) for bath correlation ($\tilde{\mathcal{J}}_{\text{cross}}$), different functionals and solvent models strongly influence the energy gap between electronic states. This can dramatically increase fast non-adiabatic transitions from the S$_{1}$ state and increase shoulder intensity. Lineshapes have been shifted by the left legend value to match experiment in eV. Legend values on the right are the average gap $\omega^\text{av}_{12}$ in eV. }
  \label{fig:variable_gaps}
\end{figure*}

The computed absolute vertical excitation energies of S$_{1}$ and S$_{2}$ vary by 0.6 eV depending on the density functional used (see SI Sec.~\ref{SI:vertical_func}). We also find that the diabatic S$_{2}$ state shows a larger variance in energy depending on the amount of exact exchange in the functional due to having greater charge-transfer character than S$_{1}$. This leads to a strong functional sensitivity in the  S$_{1}$-S$_{2}$ gap. In principle, it would be desirable to determine an accurate S$_1$-S$_2$ energy gap using higher-level electronic structure methods. However, both the size of the MB molecule and the fact that explicit solvent environment plays a significant role in the value of the energy gap makes precise evaluation prohibitive. Instead, we proceed by calculating the MDCC MPS/T-TEDOPA dynamics and lineshape for parameters calculated in different solvent environments (QM vs MM) and for B3LYP/TDDFT vs CAM-B3LYP/TDA to examine the influence of the LVC parameter choice on the computed lineshape. Fig. \ref{fig:variable_gaps} summarizes these results, with the legend of the population dynamics (right) indicating the average energy gap for the given LVC parameterization. The average energy gap $\omega_{12}^\text{av}$ is the energy difference between diabatic S$_1$ and S$_2$ averaged over our molecular dynamics configuration sampling around the ground state equilibrium ($\omega_{12}^\text{av}=\omega_{02}^\text{av}-\omega_{01}^\text{av}$), and is a robust measure of the gap between states in the Condon region. 

The most extreme average energy gap $\omega_{12}^\text{av}$ values are CAM-B3LYP QM (0.29 eV) and B3LYP MM (0.12 eV). Here, using point charge models for solvent reduces $\omega_{12}^\text{av}$ by 0.05 eV. These values directly translate into the population dynamics; we find that large $\omega_{12}^\text{av}$ leads to a retention of population in S$_1$, whereas the smallest value leads to the greatest population transfer. For the first 20 fs the population transfer is almost four times larger for B3LYP MM over CAM-B3LYP QM, which directly manifests in the absorption lineshape. Although the trend in population transfer is gap sensitive, the structure and oscillations are remarkably consistent when using MDCC. This consistency stems from the similarity in spectral densities and cross-correlation between data sets. 

Also in this figure we present the MDCC and UC populations and lineshape for the CAM-B3LYP/QM data but with the average energy gap artificially reduced by 40\%., to match the B3LYP/QM system. For the reduced gap MDCC, the structure of the dynamics is conserved but the population transfer is greater. The short time dynamics are equivalent to B3LYP/QM, but at longer times we see that reducing the gap leads to greater population transfer, with S$_2$ becoming the majority populated state. However, reducing the energy gap alone is insufficient for increasing the lineshape shoulder. If the bath correlation is ignored (UC) then the dynamics (particularly oscillations) are dampened, and whilst the general population transfer is large, the shoulder is weaker and decays slowly. This finding indicates that the reproducing the shoulder in the lineshape is only possible with an exact description of $\mathcal{J}_\text{cross}$ as well as accurate evaluation of the solvent polarized electronic energies. 

\subsubsection{Controlling the average energy gap}

\begin{figure*}
  \centering
  \includegraphics[width=\columnwidth]{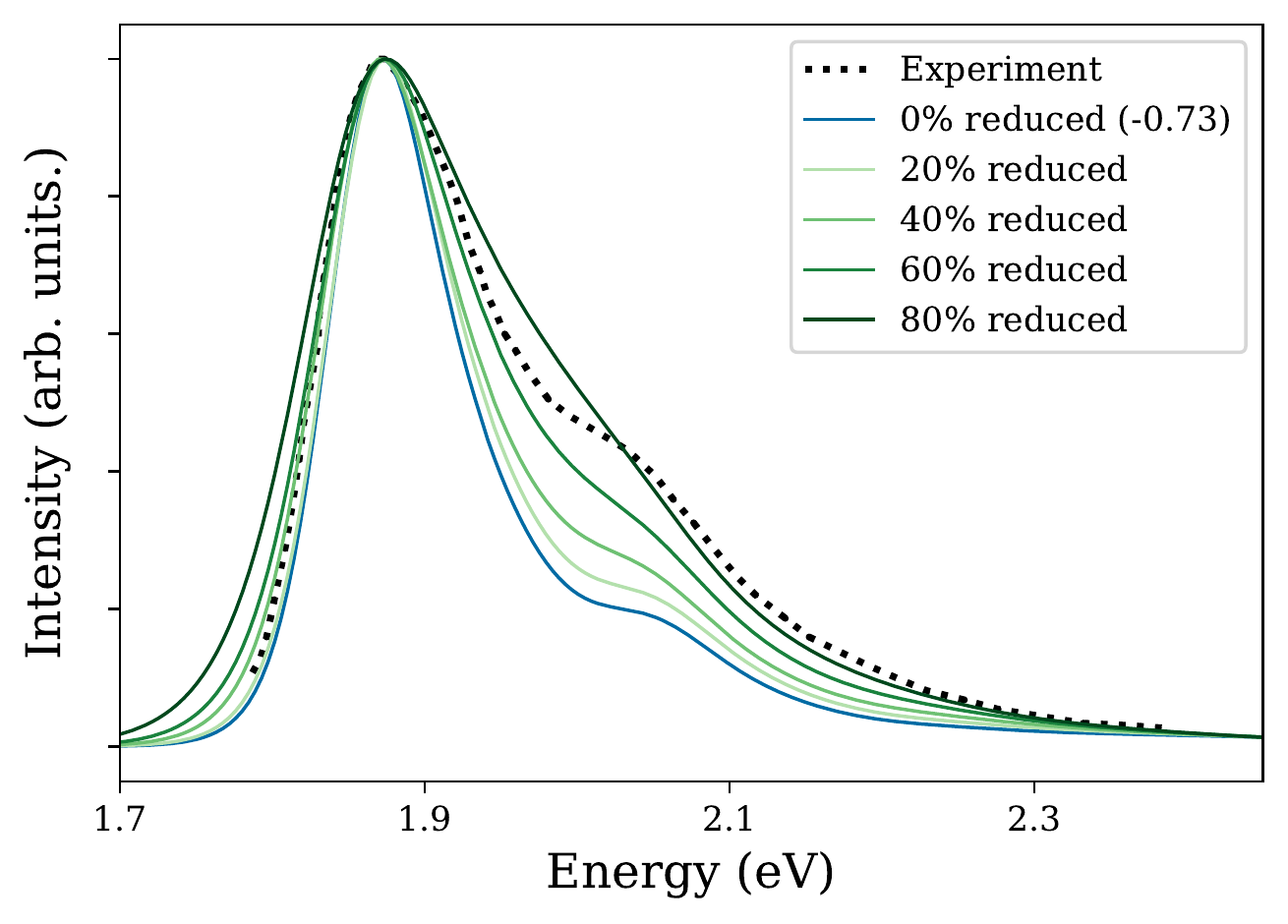}
  \includegraphics[width=\columnwidth]{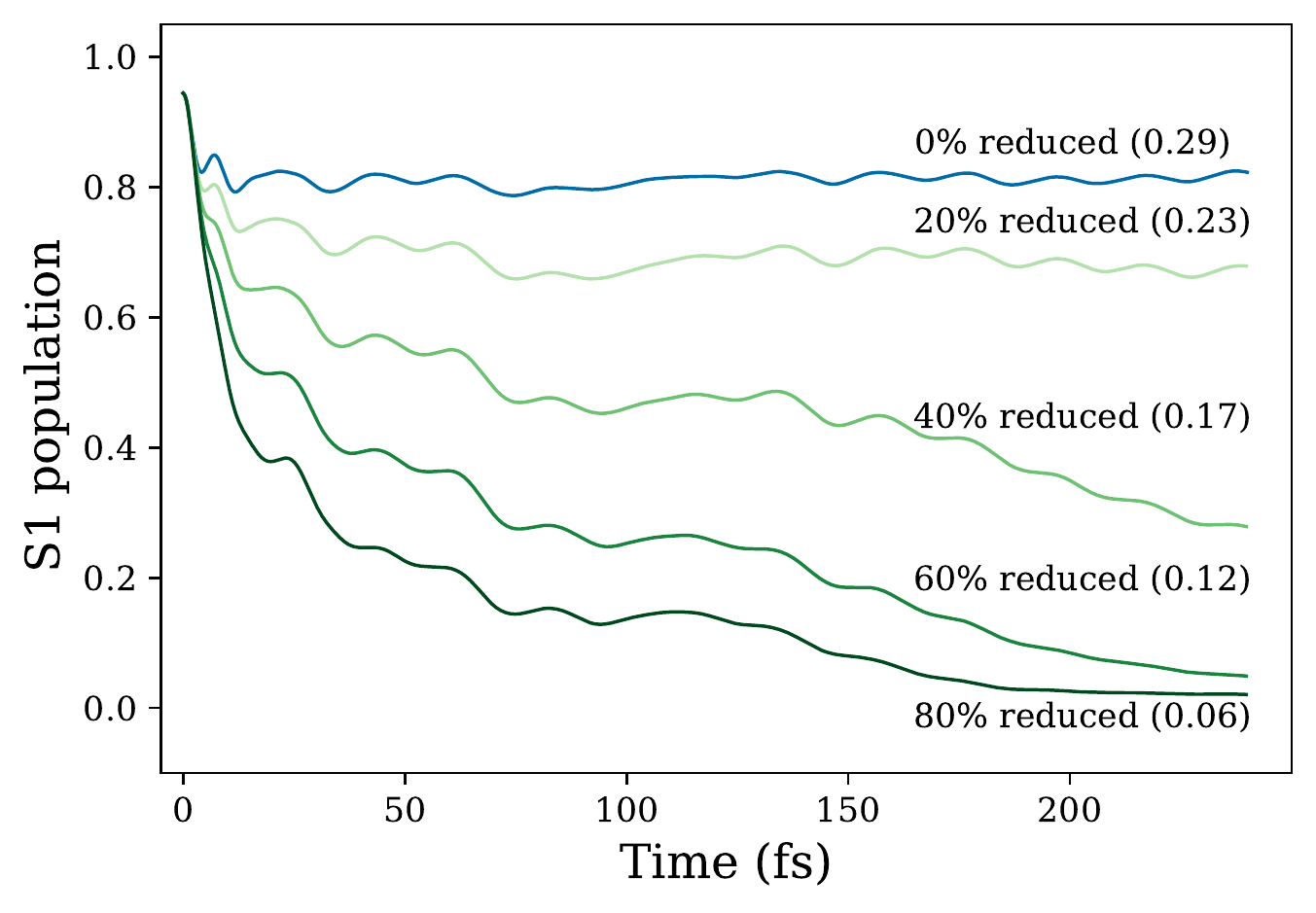}
  \caption{Using molecular dynamics cross-correlation (MDCC) for bath correlation ($\tilde{\mathcal{J}}_{\text{cross}}$), the reduced average S$_{1}$-S$_{2}$ gap may be set to a reduced value. For the 6\AA \, QM water TDA/CAM-B3LYP/6-31+G* spectral densities, reducing this value significantly increases the population transfer to the dark S$_{2}$ state and increases the absorption shoulder intensity. Lineshapes have been shifted by the left legend value (eV) to match experiment. Legend values on the right are the average gap $\omega^\text{av}_{12}$ in eV. }
  \label{fig:controlling_gaps}
\end{figure*}

\noindent
Given that our calculations show a large variance in average energy gap and that no singular \textit{ab initio} method presents itself as unequivocally better for the calculation of excited states, we examine how population dynamics and lineshapes change for a range of average gap values. Using the CAM-B3LYP/TDA/6\AA~QM data set we present in Fig. \ref{fig:controlling_gaps} these results for increments of 20\% average energy gap reduction. This range spans an average energy gap value from 0.29 eV to 0.06 eV. This small value is equivalent to the S$_1$-S$_2$ vertical excitation energy gap as calculated with TDA/B3LYP/PCM, which is presented in SI Sec.~\ref{SI:vertical_func}. 

Reducing the gap and utilizing the MDCC model is highly effective at progressing the calculated lineshape towards the broad experimental shoulder, with the best result occurring for the energy gap reduced by 60\%. All population dynamics have similar structure, but the driven population transfer to the S$_2$ state grows increasingly large as the gap closes. This population transfer mostly occurs in the first 25 fs. The intermediate time dynamics (100-150fs) show very similar plateaus in population transfer. The long time oscillations are strongly dampened for gap reduction $>40\%$.

We conclude that fast S${_1}\rightarrow$S$_2$ population transfer is the main driver of the broad linear absorption spectral shoulder of the MB monomer in water. However, even with inclusion of non-adiabatic effects using the exact solution to the LVC Hamiltonian and the explicit solvent environment, we are unable to completely recreate the experimental spectra. This discrepancy can be related to several factors. The most practical of these considerations is that although reducing the energy gap in an \textit{a posteriori} manner improves the lineshape, it does not correctly account for changes in the structure of the cross-correlation. Clearly the structure of the cross-correlation is important in driving fast dynamics and spectral shoulder intensity, and even moderate changes can reduce low frequency broadening and more strongly couple high frequency vibrations. Better electronic structure methodologies may lead to more accurate properties, in particular the coupling spectral density and the cross-correlation. Another source of improvement may be the inclusion of non-linear terms beyond our LVC model. For example, although the energy gap fluctuations of the diabatic S$_1$ and S$_2$ states seem well described by a linear coupling model given the small corrections the third-order cumulant lineshape provides over the second-order cumulant approach, we are unable to determine how accurately the coupling between S$_1$ and S$_2$ is modeled by the LVC Hamiltonian for this system.  

\section{Conclusion}

\noindent
We have presented a novel methodology for the calculation of the LVC Hamiltonian using a tensor network based approach for quantum dynamics and parameterized using data from \textit{ab initio} molecular dynamics and TDDFT. This method has been applied to the large and curious shoulder in the linear absorption spectra of aqueous methylene blue (MB), demonstrating the role of vibrationally driven population transfer from the bright S$_{1}$ to the dark S$_{2}$.

This methodology is highly attractive as it may be applied to arbitrarily large chromophores in complex environments; capturing the influence of solvent polarization and dynamics, and some anharmonicity, in the spectral densities, whilst still giving exact quantum dynamics. Indeed, the MPS/T-TEDOPA method presents itself as exceptionally computationally efficient and affordable for the evaluation of non-adiabatic dynamics between two electronic states with exact inclusion of cross-correlation by introducing an additional bath. It is readily extendable to a greater number of states, and is particularly appealing for complex linear absorption lineshapes, where only short time propagation is required to compute converged spectra. Further investigation of T-TEDOPA, particularly pertaining to the nature of many-time correlation functions and the retention of bath dynamics, is highly promising for the calculation of quantum dynamics and their interpretation in spectroscopy. Some recent studies have made use of bath observables, \cite{Schroder2019, schnedermann2019molecular} and this is to be examined in the case of MB. 

We note several key results for MB. Firstly, the mapping to the BOM, and the extension to the LVC is effective, as measured by the fluctuation statistics. The S$_{1}$ and S$_{2}$ both couple to high and low frequency modes with A$_1$ symmetry, whilst the diabatic coupling is due to B$_2$ modes. Upon calculating the LVC dynamics and lineshape we find it essential to accurately account for both; the correlations between fluctuations in S$_{1}$ and S$_{2}$, and the average energy gap. The structure of the correlations lead to more effective mixing and show richer oscillatory population dynamics, and the energy gap significantly influences the general population transfer. Most spectral densities and parameters for the LVC are fairly insensitive to the solvent model or the parameters of the TDDFT calculation, with the exception of the average S$_{1}$-S$_{2}$ gap which is very sensitive and difficult to appraise accurately. Choosing a reasonable value for this parameter when calculating the LVC lineshapes leads to significantly better agreement with experiment. Intriguingly, the excited state population dynamics show signs of a transition in the dominant character of S$_1$, as the average energy gap between the bright S$_1$ and dark S$_2$ - at the ground state geometry -  is reduced. While unimportant at room temperature for the absorption spectrum, the emission spectrum will be highly sensitive to the detailed, non-adiabatic dynamics of the excited states, as they relax in solvent. The present numerically exact methodology can be extended to these problems, and will be pursued in future work.

\begin{acknowledgments}  
  \noindent
  Calculations were performed using the MERCED computational resource, supported by the National Science Foundation Major
  Research Instrumentation program (ACI-1429783), as well as the LLNL Pascal supercomputer. A.J.D. acknowledges support
  from Ecole Doctorale Physique en Ile-de-France (EDPIF ED564). A.W.C. acknowledges support from ANR project No.
  195608/ACCEPT. T. J. Z. acknowledges startup funding provided by Oregon State University. C.M.I. was supported by the
  U.S. Department of Energy, Office of Science, Basic Energy Sciences under Award Number DE-SC0020203.
\end{acknowledgments}

\section{Data Availability}

\noindent
The data that support the findings of this study are available from the corresponding author upon reasonable request. The cumulant spectra reported in this work were computed with the MolSpeckPy package available at https://github.com/tjz21/Spectroscopy\_python\_code. All tensor network simulations were performed using the MPSDynamics.jl package available at https://github.com/angusdunnett/MPSDynamics.  

\makeatletter\@input{supaux.tex}\makeatother
\bibliography{bibliography}
\end{document}


\title[]{Supplementary material for ``Influence of non-adiabatic effects on linear absorption spectra in the condensed phase: Methylene blue''}

\author{Angus J. Dunnett}
\affiliation{Sorbonne Universit\'{e}, CNRS, Institut des NanoSciences de Paris, 4 place Jussieu, 75005 Paris, France}

\author{Duncan Gowland}%
\affiliation{Department of Physics, King's College London, London WC2R 2LS, United Kingdom}

\author{Christine M. Isborn}%
\affiliation{Chemistry and Chemical Biology, University of California Merced, Merced, CA 95343, USA}

\author{Alex W. Chin}%
\affiliation{Sorbonne Universit\'{e}, CNRS, Institut des NanoSciences de Paris, 4 place Jussieu, 75005 Paris, France}

\author{Tim J. Zuehlsdorff}
\email{zuehlsdt@oregonstate.edu}
\affiliation{Department of Chemistry, Oregon State University, Corvallis, Oregon 97331, USA}

\date{\today}
\maketitle

\tableofcontents
\newpage

\section{LVC model parameters for all data sets}\label{SI:all_parameters}

\noindent
The data for the linear absorption spectrum of MB presented in the main text is the result of a number of different parameterizations of the MB Hamiltonian. Specifically, we consider parameterizations from B3LYP TDDFT and CAM-B3LYP TDA vertical excitation energy calculations, both for a pure MM and a 6~\AA~QM representation of the solvent environment, all based on the same CAM-B3LYP ground state dynamics using an MM solvent environment. 

\begin{figure}[h]
    \centering
    \includegraphics[width=1\textwidth]{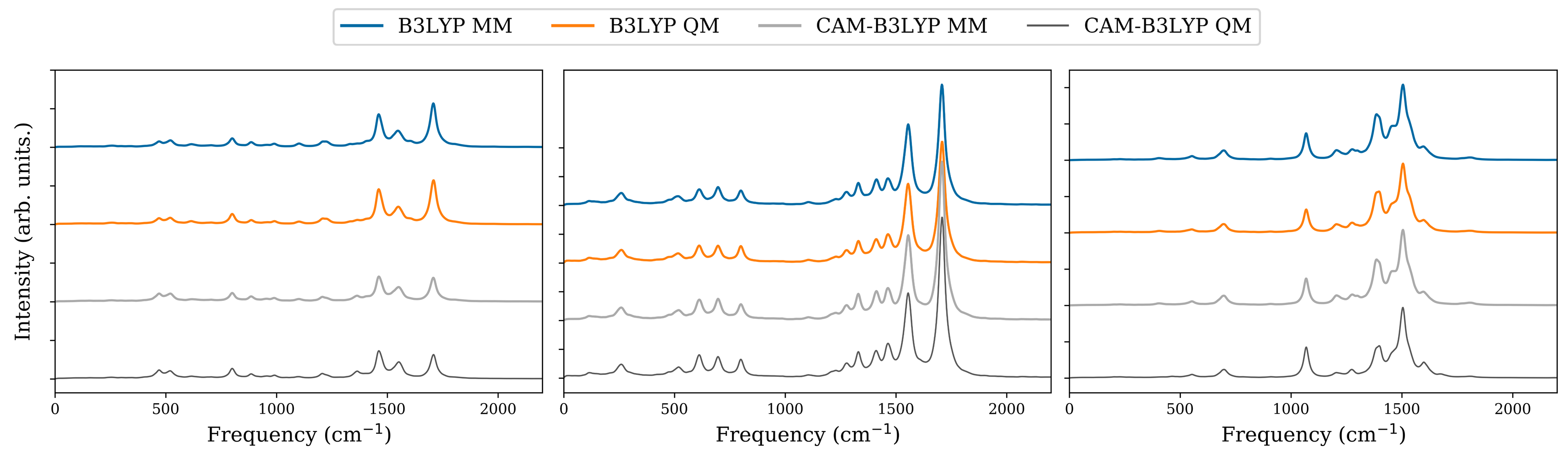}
    \caption{Spectral densities of CAM-B3LYP dynamics with B3LYP vertical excitation calculations in MM (blue) and 6\AA QM (orange) solvent.Spectral densities of CAM-B3LYP dynamics with CAM-B3LYP vertical excitation calculations in MM (light gray) and 6\AA QM (dark gray) solvent. $\mathcal{J}_{01}$ left, $\mathcal{J}_{02}$ middle, $\mathcal{J}_{12}$ right.   }
   \label{fig:SDs_CAMB3LYP_B3LYPs}
\end{figure}

The resulting $\mathcal{J}_{01}$, $\mathcal{J}_{02}$, and $\mathcal{J}_{12}$ spectral densities for B3LYP and CAMB3LYP excited state calculations can be found in Fig.~\ref{fig:SDs_CAMB3LYP_B3LYPs}. 

The remaining parameters needed to specify the LVC Hamiltonian are the average transition dipole moments of the diabatic S$_1$ and S$_2$ states, as well as the adiabatic energy gaps $\Delta_{01}$ and $\Delta_{02}$ for the S$_1$ and the S$_2$ states respectively. These parameters, together with the coupling reorganization energy $\lambda_{12}^\text{R}$, a measure of the non-adiabatic coupling strength between S$_1$ and S$_2$, as well as the adiabatic energy gap between the first and the second excited state $\Delta_{12}$ for all parameter sets considered in this work can be found in Table~\ref{table:lvc_parameters}. 

\begin{table} [h]
\begin{tabular}{c|c|c|c|c}
    & \multicolumn{2}{c}{CAM-B3LYP} &  \multicolumn{2}{|c}{B3LYP} \\ \hline
     &  MM & 6\AA~QM  & MM & 6\AA~QM\\ \hline \hline
     $|V_{01}|$ (a.u) & 4.4027 & 4.3762 &  3.6116 & 3.6355 \\
     $|V_{02}|$ (a.u) & 0.3752 & 0.3635 &  0.2147 & 0.2024 \\ \hline
     $\Delta_{01}$ (eV) & 2.7371 & 2.6168 &  2.3543 & 2.2353 \\
     $\Delta_{02}$ (eV) & 2.7983 & 2.7040 &  2.3286 & 2.2400 \\
     $\Delta_{12}$ (eV) & 0.0612 & 0.0872 &  -0.0257 & 0.0046 \\ \hline
     $\lambda_{12}^\textrm{R}$ (eV) & 0.0701 & 0.0698 & 0.0746 & 0.0725
\end{tabular}
\caption{Selected LVC Hamiltonian parameters obtained from the different data sets considered in this work.}
\label{table:lvc_parameters}
\end{table}

 Table~\ref{table:lvc_parameters} shows that the B3LYP data set predicts lower transition dipole moments for the diabatic S$_1$ and S$_2$ states as compared to the CAM-B3LYP data sets, a discrepancy that can be ascribed to the effects of the Tamm-Dancoff approximation applied in the CAM-B3LYP results. Further, B3LYP predicts significantly lower adiabatic energy gaps for both S$_1$ and S$_2$, likely due to the different treatment of long-range exchange between the two functionals. More importantly, the gap between the S$_1$ and the S$_2$ minima is only 5~meV in B3LYP/6\AA~QM, as compared to 87~meV for CAM-B3LYP/6\AA~QM. For the B3LYP/MM data set, $\Delta_{12}$ becomes negative, indicating that the minimum of the S$_2$ surface, as predicted from an LVC mapping of energy gap fluctuations measured around the ground state minimum, is actually lower than the S$_1$ minimum. It should be noted that the LVC parameterization obtained in this work is constructed by mapping fluctuations in the Condon region to a fictitious displaced harmonic oscillator model to model the linear absorption spectrum. This paramterization is only expected to be valid within the Condon region and ignores excited state solvent relaxations. Thus, a negative $\Delta_{12}$ does not indicate that the S$_2$ minimum is lower than the S$_1$ minimum in the potential energy surface of the real molecule, but it can be taken as an indication that the two surfaces are far closer together in the Condon region. When comparing MM and 6\AA~QM data sets for both CAM-B3LYP and B3LYP, it becomes clear that treating the solvent environment leads to a slight increase in the S$_1$-S$_2$ energy gap as compared to the MM model of the solvent environment, of the order of 26 to 29 meV for CAM-B3LYP and B3LYP respectively.  

\section{The normalized cross-correlation function for MB}

\begin{figure}
    \centering
    \includegraphics[width=0.6\columnwidth]{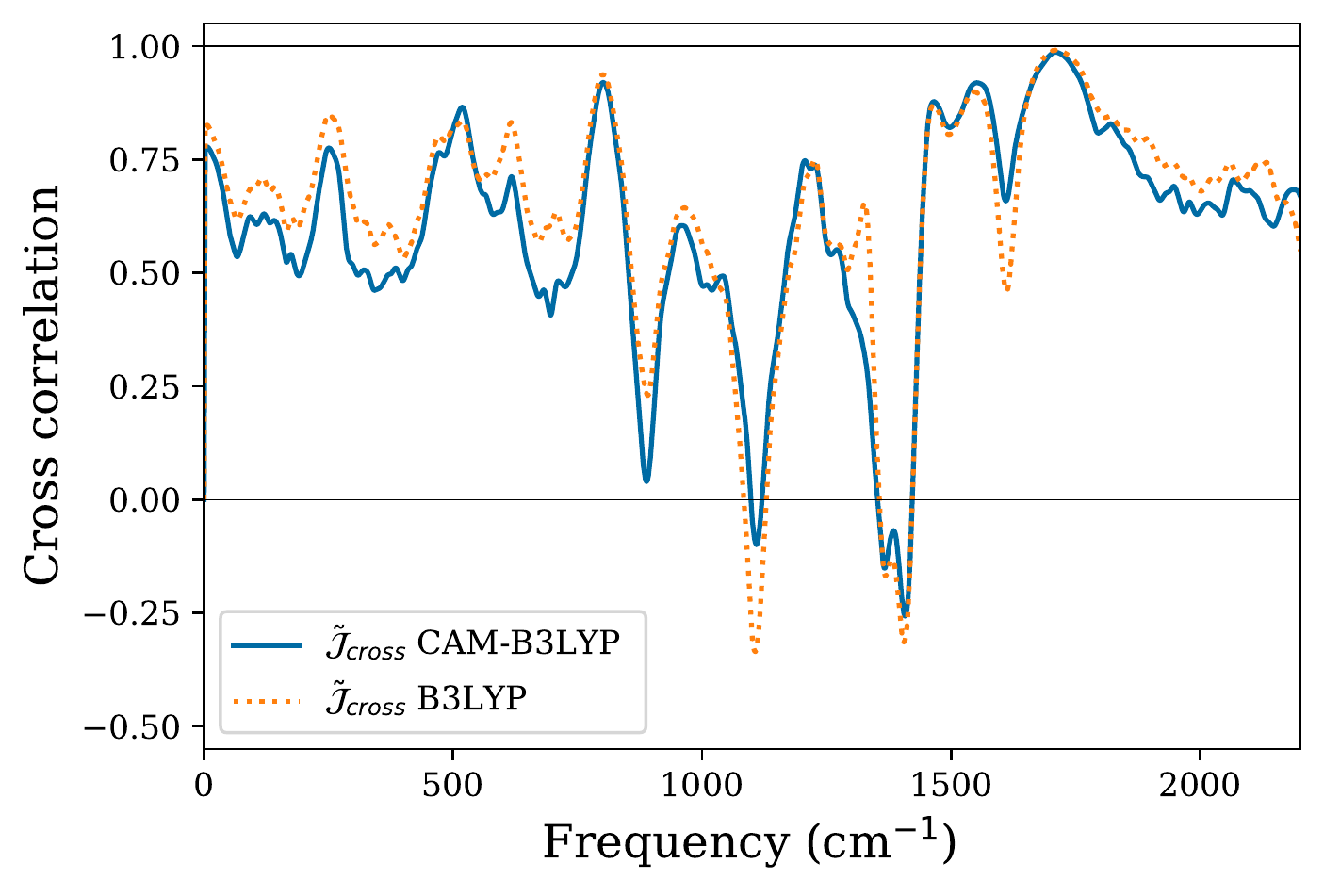}
    \caption{The normalized cross correlation spectral density, defined by $\tilde{\mathcal{J}}_{\text{cross}}=\mathcal{J}_{\text{cross}}/ \sqrt{\mathcal{J}_{01}\mathcal{J}_{02}}$, shows the strength and parity of the correlations in the S$_{1}$ and S$_{2}$ energy fluctuations across the bath spectrum. }
    \label{fig:normcross}
\end{figure}

\noindent
The normalized cross-correlation spectral density for MB, as computed with the B3LYP and the CAM-B3LYP functional for the 6~\AA~QM region is plotted in Fig.~\ref{fig:normcross}. We find that, except at 890, 1120, and 1370/1400 cm$^{-1}$, all environmental modes are between 40\% and 100\% positively correlated, with the most strongly coupled mode at 1690 cm$^{-1}$ being almost fully positively correlated (further analysis is presented in SI Sec.\ref{SI:mode_assign}).

\section{Parameterizing the correlated bath Hamiltonian}\label{SI:bath_parameters}

\noindent
In this section we describe the procedure for determining the elements $g_{k}^{\alpha\beta}$ of the Hamiltonian $H_\text{I}^\text{EL}$ defined as 

\begin{equation}
\label{eq:HI_EL_SI}
    H_\text{I}^\text{EL}=\sum_{k}\sum_{\alpha\beta}g_{k}^{\alpha\beta}\ket{S_{\alpha}}\bra{S_{\alpha}}(a_{\beta k}^{\dagger}+a_{\beta k}).
\end{equation}

The task is to determine the coupling coefficients $g_{k}^{\alpha\beta}$ in terms of the known spectral densities $\mathcal{J}_{01}$, $\mathcal{J}_{02}$ and $\mathcal{J}_{\text{cross}}$. The following derivation is independent of the T-TEDOPA thermal mapping, thus, to include the effects of temperature, one can simply apply the following to the thermal spectral densities $\mathcal{J}_{01}^{\beta}$ etc.  

We note first of all that since these three spectral densities offer a complete description of the reduced system dynamics, there exists a redundancy in our general prescription of $H_\text{I}^\text{EL}$, which contains four free parameters. We therefore make the following simplification without loss of generality:
\begin{equation}
     g_{k}^{12}=c_{k}g_{1k}, \hspace{5pt} \text{where}  \hspace{5pt} g_{1k}=g_{k}^{11} \hspace{5pt} \text{and}
\end{equation}
\begin{equation}
     g_{k}^{21}=c_{k}g_{2k}, \hspace{5pt} \text{where}  \hspace{5pt} g_{2k}=g_{k}^{22}.
\end{equation}
The parameters $c_{k}$ control the strength and parity of the correlations. By setting $c_{k}=0$ we recover the uncorrelated interaction Hamiltonian 

\begin{equation}
\label{eq:HI_EL_uncorrelated_SI}
    H_\text{I}^\text{EL}=-\frac{1}{\sqrt{2}}\sum_{j,\alpha=1}^{2}K_{j}^{\{\alpha\}}\omega_{j}^{\frac{3}{2}}(a_{j\alpha}^{\dagger}+a_{j\alpha})\ket{S_{\alpha}}\bra{S_{\alpha}}.
\end{equation}

Similarly, the FPC and FNC limits are obtained when the $c_{k}$s take values of 1 and -1 respectively.

We define continuous spectral densities based on these parameters:
\begin{equation}
    \mathcal{G}_{\alpha}(\omega)=\sum_{k}g_{\alpha k}^{2}\delta(\omega-\omega_{k}),
\end{equation}
\begin{equation}
    \mathcal{C}(\omega)=\sum_{k}c_{k}^{2}\delta(\omega-\omega_{k}).
\end{equation}

The energy gap fluctuation operators for this model are given by
\begin{equation}
    \delta U_{01}=\int \text{d}\omega\sqrt{\mathcal{G}_{1}(\omega)}(a_{1}^{\dagger}(\omega)+a_{1}(\omega))+ \sqrt{\mathcal{C}(\omega)\mathcal{G}_{1}(\omega)}(a_{2}^{\dagger}(\omega)+a_{2}(\omega)),
\end{equation}  
\begin{equation}
    \delta U_{02}=\int \text{d}\omega\sqrt{\mathcal{G}_{2}(\omega)}(a_{2}^{\dagger}(\omega)+a_{2}(\omega))+\sqrt{\mathcal{C}(\omega)\mathcal{G}_{2}(\omega)}(a_{1}^{\dagger}(\omega)+a_{1}(\omega)).
\end{equation}

Recalling the links between the spectral densities and gap autocorrelation functions given in the main text, namely
\begin{equation}\label{eq:spectral-density2}
    \mathcal{J}_{01}(\omega) = \textrm{i}\theta(\omega) \int \textrm{d}t\ e^{\textrm{i} \omega t} \ \mathrm{Im}\ C_{01}(t),
\end{equation}
where $\theta(\omega)$ is the Heaviside step function and the quantum autocorrelation function of the energy gap fluctuation operator is given by
\begin{equation}\label{eq:energy-gap-fluctuation-autocorrelation-function2}
    C_{01}(t) = \langle \delta U_{01}(\hat{\mathbf{q}}, t) \delta U_{01}(\hat{\mathbf{q}}, 0) \rangle,
\end{equation}
and $\delta U_{01}=\left(H_1-H_0\right)-\omega_{01}^\textrm{av}=U_{01}-\omega_{01}^\textrm{av}$,
we can then define an analogous cross correlation spectral density, yielding

\begin{equation}
\label{eq:J0aplha}
     \mathcal{J}_{0\alpha}(\omega)=\mathcal{G}_{\alpha}(\omega)(1+\mathcal{C}(\omega)),
\end{equation}
\begin{equation}
\label{eq:Jcross}
     \mathcal{J}_{\text{cross}}(\omega)=2\sqrt{\mathcal{G}_{1}(\omega)\mathcal{G}_{2}(\omega)\mathcal{C}(\omega)}.
\end{equation}
Eqs.~\ref{eq:J0aplha} and \ref{eq:Jcross} constitute a system of non-linear equations with the functions $\mathcal{G}_{1}$, $\mathcal{G}_{2}$ and $\mathcal{C}$ as unknowns. Solving for $\mathcal{C}$, we obtain two solutions
\begin{equation}
    \mathcal{C}(\omega)=\frac{2\mathcal{J}_{01}(\omega)\mathcal{J}_{02}(\omega)}{\mathcal{J}_{\text{cross}}(\omega)}\left( 1\pm\sqrt{1-\frac{\mathcal{J}_{\text{cross}}^{2}(\omega)}{\mathcal{J}_{01}(\omega)\mathcal{J}_{02}(\omega)}}\right)-1,
\end{equation}
of which, the solution with the positive root may be discarded as it diverges in the uncorrelated limit ($\mathcal{J}_{\text{cross}}=0$). With the function $\mathcal{C}(\omega)$ determined, we obtain the spectral densities $\mathcal{G}_{\alpha}(\omega)$ via the expression
\begin{equation}
    \mathcal{G}_{\alpha}(\omega) = \frac{\mathcal{J}_{\text{cross}}(\omega)}{2\mathcal{J}_{0\bar{\alpha}}(\omega)\left( 1-\sqrt{1-\frac{\mathcal{J}_{\text{cross}}(\omega)^{2}}{\mathcal{J}_{01}(\omega)\mathcal{J}_{02}(\omega)}}\right)},
\end{equation}
where $\bar{\alpha}=1(2)$ when $\alpha=2(1)$. 
This is an important step as the functions $\mathcal{G}_{\alpha}(\omega)$ will form the basis of the chain mapping which will be described in the following section.
\section{The chain mapping with correlated baths}\label{SI:chain_mapping}

\noindent
In order to render the LVC Hamiltonian amenable to efficient MPS/MPO methods, we are required to perform a transformation which will result in a 1D or quasi 1D topology. The transformation that achieves this is known as the chain mapping. The chain mapping possesses a natural representation in terms of orthogonal polynomials. Starting from an OQS Hamiltonian with linear coupling, one typically proceeds by finding the set of polynomials $\{\tilde{p}_{n} \in \mathbb{P}_{n}, n=0,1,2,...\}$ which are orthonormal with respect to the measure $\mathcal{J}(\omega)\text{d}\omega$, where $\mathcal{J}(\omega)$ is the spectral density. One then constructs a unitary transformation via $U_{n}(\omega)=\sqrt{\mathcal{J}(\omega)}\tilde{p}_{n}(\omega)$, which is applied to the bath modes to obtain a discrete set of chain modes labeled by $n$. Thanks to the special properties of the orthogonal polynomials, namely their orthogonality and the fact that they satisfy a three term recurrence relation, the resulting chain modes couple only to their nearest neighbours and only the first chain mode ($n$=0) couples to the system. The site energies and hopping strengths are determined from the recurrence coefficients of the orthogonal polynomials which are calculated using the ORTHPOL package \cite{gautschi_algorithm_1994}. We refer to Ref.~\cite{Chin2010} for the full details of the chain mapping. 

Due to the off-diagonal couplings in $H_\text{I}^\text{EL}$ in the case of correlated energy level fluctuations, the chain mapping, while still giving a 1D or quasi 1D topology, will contain non-local couplings between the system and the chain modes.

The energy level interaction Hamiltonian, for the general case of correlations determined by a cross correlation spectral density $\mathcal{J}_{\text{cross}}$, is given by
\begin{equation}
    H_\text{I}^\text{EL}=\sum_{\alpha=1}^{2}\int \textrm{d}\omega\left(\sqrt{\mathcal{G}_{\alpha}(\omega)}\ket{S_{\alpha}}\bra{S_{\alpha}}+\sqrt{\mathcal{C}(\omega)\mathcal{G}_{\bar{\alpha}}(\omega)}\ket{S_{\bar{\alpha}}}\bra{S_{\bar{\alpha}}}\right)\left(a_{\alpha}^{\dagger}(\omega)+a_{\alpha}(\omega)\right).
\end{equation}
This interaction Hamiltonian differs from the one normally considered for a chain mapping as each bath couples to two system operators with different couplings. We choose, nonetheless, to perform chain mappings with respect to $\mathcal{G}_{\alpha}(\omega)$. The new chain modes are defined as $b_{\alpha,n}^{(\dagger)}=\int \text{d}\omega U_{\alpha,n}(\omega)a_{\alpha}^{\dagger}(\omega)$, where the unitary transformation is given by $U_{\alpha,n}(\omega)=\sqrt{\mathcal{G}_{\alpha}(\omega)}\tilde{p}_{\alpha,n}(\omega)$ and $\tilde{p}_{\alpha,n}(\omega)$ are orthonormal polynomials satisfying $\int \text{d}\omega\mathcal{G}_{\alpha}(\omega) \tilde{p}_{\alpha,n}(\omega)\tilde{p}_{\alpha,m}(\omega)=\delta_{n,m}$.

This yields
\begin{equation}
\label{eq:HIEL_chainmapped}
    H_\text{I}^\text{EL}=\sum_{\alpha=1}^{2}\sum_{n=0}^{\infty}\left(\underbrace{\int \text{d}\omega\mathcal{G}_{\alpha}(\omega)\tilde{p}_{\alpha,n}(\omega)}_{=\delta_{n,0}/\tilde{p}_{\alpha,0}}\ket{S_{\alpha}}\bra{S_{\alpha}} + \underbrace{\int \text{d}\omega \mathcal{J}_{\text{cross}}(\omega)\tilde{p}_{\alpha,n}(\omega)}_{\vcentcolon=\kappa_{\alpha,n}}\ket{S_{\bar{\alpha}}}\bra{S_{\bar{\alpha}}}\right) \left(b_{\alpha,n}^{\dagger}+b_{\alpha,n}\right),
\end{equation}
where we have used Eq.~\ref{eq:Jcross} to substitute in $\mathcal{J}_{\text{cross}}(\omega)$. Note that it is important to use the original $\mathcal{J}_{\text{cross}}(\omega)$ in this expression rather than substituting the solution for $\mathcal{C}(\omega)$ into Eq.~\ref{eq:Jcross}, as in doing so one would lose the information pertaining to the parity of the correlations.

As indicated in Eq.~\ref{eq:HIEL_chainmapped}, the diagonal couplings reduce to a local interaction between the system energy level and the first site on the chain. For the off-diagonal couplings however, no such simplification is possible as $\mathcal{J}_{\text{cross}}$ bears no relationship in general to the polynomials $\tilde{p}_{\alpha,n}(\omega)$. Instead, the energy level couples to every chain mode via the long-range coupling coefficients $\kappa_{\alpha,n}$, determined via the integrals.

The magnitude of the long-range coupling coefficients depends on the cross correlation spectral density $\mathcal{J}_{\text{cross}}(\omega)$. We note that in the special case where $\mathcal{J}_{\text{cross}}(\omega)=0$, the long-range couplings vanish and one recovers a Hamiltonian with nearest neighbour interactions. We further note that in the FPC(FNC) limit, where $\mathcal{J}_{\text{cross}}(\omega)=+(-)\sqrt{\mathcal{J}_{01}(\omega)\mathcal{J}_{02}(\omega)}$, the interaction Hamiltonian factorizes, so as to reduce to the coupling of one bath with a system operator involving both S$_{1}$ and S$_{2}$ levels. In this latter case, the long-range couplings are still present.

\section{Functional dependence in single molecule calculations }\label{SI:vertical_func}

\noindent
In the main text we explicitly consider the influence of the second singlet state and the quantum mechanical environment on the spectral lineshape. The need for this may be illustrated by the results from more traditional single molecule approaches, such as vertical excitation energies in polarizable continuum modeled (PCM) water or Franck-Condon lineshape calculations using the overlap of normal modes between the ground and excited state. Due to the charge-transfer nature of the excitations we find strong sensitivity to the choice of functional and PCM formalism in terms of transition intensity and energy, as well as the intensity of shoulder features. Despite this sensitivity, no single molecule calculation result appears close to producing the experimental lineshape broadness. The Gaussian 16 package was used for all single molecule calculations. \cite{g16}

\subsection{Vertical excitation}

\noindent
Using a range of functionals (B3LYP, CAM-B3LYP, M06, M06-2X, PBE0, and $\omega$B97X-D)\cite{omegab97,m06funcs,b3lyp,pbe0,Yanai2004} and the 6-31+G* basis, the ground state geometry of the closed-shell methylene blue (MB) cation (+1) was optimized in IEFPCM water and the vertical excitation energies calculated using time-dependent density functional theory (TDDFT).  

The geometry optimization of the ground state was validated through the calculation of vibrational modes, which were later used in the calculation of Franck-Condon lineshapes. No imaginary modes were found for any electronic structure method used, and the calculated vibrational eigenmodes and eigenvalues were comparable to previous experiment and calculation. \cite{Ovchinnikov2016} The ordering of some modes varied between functionals. In this SI we compare the single molecule vibrations of MB with CAM-B3LYP to the spectral densities from molecular dynamics sampling. The underlying frequencies are not exactly equivalent due to the difference in environment modelling and the anharmonicity included in the molecular dynamics approach, however many features are shared, and so features of the spectral density can be described in terms of the ground state normal modes. 

Calculation of the first five singlet vertical excited states was performed using both equilibrium and non-equilibrium PCM formalisms, as well as both full TDDFT and Tamm-Dancoff Approximated TDDFT (TDA). We present results for the first three singlet states, which generally span $\sim 1$ eV, well beyond the window of the MB experimental lineshape. This indicates the relevance of only the first two singlet states (S$_1$ and S$_2$), as the third and beyond are energetically very well removed. 

For all calculations we report a bright first singlet state (oscillator strength $\sim 1.5 $ within the Tamm-Dancoff approximation) and a low intensity (oscillator strength $\sim 0.01)$ second and third singlet excitation. Under the C$_\text{2v}$ symmetry of the ground state, the first singlet excited state state has B$_2$ symmetry, the second A$_1$, and the third B$_1$. The state ordering is generally robust. Depending on the functional and PCM formalism, the excitation energy of the S$_1$ has a range from 2.09 to 2.65, whilst the S$_2$ has a range from 2.47 to 3.07 eV. Generally, using TDDFT over TDA in non-equilibrium PCM leads to a red shift of $\sim 0.2 $ eV for the S$_1$, and $\sim 0.1$ eV for the S$_2$, widening the gap between the electronic excited states. For equilibrium PCM, this shift due to changing from TDA to TDDFT is less pronounced, and the gap between S$_1$ and S$_2$ is consistently larger. We summarize the energy gap between S$_1$ and S$_2$ in SI tables \ref{tab:s1_s2_gap_neqpcm} and \ref{tab:s1_s2_gap_eqpcm}, showing that the values range from 0.06 to 0.67 eV. 

The transition dipole moments of both S$_1$ and S$_2$ excitation in non-equilibrium PCM are significantly (30 - 50 \%) lower than equilibrium PCM. The TDDFT results also have lower transition dipole moments than their TDA counterparts. The most physically motivated results are non-equilibrium PCM TDDFT due to correct solvent response and the fulfillment of the Thomas-Reiche-Kuhn Sum Rules. Here the S$_2$ is consistently 0.5\% the intensity of the S$_1$ excitation. Results from optimizing the S$_1$ and S$_2$ excited state molecular geometry suggest the S$_2$ intensity does not vary significantly at those respective minima. 

\newpage
\begin{table}[h!]
\centering
\resizebox{9cm}{!}{
\begin{tabular}{|c|l|l|l|l|l|l|l|c|l|l|l|l|l|l|}
\hline
      & \multicolumn{6}{c|}{TDA/B3LYP}                                                                                                                                  & \multicolumn{1}{c|}{} &       & \multicolumn{6}{c|}{TDDFT/B3LYP}                                                                                                                                \\ \hline
State & \multicolumn{1}{c|}{eV} & \multicolumn{1}{c|}{Osc. Str.} & \multicolumn{1}{c|}{X} & \multicolumn{1}{c|}{Y} & \multicolumn{1}{c|}{Z} & \multicolumn{1}{c|}{Dip.} & \multicolumn{1}{c|}{} & State & \multicolumn{1}{c|}{eV} & \multicolumn{1}{c|}{Osc. Str.} & \multicolumn{1}{c|}{X} & \multicolumn{1}{c|}{Y} & \multicolumn{1}{c|}{Z} & \multicolumn{1}{c|}{Dip.} \\ \hline
S$_1$    & 2.598                   & 1.488                          & 4.834                  & 0.000                  & 0.000                  & 23.370                    &                       & S$_1$    & 2.323                   & 0.985                          & 4.161                  & 0.000                  & 0.000                  & 17.315                    \\ \hline
S$_2$    & 2.661                   & 0.008                          & 0.000                  & -0.356                 & 0.000                  & 0.127                     &                       & S$_2$    & 2.535                   & 0.004                          & 0.000                  & -0.248                 & 0.000                  & 0.061                     \\ \hline
S3    & 3.402                   & 0.002                          & 0.000                  & 0.000                  & 0.159                  & 0.025                     &                       & S3    & 3.365                   & 0.001                          & 0.000                  & 0.000                  & 0.127                  & 0.016                     \\ \hline
      & \multicolumn{6}{c|}{TDA/CAM-B3LYP}                                                                                                                              & \multicolumn{1}{c|}{} &       & \multicolumn{6}{c|}{TDDFT/CAM-B3LYP}                                                                                                                            \\ \hline
State & \multicolumn{1}{c|}{eV} & \multicolumn{1}{c|}{Osc. Str.} & \multicolumn{1}{c|}{X} & \multicolumn{1}{c|}{Y} & \multicolumn{1}{c|}{Z} & \multicolumn{1}{c|}{Dip.} & \multicolumn{1}{c|}{} & State & \multicolumn{1}{c|}{eV} & \multicolumn{1}{c|}{Osc. Str.} & \multicolumn{1}{c|}{X} & \multicolumn{1}{c|}{Y} & \multicolumn{1}{c|}{Z} & \multicolumn{1}{c|}{Dip.} \\ \hline
S$_1$    & 2.669                   & 1.506                          & 4.799                  & 0.000                  & 0.000                  & 23.029                    &                       & S$_1$    & 2.452                   & 1.141                          & 4.359                  & 0.000                  & 0.000                  & 19.005                    \\ \hline
S$_2$    & 3.058                   & 0.014                          & 0.000                  & -0.428                 & 0.000                  & 0.183                     &                       & S$_2$    & 2.919                   & 0.007                          & 0.000                  & -0.309                 & 0.000                  & 0.095                     \\ \hline
S3    & 3.688                   & 0.003                          & 0.000                  & 0.000                  & 0.175                  & 0.031                     &                       & S3    & 3.633                   & 0.002                          & 0.000                  & 0.000                  & 0.141                  & 0.020                     \\ \hline
      & \multicolumn{6}{c|}{TDA/M06}                                                                                                                                    & \multicolumn{1}{c|}{} &       & \multicolumn{6}{c|}{TDDFT/M06}                                                                                                                                  \\ \hline
State & \multicolumn{1}{c|}{eV} & \multicolumn{1}{c|}{Osc. Str.} & \multicolumn{1}{c|}{X} & \multicolumn{1}{c|}{Y} & \multicolumn{1}{c|}{Z} & \multicolumn{1}{c|}{Dip.} & \multicolumn{1}{c|}{} & State & \multicolumn{1}{c|}{eV} & \multicolumn{1}{c|}{Osc. Str.} & \multicolumn{1}{c|}{X} & \multicolumn{1}{c|}{Y} & \multicolumn{1}{c|}{Z} & \multicolumn{1}{c|}{Dip.} \\ \hline
S$_1$    & 2.594                   & 1.433                          & -4.749                 & 0.000                  & 0.000                  & 22.554                    &                       & S$_1$    & 2.348                   & 0.989                          & -4.145                 & 0.000                  & 0.000                  & 17.185                    \\ \hline
S$_2$    & 2.740                   & 0.010                          & 0.000                  & -0.381                 & 0.000                  & 0.145                     &                       & S$_2$    & 2.611                   & 0.005                          & 0.000                  & -0.275                 & 0.000                  & 0.076                     \\ \hline
S3    & 3.400                   & 0.002                          & 0.000                  & 0.000                  & 0.159                  & 0.025                     &                       & S3    & 3.353                   & 0.001                          & 0.000                  & 0.000                  & 0.126                  & 0.016                     \\ \hline
      & \multicolumn{6}{c|}{TDA/M06-2X}                                                                                                                                 & \multicolumn{1}{c|}{} &       & \multicolumn{6}{c|}{TDDFT/M06-2X}                                                                                                                               \\ \hline
State & \multicolumn{1}{c|}{eV} & \multicolumn{1}{c|}{Osc. Str.} & \multicolumn{1}{c|}{X} & \multicolumn{1}{c|}{Y} & \multicolumn{1}{c|}{Z} & \multicolumn{1}{c|}{Dip.} & \multicolumn{1}{c|}{} & State & \multicolumn{1}{c|}{eV} & \multicolumn{1}{c|}{Osc. Str.} & \multicolumn{1}{c|}{X} & \multicolumn{1}{c|}{Y} & \multicolumn{1}{c|}{Z} & \multicolumn{1}{c|}{Dip.} \\ \hline
S$_1$    & 2.665                   & 1.526                          & -4.833                 & 0.000                  & 0.000                  & 23.361                    &                       & S$_1$    & 2.443                   & 1.141                          & -4.365                 & 0.000                  & 0.000                  & 19.056                    \\ \hline
S$_2$    & 3.066                   & 0.012                          & 0.000                  & -0.395                 & 0.000                  & 0.156                     &                       & S$_2$    & 2.928                   & 0.006                          & 0.000                  & -0.281                 & 0.000                  & 0.079                     \\ \hline
S3    & 3.573                   & 0.003                          & 0.000                  & 0.000                  & 0.172                  & 0.030                     &                       & S3    & 3.471                   & 0.002                          & 0.000                  & 0.000                  & 0.134                  & 0.018                     \\ \hline
      & \multicolumn{6}{c|}{TDA/PBE0}                                                                                                                                & \multicolumn{1}{c|}{} &       & \multicolumn{6}{c|}{TDDFT/PBE0}                                                                                                                              \\ \hline
State & \multicolumn{1}{c|}{eV} & \multicolumn{1}{c|}{Osc. Str.} & \multicolumn{1}{c|}{X} & \multicolumn{1}{c|}{Y} & \multicolumn{1}{c|}{Z} & \multicolumn{1}{c|}{Dip.} & \multicolumn{1}{c|}{} & State & \multicolumn{1}{c|}{eV} & \multicolumn{1}{c|}{Osc. Str.} & \multicolumn{1}{c|}{X} & \multicolumn{1}{c|}{Y} & \multicolumn{1}{c|}{Z} & \multicolumn{1}{c|}{Dip.} \\ \hline
S$_1$    & 2.636                   & 1.499                          & -4.817                 & 0.000                  & 0.000                  & 23.204                    &                       & S$_1$    & 2.376                   & 1.018                          & -4.182                 & 0.000                  & 0.000                  & 17.488                    \\ \hline
S$_2$    & 2.759                   & 0.009                          & 0.000                  & -0.356                 & 0.000                  & 0.126                     &                       & S$_2$    & 2.631                   & 0.004                          & 0.000                  & -0.248                 & 0.000                  & 0.061                     \\ \hline
S3    & 3.480                   & 0.002                          & 0.000                  & 0.000                  & 0.162                  & 0.026                     &                       & S3    & 3.439                   & 0.001                          & 0.000                  & 0.000                  & 0.130                  & 0.017                     \\ \hline
      & \multicolumn{6}{c|}{TDA/wB97X-D}                                                                                                                                & \multicolumn{1}{c|}{} &       & \multicolumn{6}{c|}{TDDFT/wB97X-D}                                                                                                                              \\ \hline
State & \multicolumn{1}{c|}{eV} & \multicolumn{1}{c|}{Osc. Str.} & \multicolumn{1}{c|}{X} & \multicolumn{1}{c|}{Y} & \multicolumn{1}{c|}{Z} & \multicolumn{1}{c|}{Dip.} & \multicolumn{1}{c|}{} & State & \multicolumn{1}{c|}{eV} & \multicolumn{1}{c|}{Osc. Str.} & \multicolumn{1}{c|}{X} & \multicolumn{1}{c|}{Y} & \multicolumn{1}{c|}{Z} & \multicolumn{1}{c|}{Dip.} \\ \hline
S$_1$    & 2.654                   & 1.453                          & 4.727                  & 0.000                  & 0.000                  & 22.344                    &                       & S$_1$    & 2.446                   & 1.139                          & 4.359                  & 0.000                  & 0.000                  & 19.003                    \\ \hline
S$_2$    & 3.071                   & 0.015                          & 0.000                  & -0.453                 & 0.000                  & 0.205                     &                       & S$_2$    & 2.924                   & 0.008                          & 0.000                  & -0.329                 & 0.000                  & 0.108                     \\ \hline
S3    & 3.648                   & 0.003                          & 0.000                  & 0.000                  & 0.174                  & 0.030                     &                       & S3    & 3.592                   & 0.002                          & 0.000                  & 0.000                  & 0.140                  & 0.020                     \\ \hline
\end{tabular}
}
\caption{Functional dependence of vertical excitation energies of the methylene blue cation with the 6-31+G* basis in non-equilibrium PCM water. Excitation energies are given in eV, and transition dipole moments are in atomic units.}
\end{table}
\begin{table}[h!]
\centering
\resizebox{7cm}{!}{

\begin{tabular}{|c|r|c|r|}
\hline
              & \multicolumn{1}{c|}{S$_1$-S$_2$ gap (eV)} &                 & \multicolumn{1}{c|}{S$_1$-S$_2$ gap (eV)} \\ \hline
TDA/B3LYP     & 0.0632                              & TDDFT/B3LYP     & 0.2122                              \\ \hline
TDA/CAM-B3LYP & 0.3885                              & TDDFT/CAM-B3LYP & 0.4675                              \\ \hline
TDA/M06-2X    & 0.4003                              & TDDFT/M06-2X    & 0.4846                              \\ \hline
TDA/M06       & 0.1462                              & TDDFT/M06       & 0.2634                              \\ \hline
TDA/PBE0   & 0.1225                              & TDDFT/PBE0   & 0.2549                              \\ \hline
TDA/wB97X-D   & 0.4179                              & TDDFT/wB97X-D   & 0.4773                              \\ \hline
\end{tabular}
}
\caption{Functional dependence of S$_1$-S$_2$ energy gap at the optimized ground state geometry for 6-31+G*/\emph{non}-equilibrium PCM water.  Excitation energies are given in eV, and transition dipole moments are in atomic units.}
\label{tab:s1_s2_gap_neqpcm}
\end{table}
\newpage

\newpage
\begin{table}[h!]
\centering
\resizebox{9cm}{!}{
\begin{tabular}{|c|r|r|r|r|r|r|l|c|r|r|r|r|r|r|}
\hline
      & \multicolumn{6}{c|}{TDA/B3LYP}                                                                                                                                                 & \multicolumn{1}{c|}{} &       & \multicolumn{6}{c|}{TDDFT/B3LYP}                                                                                                                        \\ \hline
State & \multicolumn{1}{c|}{eV}    & \multicolumn{1}{c|}{Osc. Str.}     & \multicolumn{1}{c|}{X}      & \multicolumn{1}{c|}{Y}      & \multicolumn{1}{c|}{Z}     & \multicolumn{1}{c|}{Dip.}   & \multicolumn{1}{c|}{} & State & \multicolumn{1}{c|}{eV} & \multicolumn{1}{c|}{Osc. Str.} & \multicolumn{1}{c|}{X} & \multicolumn{1}{c|}{Y} & \multicolumn{1}{c|}{Z} & \multicolumn{1}{c|}{Dip.} \\ \hline
S$_1$    & 2.278                      & 1.726                      & 5.560                       & 0.000                       & 0.000                      & 30.916                      &                       & S$_1$    & 2.087                   & 1.310                  & 5.062                  & 0.000                  & 0.000                  & 25.622                    \\ \hline
S$_2$    & 2.581                      & 0.012                      & 0.000                       & -0.443                      & 0.000                      & 0.196                       &                       & S$_2$    & 2.473                   & 0.007                  & 0.000                  & -0.343                 & 0.000                  & 0.118                     \\ \hline
S3    & 3.389                      & 0.003                      & 0.000                       & 0.000                       & 0.192                      & 0.037                       &                       & S3    & 3.355                   & 0.002                  & 0.000                  & 0.000                  & 0.168                  & 0.028                     \\ \hline
      & \multicolumn{6}{c|}{TDA/CAM-B3LYP}                                                                                                                                             & \multicolumn{1}{c|}{} &       & \multicolumn{6}{c|}{TDDFT/CAM-B3LYP}                                                                                                                    \\ \hline
State & \multicolumn{1}{c|}{eV}    & \multicolumn{1}{c|}{Osc. Str.}     & \multicolumn{1}{c|}{X}      & \multicolumn{1}{c|}{Y}      & \multicolumn{1}{c|}{Z}     & \multicolumn{1}{c|}{Dip.}   & \multicolumn{1}{c|}{} & State & \multicolumn{1}{c|}{eV} & \multicolumn{1}{c|}{Osc. Str.} & \multicolumn{1}{c|}{X} & \multicolumn{1}{c|}{Y} & \multicolumn{1}{c|}{Z} & \multicolumn{1}{c|}{Dip.} \\ \hline
S$_1$    & 2.348                      & 1.645                      & 5.347                       & 0.000                       & 0.000                      & 28.590                      &                       & S$_1$    & 2.185                   & 1.464                  & 5.229                  & 0.000                  & 0.000                  & 27.344                    \\ \hline
S$_2$    & 2.959                      & 0.021                      & 0.000                       & -0.533                      & 0.000                      & 0.284                       &                       & S$_2$    & 2.840                   & 0.012                  & 0.000                  & -0.420                 & 0.000                  & 0.176                     \\ \hline
S3    & 3.672                      & 0.004                      & 0.000                       & 0.000                       & 0.213                      & 0.045                       &                       & S3    & 3.621                   & 0.003                  & 0.000                  & 0.000                  & 0.186                  & 0.034                     \\ \hline
      & \multicolumn{6}{c|}{TDA/M06}                                                                                                                                                   & \multicolumn{1}{c|}{} &       & \multicolumn{6}{c|}{TDDFT/M06}                                                                                                                          \\ \hline
State & \multicolumn{1}{c|}{eV}    & \multicolumn{1}{c|}{Osc. Str.}     & \multicolumn{1}{c|}{X}      & \multicolumn{1}{c|}{Y}      & \multicolumn{1}{c|}{Z}     & \multicolumn{1}{c|}{Dip.}   & \multicolumn{1}{c|}{} & State & \multicolumn{1}{c|}{eV} & \multicolumn{1}{c|}{Osc. Str.} & \multicolumn{1}{c|}{X} & \multicolumn{1}{c|}{Y} & \multicolumn{1}{c|}{Z} & \multicolumn{1}{c|}{Dip.} \\ \hline
S$_1$    & 2.279                      & 1.650                      & -5.436                      & 0.000                       & 0.000                      & 29.545                      &                       & S$_1$    & 2.108                   & 1.316                  & -5.047                 & 0.000                  & 0.000                  & 25.475                    \\ \hline
S$_2$    & 2.658                      & 0.015                      & 0.000                       & -0.479                      & 0.000                      & 0.229                       &                       & S$_2$    & 2.547                   & 0.009                  & 0.000                  & -0.383                 & 0.000                  & 0.147                     \\ \hline
S3    & 3.387                      & 0.003                      & 0.000                       & 0.000                       & 0.194                      & 0.038                       &                       & S3    & 3.343                   & 0.002                  & 0.000                  & 0.000                  & 0.167                  & 0.028                     \\ \hline
      & \multicolumn{6}{c|}{TDA/M06-2X}                                                                                                                                                & \multicolumn{1}{c|}{} &       & \multicolumn{6}{c|}{TDDFT/M06-2X}                                                                                                                       \\ \hline
State & \multicolumn{1}{c|}{eV}    & \multicolumn{1}{c|}{Osc. Str.}     & \multicolumn{1}{c|}{X}      & \multicolumn{1}{c|}{Y}      & \multicolumn{1}{c|}{Z}     & \multicolumn{1}{c|}{Dip.}   & \multicolumn{1}{c|}{} & State & \multicolumn{1}{c|}{eV} & \multicolumn{1}{c|}{Osc. Str.} & \multicolumn{1}{c|}{X} & \multicolumn{1}{c|}{Y} & \multicolumn{1}{c|}{Z} & \multicolumn{1}{c|}{Dip.} \\ \hline
S$_1$    & \multicolumn{1}{l|}{2.341} & \multicolumn{1}{l|}{1.680} & \multicolumn{1}{l|}{-5.412} & \multicolumn{1}{l|}{0.000}  & \multicolumn{1}{l|}{0.000} & \multicolumn{1}{l|}{29.293} &                       & S$_1$    & 2.177                   & 1.470                  & -5.249                 & 0.000                  & 0.000                  & 27.549                    \\ \hline
S$_2$    & \multicolumn{1}{l|}{2.965} & \multicolumn{1}{l|}{0.018} & \multicolumn{1}{l|}{0.000}  & \multicolumn{1}{l|}{-0.494} & \multicolumn{1}{l|}{0.000} & \multicolumn{1}{l|}{0.244}  &                       & S$_2$    & 2.846                   & 0.010                  & 0.000                  & -0.384                 & 0.000                  & 0.148                     \\ \hline
S3    & \multicolumn{1}{l|}{3.558} & \multicolumn{1}{l|}{0.004} & \multicolumn{1}{l|}{0.000}  & \multicolumn{1}{l|}{0.000}  & \multicolumn{1}{l|}{0.209} & \multicolumn{1}{l|}{0.044}  &                       & S3    & 3.459                   & 0.003                  & 0.000                  & 0.000                  & 0.176                  & 0.031                     \\ \hline
      & \multicolumn{6}{c|}{TDA/PBE0}                                                                                                                                               & \multicolumn{1}{c|}{} &       & \multicolumn{6}{c|}{TDDFT/PBE0}                                                                                                                      \\ \hline
State & \multicolumn{1}{c|}{eV}    & \multicolumn{1}{c|}{Osc. Str.}     & \multicolumn{1}{c|}{X}      & \multicolumn{1}{c|}{Y}      & \multicolumn{1}{c|}{Z}     & \multicolumn{1}{c|}{Dip.}   & \multicolumn{1}{c|}{} & State & \multicolumn{1}{c|}{eV} & \multicolumn{1}{c|}{Osc. Str.} & \multicolumn{1}{c|}{X} & \multicolumn{1}{c|}{Y} & \multicolumn{1}{c|}{Z} & \multicolumn{1}{c|}{Dip.} \\ \hline
S$_1$    & 2.317                      & 1.725                      & -5.513                      & 0.000                       & 0.000                      & 30.396                      &                       & S$_1$    & 2.136                   & 1.349                  & -5.077                 & 0.000                  & 0.000                  & 25.775                    \\ \hline
S$_2$    & 2.674                      & 0.013                      & 0.000                       & -0.440                      & 0.000                      & 0.194                       &                       & S$_2$    & 2.565                   & 0.007                  & 0.000                  & -0.341                 & 0.000                  & 0.116                     \\ \hline
S3    & 3.467                      & 0.003                      & 0.000                       & 0.000                       & 0.196                      & 0.038                       &                       & S3    & 3.429                   & 0.002                  & 0.000                  & 0.000                  & 0.170                  & 0.029                     \\ \hline
      & \multicolumn{6}{c|}{TDA/wB97X-D}                                                                                                                                               & \multicolumn{1}{c|}{} &       & \multicolumn{6}{c|}{TDDFT/wB97X-D}                                                                                                                      \\ \hline
State & \multicolumn{1}{c|}{eV}    & \multicolumn{1}{c|}{Osc. Str.}     & \multicolumn{1}{c|}{X}      & \multicolumn{1}{c|}{Y}      & \multicolumn{1}{c|}{Z}     & \multicolumn{1}{c|}{Dip.}   & \multicolumn{1}{c|}{} & State & \multicolumn{1}{c|}{eV} & \multicolumn{1}{c|}{Osc. Str.} & \multicolumn{1}{c|}{X} & \multicolumn{1}{c|}{Y} & \multicolumn{1}{c|}{Z} & \multicolumn{1}{c|}{Dip.} \\ \hline
S$_1$    & 2.340                      & 1.584                      & 5.257                       & 0.000                       & 0.000                      & 27.637                      &                       & S$_1$    & 2.178                   & 1.460                  & 5.230                  & 0.000                  & 0.000                  & 27.355                    \\ \hline
S$_2$    & 2.971                      & 0.023                      & 0.000                       & -0.565                      & 0.000                      & 0.319                       &                       & S$_2$    & 2.843                   & 0.014                  & 0.000                  & -0.448                 & 0.000                  & 0.201                     \\ \hline
S3    & 3.632                      & 0.004                      & 0.000                       & 0.000                       & 0.212                      & 0.045                       &                       & S3    & 3.580                   & 0.003                  & 0.000                  & 0.000                  & 0.185                  & 0.034                     \\ \hline
\end{tabular}
}
\caption{Functional dependence of vertical excitation energies of the methylene blue cation with the 6-31+G* basis in equilibrium PCM water}
\end{table}
\begin{table}[h!]
\centering
\resizebox{7cm}{!}{
\begin{tabular}{|c|r|c|r|}
\hline
              & \multicolumn{1}{c|}{S$_1$-S$_2$ gap (eV)} &                 & \multicolumn{1}{c|}{S$_1$-S$_2$ gap (eV)} \\ \hline
TDA/B3LYP     & 0.302                                & TDDFT/B3LYP     & 0.385                                \\ \hline
TDA/CAM-B3LYP & 0.612                                & TDDFT/CAM-B3LYP & 0.655                                \\ \hline
TDA/M06-2X    & 0.624                                & TDDFT/M06-2X    & 0.669                                \\ \hline
TDA/M06       & 0.378                                & TDDFT/M06       & 0.439                                \\ \hline
TDA/PBE0   & 0.358                                & TDDFT/PBE0   & 0.429                                \\ \hline
TDA/wB97X-D   & 0.632                                & TDDFT/wB97X-D   & 0.666                                \\ \hline
\end{tabular}
}
\caption{Functional dependence of S$_1$-S$_2$ energy gap at the optimized ground state geometry for 6-31+G*/equilibrium PCM water. }
\label{tab:s1_s2_gap_eqpcm}
\end{table}

\subsection{Lineshapes from adiabatic Hessian Franck-Condon calculations} \label{SI:AHFC_func}

\begin{figure}[h]
    \centering
    \includegraphics[width=1\textwidth]{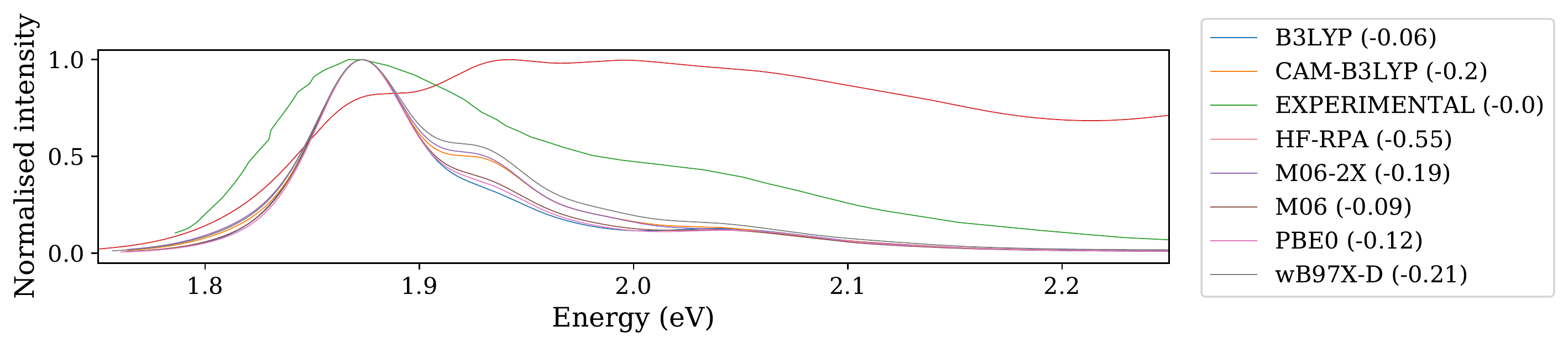}
    \includegraphics[width=1\textwidth]{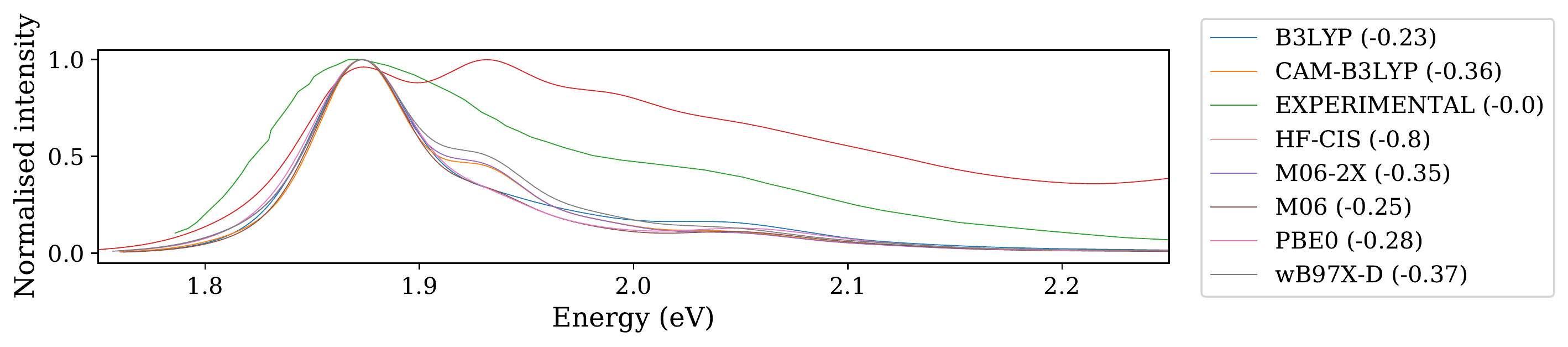}   
    \caption{Functional dependence of single molecule Franck-Condon lineshapes, calculated with TDDFT (top) and TDA (bottom), the 6-31+G* basis, and equilibrium formalism PCM water. Shifts to match experimental maxima are labeled on the right in eV.  }
    \label{tdfc_eqpcm}
\end{figure}

\noindent
We calculated optimized geometries of the ground and S$_1$ electronic excited state of the MB cation before evaluating vibrational properties. The vibrational eigenmodes and frequencies were then used in calculating Franck-Condon lineshapes. Here (SI fig.~\ref{tdfc_eqpcm}) we present adiabatic Hessian Franck-Condon lineshapes calculated by the time-dependent formalism implemented in Gaussian 16. This is shown for a range of DFT functionals for both TDA and TDDFT. We also present Franck-Condon results from a Hartree-Fock ground state paired with a configurational interaction singles (CIS)/ time-dependent Hartree-Fock (HF-RPA). The Franck-Condon response is convoluted with a Gaussian function with a half-width half maximum of 135 cm$^{-1}$. These calculations used an equilibrium solvent formalism, which allows for analytical second derivatives of the excited state to be evaluated. 
Figure \ref{tdfc_eqpcm} shows that, for all functionals, Franck-Condon calculations significantly underestimate the broadness of the experimental lineshape. In this figure the lineshapes have been shifted and normalized to the experimental maxima. Functionals with a larger Hartree-Fock exchange contribution, such as CAM-B3LYP and M06-2X, display a strong local shoulder to the 0-0 maxima. These functionals show the largest gap in S$_1$-S$_2$ energies.  The shoulder is slightly more pronounced for TDDFT than TDA. For both HF-RPA and HF-CIS, the calculations containing pure Hartree-Fock exchange, the calculated spectra are very broad, with a vibronic shoulder that has higher intensity than the 0-0 transition, in disagreement with experiment. It can be concluded that the character of the S$_1$ state is very sensitive to the treatment of long-range exchange in the DFT functional. However, apart from the spurious spectra produced by the pure HF approach, all functionals tested consistently underestimate the pronounced shoulder present in the experimental absorption spectrum.  From here we focus on B3LYP and CAM-B3LYP due to their use in our molecular dynamics sampling/ cumulant calculations and their representative nature of the set of DFT functionals. 

\begin{figure}[h]
    \centering
    \includegraphics[width=1\textwidth]{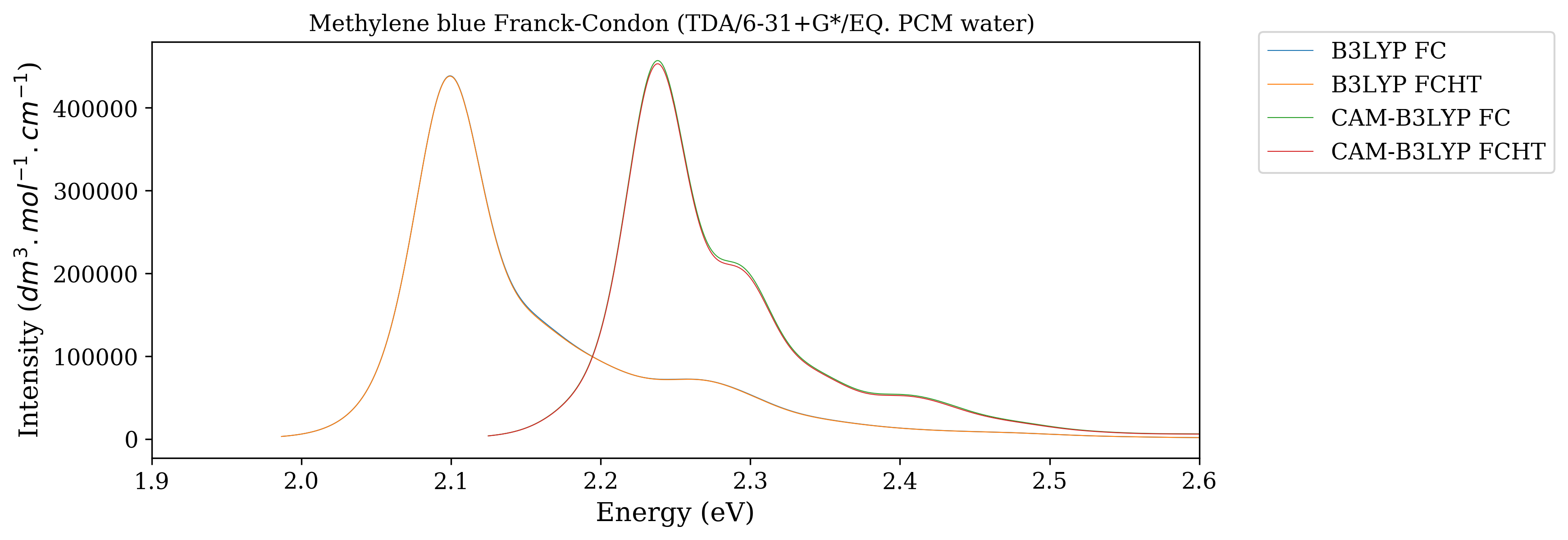}
    \caption{Influence of Herzberg-Teller effects in adiabatic Hessian vibronic calculations of absorption spectral lineshape. }
\end{figure}

For B3LYP and CAM-B3LYP, inclusion of Herzberg-Teller effects have a minimal effect on the vibronic spectral lineshape. For B3LYP, the difference cannot be distinguished, whilst CAM-B3LYP shows a very small increase in intensity, this is apparent in both the local (0.05 eV from 0-0 peak) and tail (0.2 eV from 0-0 peak) features. Overall, the high intensity nature of the S$_1$ leads to negligible contribution from the transition dipole moment derivatives.  \\

\begin{figure}[h]
    \centering
    \includegraphics[width=1\textwidth]{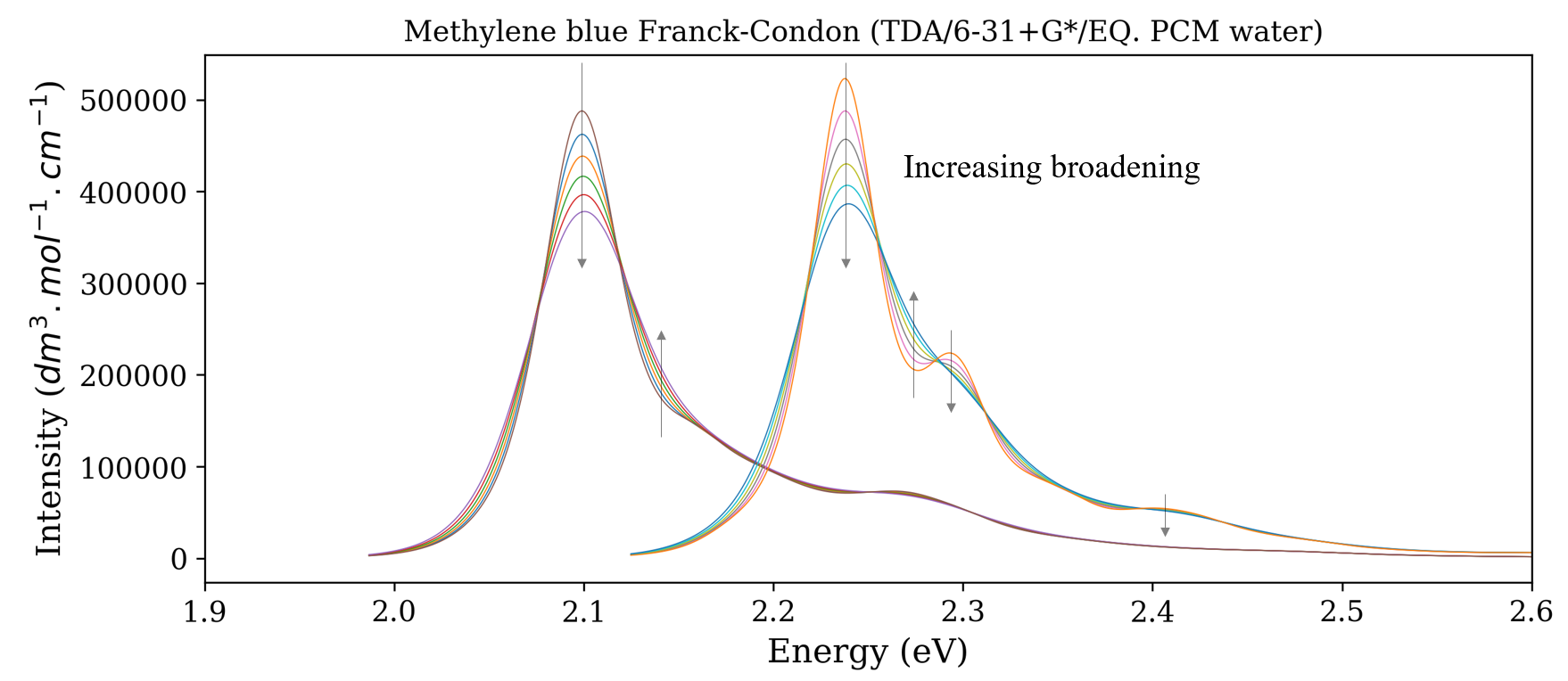}
    \caption{Influence of gaussian broadening on Franck-Condon lineshapes. B3LYP on the left, and CAM-B3LYP on the right, with broadening of 95 to 195 cm$^{-1}$ in increments of 20 cm$^{-1}$  }
\end{figure}

Changing the amount of inhomogeneous broadening can influence the spectral lineshape significantly. As can be seen, varying the gaussian broadening HWHM most significantly influences the 0-0 peak, with an increase in broadening value decreasing the 0-0 peak intensity. Reducing the gaussian broadening also allows shoulder features to be resolved, in particular the local shoulder (0.05 eV from 0-0 peak) that has a strong absorption signal for M06-2X, CAM-B3LYP and $\omega$B97X-D shows an increase in intensity. However, adjusting the broadening does little to resolve the broad experimental shoulder, which remains underestimated. Similarly, the rising edge of the experimental lineshape (SI fig.~\ref{tdfc_eqpcm}) is much rounder than any gaussian broadening gives. 

\begin{figure}[h]
    \centering
    \includegraphics[width=1\textwidth]{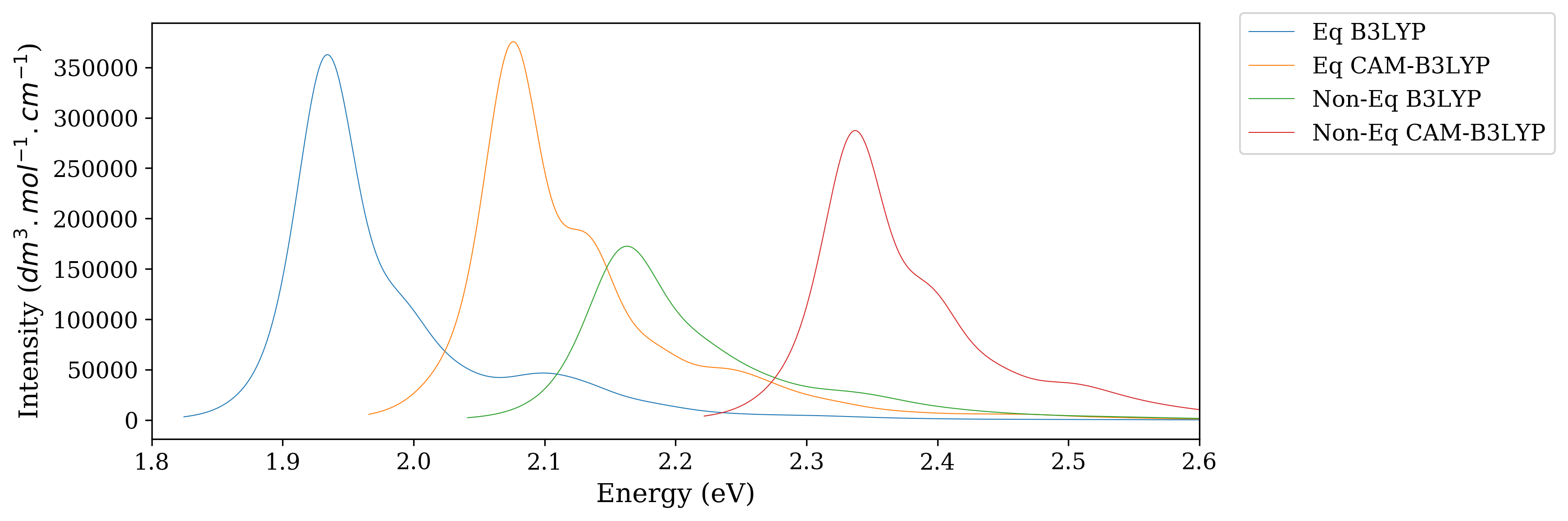}
    \caption{Lineshapes from equilibrium and non-equilibrium PCM formalisms  }
    \label{fig:S1_eq_neq_pcm}
\end{figure}
Noting the significant difference in vertical excitation properties between equilibrium and non-equilibrium solvent response we also investigated Franck-Condon calculations featuring excited states with modes calculated using the non-equilibrium formalism (SI fig.~\ref{fig:S1_eq_neq_pcm}). These spectra are less intense, which is attributable to the change in the transition dipole moment, and broader, which is likely due to the stronger coupling to a low frequency mode. 

\subsection{B3LYP TDDFT in vacuum}\label{SI:dequeiroz}

\begin{figure}[h]
    \centering
    \includegraphics[width=\textwidth]{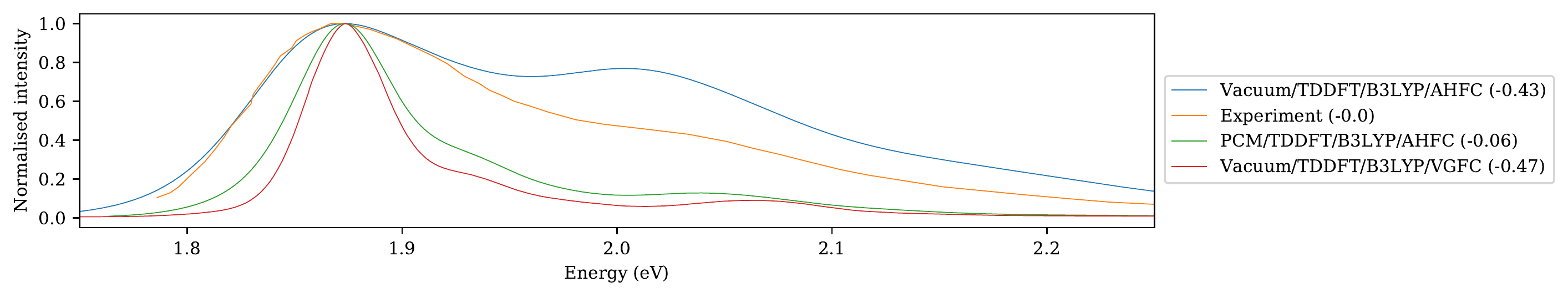}
    \caption{Linear absorption spectra computed from Franck-Condon calculations using B3LYP/TDDFT/6-31+G* }
    \label{leppert_fig}
\end{figure}

\noindent
In a recent study, de Queiroz et. al. performed AHFC calculations of MB in vacuum at the TDDFT/B3LYP/def2-SVP level, which produced a large shoulder in line with experimental results, stemming purely from vibronic transitions to the S$_1$ state.\cite{deQueiroz2021} Our calculations over a range of functionals and lineshape approaches (in the previous section) suggest that this large shoulder is anomalous, occurring only for this specific combination of full TDDFT and B3LYP in vacuum, and is a result of the excited state geometry optimization yielding a state that is of mixed S$_1$-S$_2$ character. To demonstrate this, we here present analogous lineshapes of vacuum/TDDFT/B3LYP/AHFC (but with our chosen basis set) in contrast to the same calculation in PCM solvent and using the vertical gradient (VGFC) model (SI fig.~\ref{leppert_fig}). 

The vacuum/AHFC in this figure clearly shows a dramatic increase in shoulder intensity that is not found for the other parameters presented here, nor for any of the functionals examined in the previous section. The cause of this can seen by examining the electronic states from excited state geometry optimization (SI Table \ref{leppert_table}). 
\begin{table}[h]
\resizebox{0.7\textwidth}{!}{%
\begin{tabular}{|c|c|c|c|c|}
\hline
                                          & S$_1$  energy       & S$_1$ Osc. Str.            & S$_2$  energy      & S$_2$  Osc. Str.            \\ \hline
B3LYP/TDDFT/Vacuum/S$_0$ geometry            & 2.389          & 0.773          & 2.496         & 0.003          \\ \hline
\textbf{B3LYP/TDDFT/Vacuum/S$_1$  geometry} & \textbf{2.224} & \textbf{0.382} & \textbf{2.514} & \textbf{0.351} \\ \hline
B3LYP/TDDFT/PCM/S$_0$ geometry               & 2.087          & 1.310     & 2.473         & 0.007             \\ \hline
B3LYP/TDDFT/PCM/S$_1$ geometry                & 2.007         & 1.264         & 2.357         & 0.006          \\ \hline
\end{tabular}%
}
\caption{Excitation energies (eV) and oscillator strengths as calculated with B3LYP/TDDFT at the S$_0$ and S$_1$ optimized geometries, demonstrating the failure of the Condon approximation and state mixing that occurs during vacuum excited state optimization (bold)}
\label{leppert_table}
\end{table}

The dipole lending that occurs for vacuum/B3LYP/TDDFT at the S$_1$ geometry suggest that significant mixing has occurred between the two states during optimization, and the electronic state has changed significantly. This is quite a curious effect as the energy gap between the states is consistently large. We do not find this mixing for any other parameter set, for example the same calculation in PCM. Similarly, if one performs a vertical gradient approach, which does not utilize the excited state optimized geometry, the shoulder is not present. It should also be noted that the change in oscillator strength is obscured in the normalized lineshapes. Comparing the unnormalized lineshape, the vacuum AHFC is approximately half the intensity.

\section{Adiabatic cumulant lineshapes}\label{SI:adiabatic_cumulant}

\begin{figure}[h]
    \centering
    \begin{subfigure}[b]{0.49\textwidth}
    \centering
    \includegraphics[width=\textwidth]{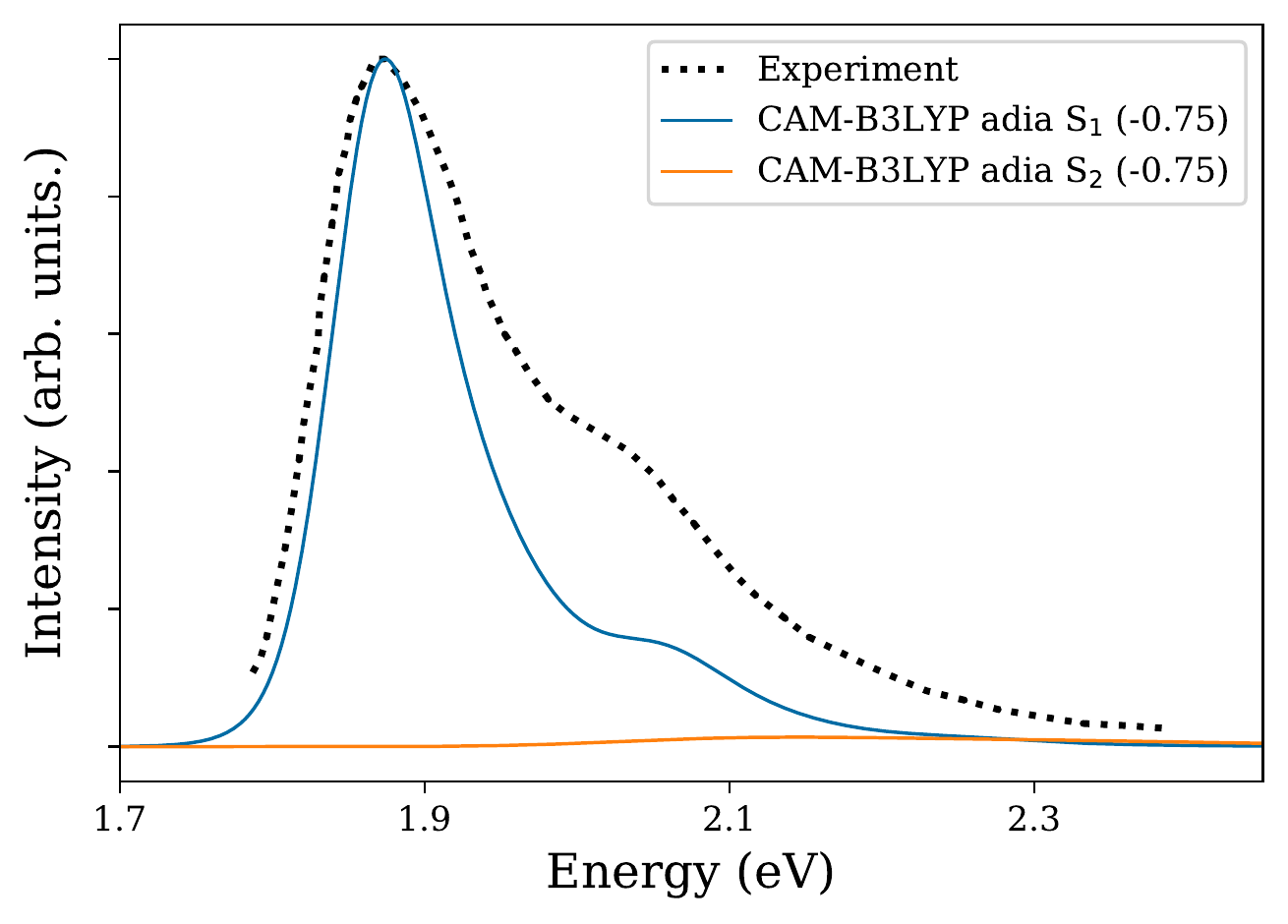}
    \end{subfigure}
    \begin{subfigure}[b]{0.49\textwidth}
    \centering
    \includegraphics[width=\textwidth]{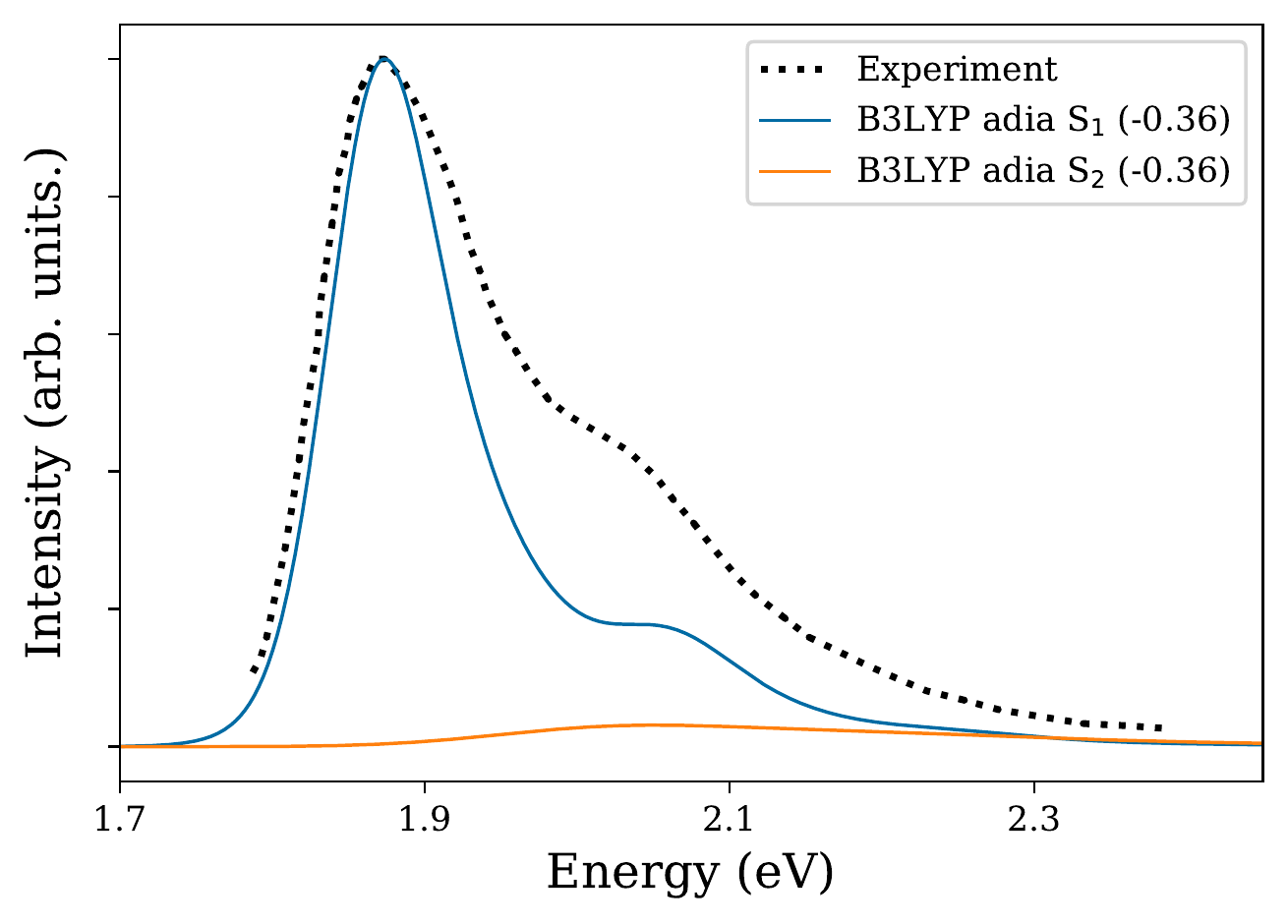}
    \end{subfigure}
    \caption{Linear absorption spectra computed from the \textit{adiabatic} energy gap fluctuations of the S$_1$ and S$_2$ state calculated with CAM-B3LYP/TDA/6-31+G* and B3LYP/TDDFT/6-31+G*, both with a 6~\AA~QM region. Spectra are calculated in the 2nd order cumulant approximation. }
    \label{fig:adiabatic_cumulant}
\end{figure}

\noindent
Calculation of the linear absorption lineshape of MB using the \emph{adiabatic} energy gap fluctuations produces a larger spectral shoulder than the diabatized approach or single molecule approaches (SI fig.~\ref{fig:adiabatic_cumulant}). Whilst this result appears attractive, this is an erroneous result of the statistical nature of the cumulant. Firstly, the electronic states are not consistently sampled here - there are fluctuations in the S$_1$ due to the S$_2$ and vice versa. Secondly, the Condon approximation is broken as the transition dipole moment of both S$_1$ and S$_2$ varies strongly from snapshot to snapshot, from fully separated configurations where S$_1$ is bright and S$_2$ is dark, to highly mixed states where both electronic excitations have comparable oscillator strength. In relation to the stated analysis in the main text, the crossings are much more frequent for a smaller energy gap, i.e. B3LYP/TDDFT crossings are much more common than CAM-B3LYP/TDA, and therefore the average dipole lending is significantly larger. This leads to the CAM-B3LYP S$_2$ being lower intensity than B3LYP - both S$_2$ states have negligible intensity of their own. 

\newpage
\section{Validity of the linear coupling model in MB}\label{SI:BOM_2nd_3rd}

\noindent
In the BOM and LVC Hamiltonians discussed in the main manuscript, it is assumed that the nuclear vibrations couple linearly to electronic degrees of freedom, and that the coupling between S$_1$ and S$_2$ is equally linear with respect to nuclear degrees of freedom. This treatment of electron-nuclear coupling is approximate, and is only expected to hold in cases where the energy-gap fluctuations obey Gaussian statistics. For example, even when approximating the nuclear degrees of freedom of a system as harmonic, as it is done in the static Franck-Condon calculations presented in the previous section, changes in curvature between ground- and excited state PESs and Duschinsky rotation effects generally lead to non-linear couplings of nuclear degrees of freedom to the energy gap fluctuations\cite{Zuehlsdorff2019}. 

Some of the authors have recently introduced an approach, that, within the simplified BOM Hamiltonian where any off-diagonal coupling between excited states is ignored, allows for the inclusion of non-linear coupling effects by computing a third order correction term to the cumulant expansion\cite{Zuehlsdorff2019}. Although this approach is only valid for a two-level system and not the LVC Hamiltonian explicitly considered in this work, computing the third order correction term can give some insight into whether the approximation of only considering linear couplings of nuclear degrees of freedom to the electronic excitations is justified. 

\begin{figure}[h]
    \centering
    \begin{subfigure}[b]{0.49\textwidth}
    \centering
    \includegraphics[width=\textwidth]{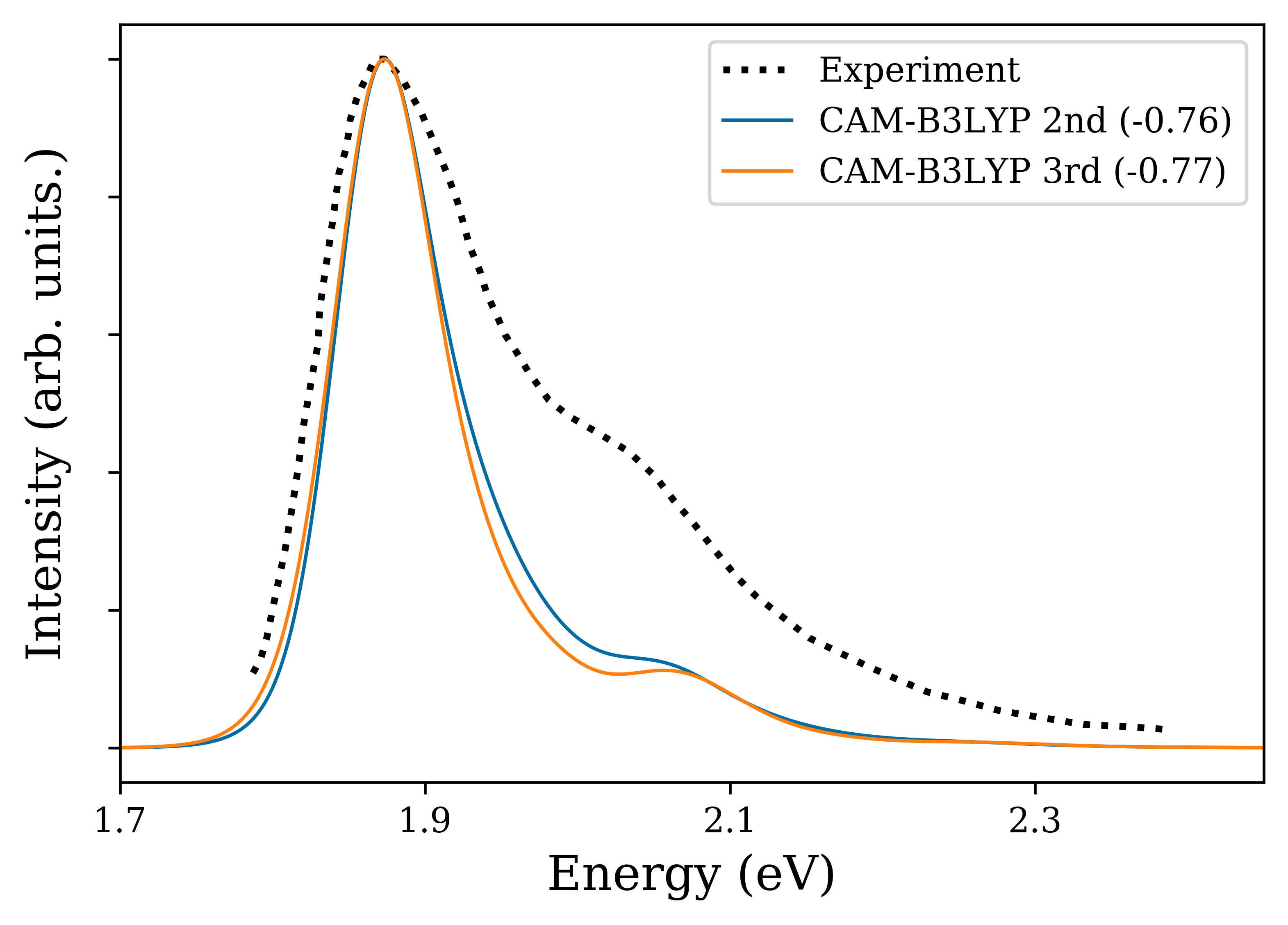}
    \end{subfigure}
    \begin{subfigure}[b]{0.49\textwidth}
    \centering
    \includegraphics[width=\textwidth]{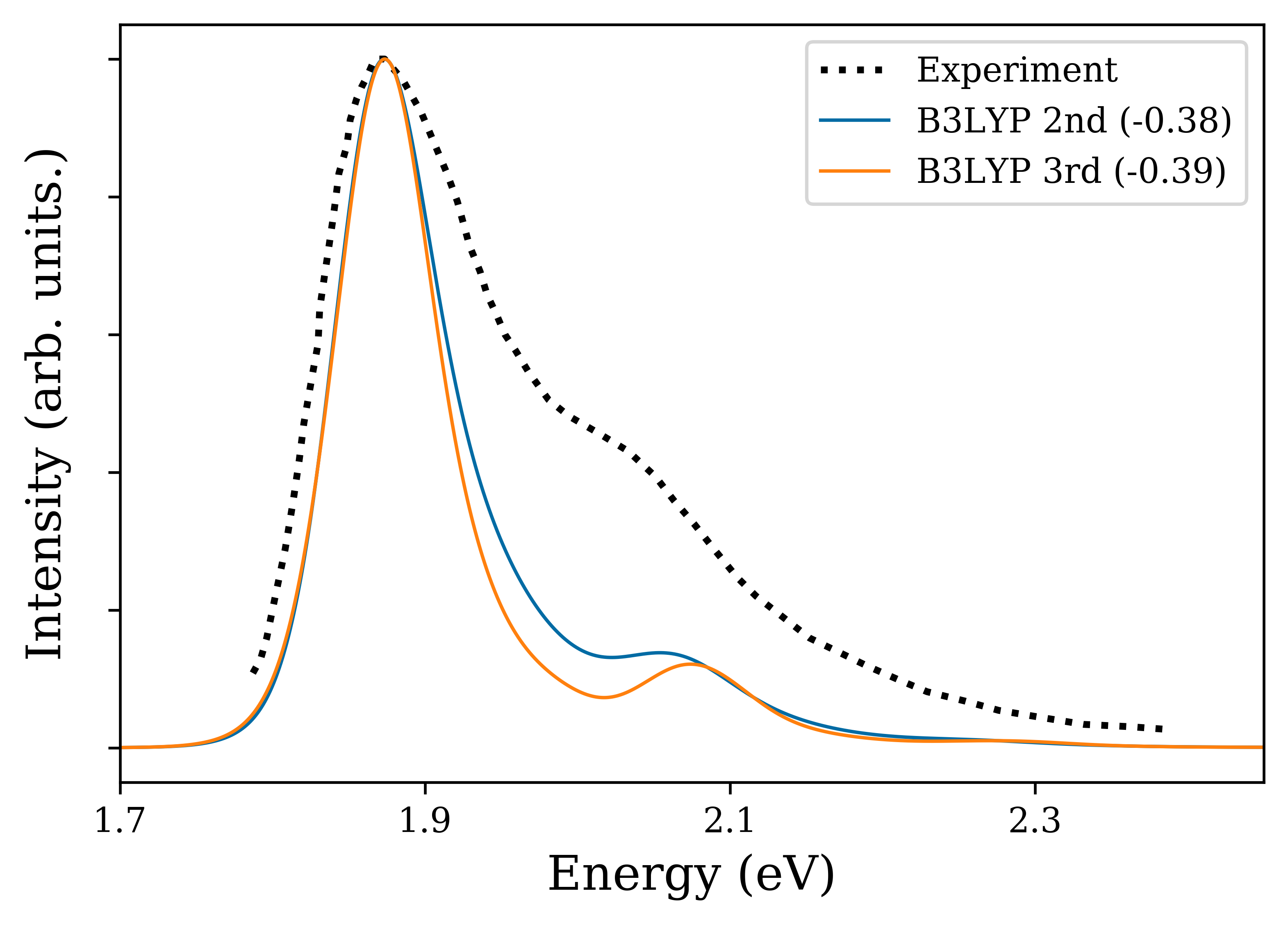}
    \end{subfigure}
    \caption{Linear absorption spectra computed from the diabatic energy gap fluctuations of the S$_1$ state calculated with CAM-B3LYP/TDA/6-31+G* and B3LYP/TDDFT/6-31+G*, both with a 6~\AA~QM region. Spectra are calculated in both the 2nd order cumulant approximation (blue) and by also including the 3rd order cumulant correction term (orange). }
    \label{fig:cumulant_camb3lyp}
\end{figure}

SI fig.~\ref{fig:cumulant_camb3lyp} shows the linear absorption spectrum as computed for the diabatized CAM-B3LYP and B3LYP data sets in a 6~\AA~QM region due to the S$_1$ state only, both in the second order cumulant approximation that is equivalent to the linear coupling model employed in the main manuscript and in the third order cumulant expansion. As can be seen, for CAM-B3LYP the two spectra are in good agreement with each other, with only slight changes in the shape of the 0-0 peak and the vibronic shoulder. This good agreement suggests that, at least for the coupling of nuclear degrees of freedom to the S$_0$-S$_1$ energy gap, the linear coupling approximation is likely valid. For the B3LYP data set, the change in going from the 2nd order cumulant to the 3rd order cumulant approximation is more pronounced but follows a similar trend, in that the shape of the vibronic shoulder is altered. However, in either case the 3rd order cumulant correction does not increase the intensity of the vibronic shoulder, suggesting that the lack of intensity in the shoulder of the computed absorption spectrum for the BOM Hamiltonian cannot be ascribed to approximating the coupling of nuclear degrees of freedom to the energy gap as linear. 

\begin{table}[]
\begin{tabular}{|c|c|c|}
\hline
Data set                                                              & S$_1$ Skewness & S$_2$ Skewness \\ \hline
CAM-B3LYP TDA MM                                                      & -0.2327     & -0.0446     \\ \hline
CAM-B3LYP TDA 6\AA \, QM & -0.0030     & -0.0001     \\ \hline
B3LYP TDDFT MM                                                        & -0.5567     & -0.1067     \\ \hline
B3LYP TDDFT 6\AA \, QM   & -0.4697     & -0.0901     \\ \hline
\end{tabular}
\caption{Skewness parameter for the diabatic energy gap fluctuations, as calculated from the different data sets used in this work. Each data set corresponds to four trajectories, with a total of 16,000 data points. }
\label{tab:skewness}
\end{table}

The results presented in Fig.~\ref{fig:cumulant_camb3lyp} can be further interpreted by quantifying the degree of non-Gaussian behavior in the energy gap fluctuations. In line with previous work of two of the authors\cite{Zuehlsdorff2019}, this can be done by calculating the skewness $\gamma$ of the data sets of energy gap fluctuations (see Table~\ref{tab:skewness}). As can be seen, the skewness $\gamma$ for the CAM-B3LYP 6\,\AA\,QM data set is close to zero. For the B3LYP data sets, the skewness values are considerably larger, suggesting that the energy gap fluctuations have some non-Gaussian character. This is directly represented by the larger discrepancy between the 2nd and the 3rd order cumulant results for the B3LYP\,6\,\AA\,QM data set. This increase in skewness for B3LYP may be due to the mismatch between the Hamiltonian for the ground state dynamics and the excitations calculations. We also note that the QM data sets have less skewed statistics, which can be attributed to a larger inhomogeneous solvent broadening in full QM environments as compared to classical point charge models, consistent with results found for other systems by some of the authors\cite{Zuehlsdorff2020}.

In summary, it is clear that mapping the energy gap fluctuations for the S$_1$ state of MB to a linear coupling model is only approximately valid. However, the analysis performed here suggests that non-linear coupling effects are likely insufficient to explain the lack of intensity in the shoulder of the S$_1$ cumulant spectrum in comparison to the experimental spectrum, and other effects such as the non-adiabatic coupling between the S$_1$ and the nearby S$_2$ state have to be accounted for. 

\section{B3LYP vs. CAM-B3LYP dynamics}\label{SI:dynamic_mismatch}

\noindent
In the datasets presented in the main manuscript, QM/MM dynamics are performed using the CAM-B3LYP functional for the chromophore and MM point charges for the environment. TDDFT calculations are then performed on the snapshots generated by the QM/MM dynamics either using the CAM-B3LYP or the B3LYP functional and different amounts of QM environment. For the B3LYP functional, this means that the dynamics are carried out with a different functional than is used for the TDDFT calculations, creating a mismatch between the Hamiltonian that drives the dynamics and the Hamiltonian that determines the energy-gap fluctuations. 

\begin{figure}
    \centering
    \includegraphics[width=0.5\textwidth]{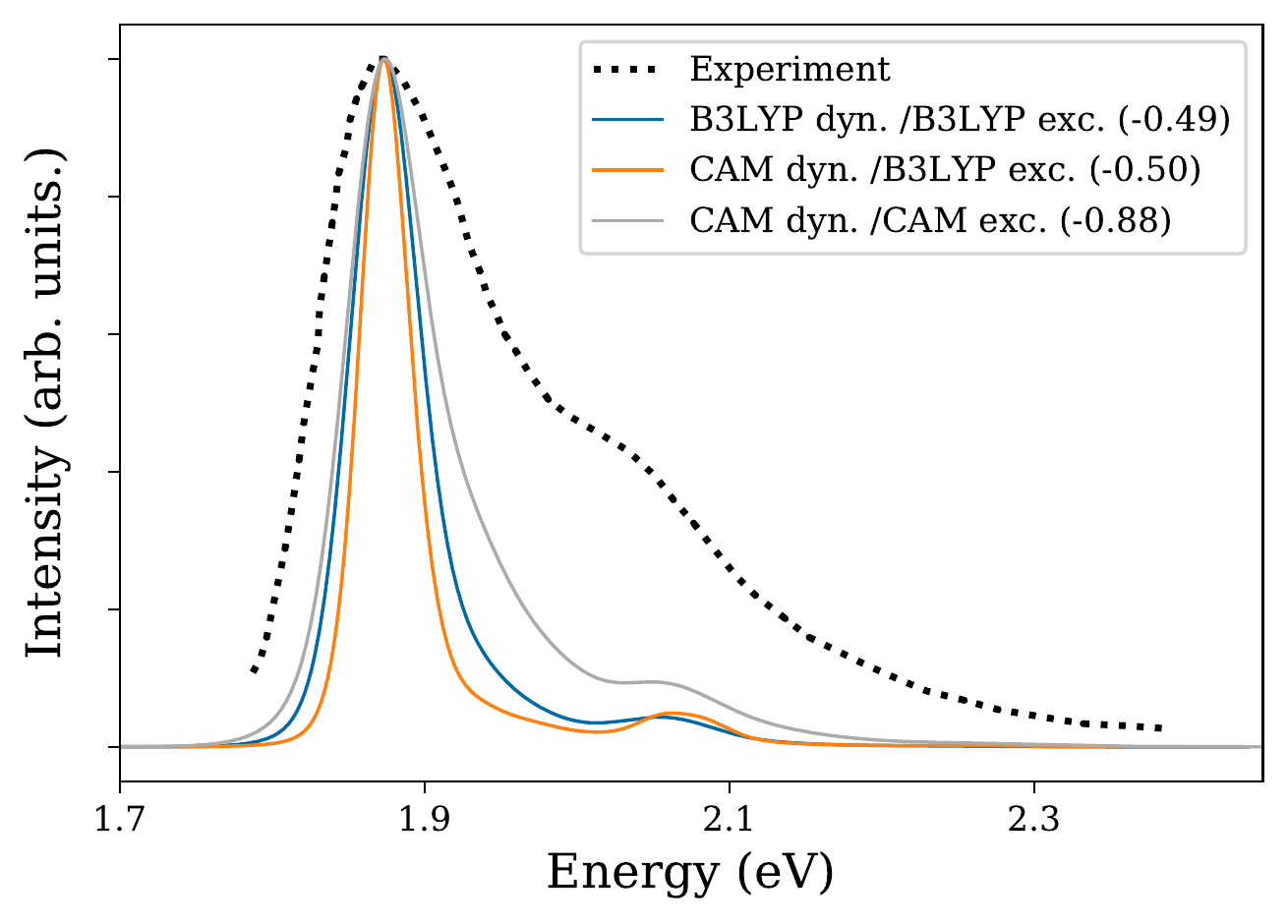}
    \caption{Linear absorption spectra computed for the diabatic S$_1$ energy gap fluctuations in the 2nd order cumulant approximation, using a pure MM representation of the solvent environment. The blue and the red lines use the four ground state CAM-B3LYP trajectories and compute vertical excitation energies either with B3LYP (orange) or CAM-B3LYP (blue). The grey line corresponds to a single 8~ps trajectory using the B3LYP functional both for the ground state dynamics and the excited state TDDFT calculations. }
    \label{fig:cumulant_camb3lyp_B3LYP}
\end{figure}

To assess whether this mismatch in the Hamiltonians is expected to have a strong influence on the computed spectra, we compute a single 10~ps QM/MM trajectory of MB in water, where the B3LYP functional is used for the ground state dynamics of the chromophore and the same simulation parameters are used as for the CAM-B3LYP trajectories discussed in the main text. For this single trajectory, the first 2~ps are discarded for equilibration and excitation energies are computed using B3LYP at the full TDDFT level, where the environment is treated through MM point charges. We compute the linear absorption spectrum of the diabatic S$_1$ state for this trajectory that is free of any mismatch between the Hamiltonian driving the dynamics and the Hamiltonian used for generating the energy gap fluctuations. Fig.~\ref{fig:cumulant_camb3lyp_B3LYP} shows the resulting spectrum, in comparison to the linear spectra based on the CAM-B3LYP ground state dynamics of a single trajectory, once where the energy gap fluctuations are evaluated using CAM-B3LYP TDA and once where the B3LYP within the full TDDFT formalism is used. The environment is treated as classical MM point charges in all energy gap calculations. 

Fig. \ref{fig:cumulant_camb3lyp_B3LYP} shows that the cumulant spectra based on CAM-B3LYP and B3LYP dynamics produce almost identical lineshape if excitations are computed at the B3LYP TDDFT level. This indicates that the driving factor between differences in the diabatic S$_1$ lineshapes between the B3LYP and CAM-B3LYP data sets shown in the main manuscript is the choice of the excited state functional and that the mismatch in Hamiltonian between ground state dynamics and excited state calculations for the B3LYP data sets has minor impact on the computed spectra. 

\begin{figure}
    \centering
    \includegraphics[width=1\textwidth]{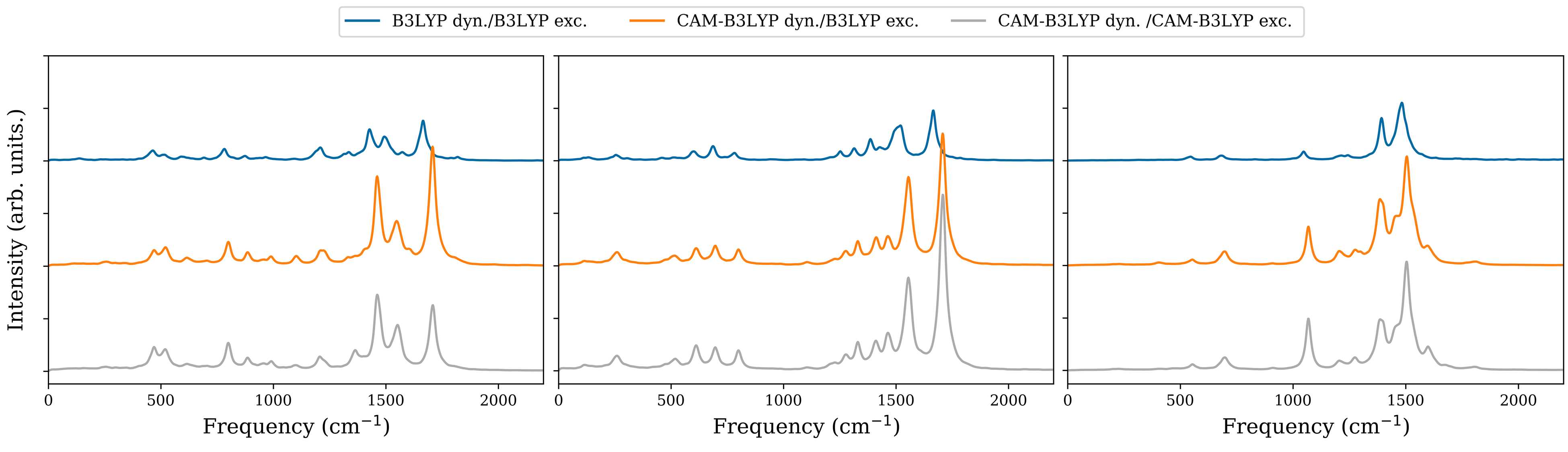}
    \caption{Spectral densities for different functional dynamics and excitations. $\mathcal{J}_{01}$ left, $\mathcal{J}_{02}$ middle, $\mathcal{J}_{12}$ right.  }
     \label{fig:cumulant_camb3lyp_b3lyp_SD}
\end{figure}

Fig.~\ref{fig:cumulant_camb3lyp_b3lyp_SD} shows the spectral densities for the linear spectra presented in Fig.~\ref{fig:cumulant_camb3lyp_b3lyp_SD}. The main difference between the B3LYP and CAM-B3LYP dynamics can be seen in the frequency shift of some high frequency vibronic peaks, indicating that the B3LYP functional predicts slightly different ground state vibrational frequencies for MB compared to the CAM-B3LYP functional. However, most features present in the spectral densities for the mismatched CAM-B3LYP dynamics/B3LYP excitation energies data set can also be found in the pure B3LYP data set with fully consistent Hamiltonians, again indicating that the mismatch in Hamiltonians in the B3LYP data sets presented in the main manuscript only has a minor influence on the computed spectra. 

\section{Identifying prominent peaks in the spectral densities} \label{SI:mode_assign}

\begin{figure}[h]
    \centering
    \includegraphics[width=1\textwidth]{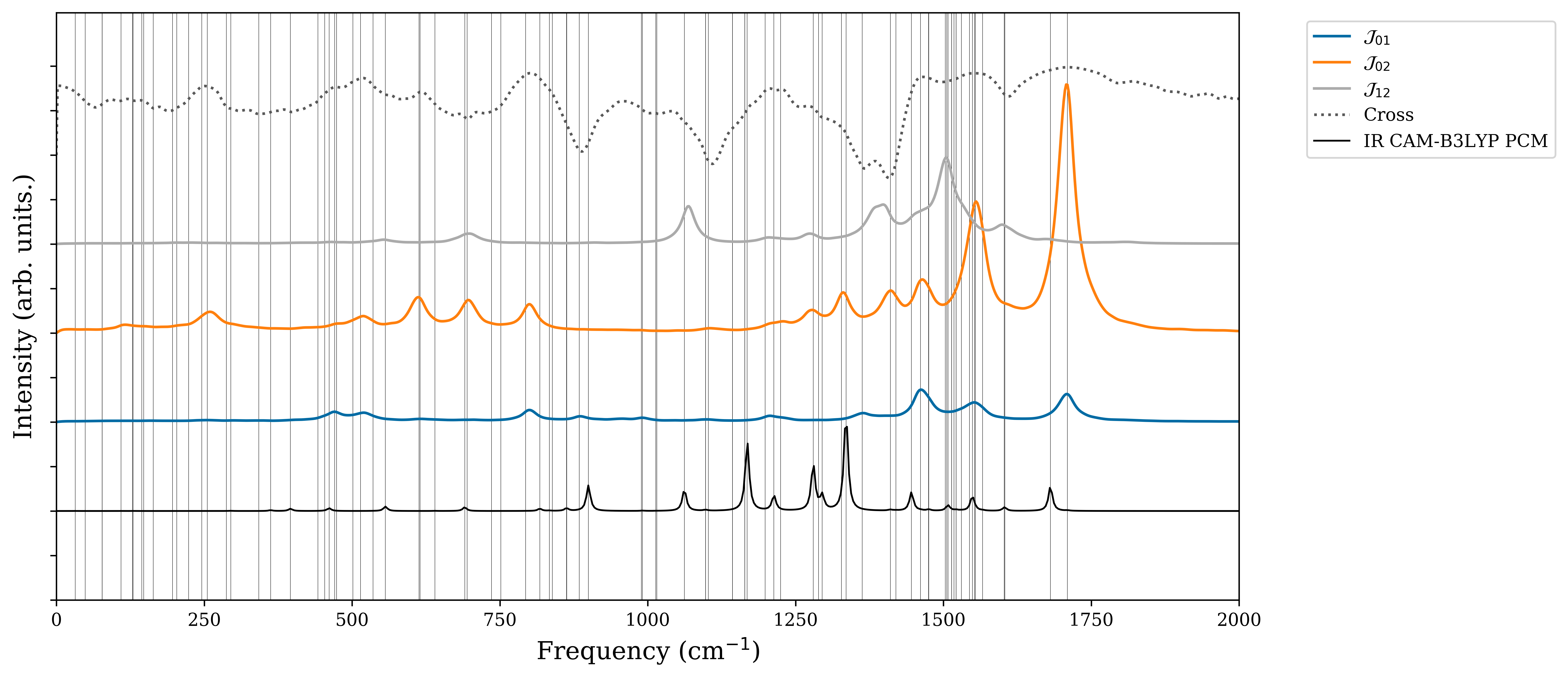}
    \caption{Spectral densities of diabatic S$_1$ and S$_2$ and their coupling $\mathcal{J}_{12}$ for the CAM-B3LYP/TDA/6\AA\, QM data set. The cross correlation of S$_1$ and S$_2$ fluctuations is also presented at the top, and the infrared spectra obtained from vibrational analysis of the nuclear hessian in PCM water (as SI section 1) is presented at the bottom. The vertical lines represent the set of normal mode frequencies. }
    \label{fig:mode_assign}
\end{figure}

\noindent
In Fig.~\ref{fig:mode_assign} we present various spectral densities of the CAM-B3LYP/TDA/6 \AA \, QM system overlaid with the normal mode frequencies from ground state CAM-B3LYP/PCM water. 

Firstly, we recall that the cross correlation has an average value of 0.8, indicating that dynamics between the S$_1$ and S$_2$ states are in general fairly strongly correlated across the entire spectral window presented. The exceptions to this are the three dips at 890, 1120, and 1370/1400 cm$^{-1}$. Surprisingly, there is no clear link between these frequencies and the other quantities presented here, for example these modes do not display strong coupling to one state but not the other. It seems likely that these frequencies relate to normal modes \#43, 52, 66, and 68 respectively (counting ordered by frequency and discounting translation and rotational modes). Whilst there is some ambiguity in this assignment due to the proximity of other normal mode frequencies, intuition for the influence of the mode on the electronic state aids in assignment. For example, the dip at 1120 cm$^{-1}$ could be reasonably assigned to the mode at 1097, 1102, or 1143 cm$^{-1}$. However, the lower and higher frequency modes here represent as/symmetric methyl rocking modes which should have little influence on the electronic states, whilst mode \#52 at 1102 cm$^{-1}$ is a strong C-S-C symmetric stretch likely to effect the electronic coupling. Mode \#43 has a strong stretch in the S- to para- position N coupled to aromatic ring stretching, and mode \#66 and 68 are para-position nitrogen strong symmetric stretching and rocking modes. All these modes are totally symmetric with respect to the C$_\text{2v}$ point group (A$_1$ symmetry). It seems rational that these modes correspond to fluctuations in the central S-N distance which, based upon the transition densities in the main text, couple to the diabatic S$_2$ state with reduced coupling to S$_1$ fluctuations.

The diabatic coupling spectral density $\mathcal{J}_{12}$ shows small features at 550(\#29) , 700(33/34), 1210(59), 1280(61), 1600(87)  cm$^{-1}$ and larger peaks at 1080(49), 1400($\sim$67), and congested around 1500(73-86) cm$^{-1}$. If one chooses to assign the peak at 700 cm$^{-1}$ to normal mode 33, then these modes all present themselves as in-plane asymmetric modes with B$_2$ symmetry. This shows that there is a set of asymmetric modes with both high and low frequency which push the states together. These modes show significant asymmetric stretching and rocking at the C-S-C site. 

$\mathcal{J}_{02}$ has both many and the most intense peaks in spectral density, indicating a strong coupling of the S$_0$-S$_2$ energy gap to motion along the specific ground state modes at those frequencies. These modes occur across the entire spectral window, as low as 110(\#5) cm$^{-1}$ and with an extremely strong feature at 1710(\#90) cm$^{-1}$. All $\mathcal{J}_{02}$ peaks correspond to modes which are in-plane symmetric modes (A$_1$) which match the symmetry of the S$_2$ electronic state. Many of these normal modes can be assigned to either symmetric stretches around the central sulphur or nitrogen atoms, or symmetric c-c stretches in the conjugated rings (as is the case for the strong feature at 1710~cm$^{-1}$). 

$\mathcal{J}_{01}$ is generally significantly less intense than $\mathcal{J}_{02}$ and they share many peaks. Assuming that they share similar dynamics, the lack of intensity may be due to the weaker charge-transfer to the sulphur atom for diabatic S$_1$ than S$_2$, leading the a less pronounced impact of the symmetric stretches of the central ring on the S$_0$-S$_1$ energy gap. Peaks can be resolved in $\mathcal{J}_{01}$, and are mostly assigned to in-plane symmetric stretches (A$_1$ symmetry). For example, 250(\#14), 520(27), 800(37), 880(43), 990(45), 1200(58), 1360(66), 1460(70), and 1710(90) cm$^{-1}$. The small peak at 470 cm$^{-1}$ is reasonably further (30 cm$^{-1}$) from any A$_1$ mode that it is possible to be coupled to a mode of alternate symmetry in relation to the electronic symmetry of S$_1$ being B$_{2}$. However, anharmonic effects and differences in molecular dynamics/environment and Hessian approaches likely account for this and this mode can similarly be assigned A$_1$. 

Therefore, we find that all spectral densities of interest for the LVC calculations can be well understood in terms of the symmetry decomposition of electronic states and discrete normal modes. The generally positive cross-correlation between the S$_1$ and S$_2$ is likely due to the shared A$_1$ symmetry of a large number of vibronically active modes spread across a large frequency window, as low as 110 cm$^{-1}$ and with strong contributions from high frequency aromatic stretches at 1710 cm$^{-1}$. Asymmetric modes with B$_2$ symmetry couple the electronic states, and a few totally symmetric A$_1$ modes strongly de-correlate the fluctuations between S$_1$ and S$_2$. 

\makeatletter\@input{mainaux.tex}\makeatother

\bibliography{bibliography}